\pdfoutput=1 
\documentclass[twocolumn,aps,pra,nofootinbib]{revtex4-2}
\usepackage[utf8]{inputenc}
\usepackage[margin=1in]{geometry}
\usepackage{amsmath}
\usepackage{amsfonts}
\usepackage[T1]{fontenc}
\usepackage{dsfont}
\usepackage{xcolor}
\definecolor{darkblue}{rgb}{0, 0, 0.8}
\definecolor{rust}{rgb}{0.8, 0.6, 0}
\usepackage{physics}
\usepackage{graphicx}
\usepackage{hyperref}[hypertexnames=false]
\hypersetup{
	colorlinks   = true, 
	urlcolor     = darkblue,
	linkcolor    = darkblue, 
	citecolor   = darkblue 
}
\usepackage{enumitem} 
\usepackage{titlesec} 
\usepackage{booktabs,tabularx}
\usepackage{multirow}
\usepackage{verbatim} 

\makeatletter
\def\maketitle{
	\@author@finish
	\title@column\titleblock@produce
	\suppressfloats[t]}

\renewcommand\onecolumngrid{
	\do@columngrid{one}{\@ne}
	\def\set@footnotewidth{\onecolumngrid}
	\def\footnoterule{\kern-6pt\hrule width 1.5in\kern6pt}%
}

\renewcommand\twocolumngrid{
	\def\footnoterule{
		\dimen@\skip\footins\divide\dimen@\thr@@
		\kern-\dimen@\hrule width.5in\kern\dimen@}
	\do@columngrid{mlt}{\tw@}
}
\makeatother

\titleformat{\section}{\bfseries\centering\uppercase}{\thesection.}{1em}{}
\titlespacing{\section}{0pt}{1em}{1em}
\titlespacing{\subsection}{0pt}{1em}{1em}

\newif\ifhyperrefSM
\hyperrefSMtrue 
\newcommand{\smref}[2]{%
	\ifhyperrefSM
	{\hyperref[#1]{#2}}%
	\else
	{#2}%
	\fi
}

\newcommand{\mainref}[1]{%
	\ifhyperrefSM
	{\ref{#1}}%
	\else
	{\ref*{#1}}%
	\fi
}

\newcommand{\ee}{\end{equation}}
\newcommand{\be}{\begin{equation}}
\newcommand{\p}[1]{\left( #1 \right)}
\newcommand{\br}[1]{\left[#1\right]}

\newcommand{\hc}[1]{#1^\dagger} 
\newcommand{\Hc}{\textrm{Hc}}
\renewcommand\bra[1]{{\langle{#1}|}}
\makeatletter
\renewcommand\ket[1]{%
	\@ifnextchar\bra{\k@t{#1}\!}{\k@t{#1}}%
}
\newcommand\k@t[1]{{|{#1}\rangle}}
\makeatother
\newcommand{\kb}[2]{\ket{#1} \bra{#2}}

\newcommand{\ra}{\rightarrow}
\newcommand{\om}{\omega}
\newcommand{\g}{\gamma}

\newcommand{\overbar}[1]{\mkern 1.5mu\overline{\mkern-1.5mu#1\mkern-1.5mu}\mkern 1.5mu}
\newcommand{\Psucc}{P_\textrm{success}}


\begin{document}
	
\title{
	Success probabilities in time-reversal based hybrid quantum state transfer
} 
\author{Kevin Randles}
\email{krandles@uoregon.edu}
\author{S.~J.~van Enk}
\email{svanenk@uoregon.edu}
\affiliation{Department of Physics and Center for Optical, Molecular and Quantum Science, University of Oregon, Eugene, OR 97403, USA}
	
\begin{abstract}
	We consider two memory nodes of a quantum network connected by flying qubits. We are particularly interested in the case where a flying qubit produced by one node has to be transformed before it can interface efficiently with the next node. 
	Such transformations can be utilized as a key part of the distribution of quantum states and hence entanglement between the nodes of a hybrid quantum network linking together different quantum technologies.
	We show how and why the probability of interfacing successfully is determined by the overlap of the spectral shape of the actual flying qubit and the ideal shape. 
	This allows us to analytically and numerically analyze how the probability of success is impacted by realistic errors, and show the utility of our scheme (in consonance with known error correction methods) in connecting hybrid nodes of a quantum network.
	We focus here on a concrete implementation in which the memory nodes consist of three-level atoms in cavities and the flying qubits are photons.
\end{abstract}

\maketitle

\section{General introduction}	
The quantum internet \cite{kimble2008quantum} is envisioned to allow the implementation of various quantum communication, computation and measurement (sensing) tasks that improve upon their classical counterparts.
Scaling the underlying networks to connect a larger numbers of quantum processor systems, often referred to as nodes in the context of quantum communication, 
improving these processors, and controlling their robustness against noise are among important current practical issues \cite{wehner2018quantum,sadhu2023practical}. The fundamental building blocks of these networks are the nodes themselves and the quantum transmission lines (also called quantum channels) that serve as the network edges. Such an edge links together two nodes so that quantum information, say the state of a qubit, can be sent between them at will. 
These nodes may range from simple devices operating on small numbers of qubits to large-scale quantum computers.
When envisaging a large quantum network, or ultimately the quantum internet, many different implementations of an edge will be needed, e.g., to reliably connect close-by units comprising a single computation node, different types of nodes,  as well as distant nodes.

Two promising research avenues for the scalability of quantum networks are in the development of \emph{distributed} and \emph{hybrid} quantum communication and computation technologies. 
The utility of distributed quantum technologies lies in the likely scenario that it is easier to connect many high-functioning modestly sized, often homogeneous, devices (which have been demonstrated), using quantum effects like entanglement as a resource, than to scale a single device to have the same net processing power \cite{grover1997quantum,cirac1999distributed,serafini2006distributed,jiang2007distributed,monroe2013scaling,beals2013efficient,van2016path}.
A complementary means of scaling quantum networks is by developing a hybrid quantum architecture, where either \cite{hybridNote}
different types of (i) qubits 
or (ii) nodes are connected so that tasks can be delegated so as to leverage the strengths of a given type of qubit or node.

Case (i) can be envisaged as a hybrid device that uses different kinds of qubits for different purposes. For instance, using separate types of qubits for local logic operations and for interfacing with quantum channels, which would allow for  communication between nodes with minimal disruption to local computation or storage processes \cite{young2022architecture,covey2023quantum,nigmatullin2016minimally,inlek2017multispecies,reiserer2016robust,stas2022robust}. 
In case (ii), heterogeneous nodes (or smaller intranodal units), that are based on different quantum technologies, are connected to form the elementary unit of a hybrid quantum network (or node). 
Such hybridization would be valuable in making more powerful large-scale nodes, say a node that integrates solid-state computation units leveraging fast nanosecond gates (at the expense of relatively short microsecond coherence times for both  depolarization and dephasing) \cite{wendin2017quantum,arute2019quantum,xue2022quantum,burkard2023semiconductor} with atom-based memory units with long coherence times on the order of milliseconds (or seconds for ions though they can be significantly longer \cite{wang2021single}, typically limited by dephasing times) at the expense of slower microsecond gate times \cite{brown2016co,bruzewicz2019trapped,morgado2021quantum}
for hybrid quantum computation \cite{ladd2010quantum,linke2017experimental,timesNote}.	
More broadly, realizing such hybrid links would increase the connectivity of quantum networks. Especially as different technologies (trapped ions, superconducting circuits, etc.) become better-established as platforms for qubit implementation,
how to interface them is an important question to address \cite{rabl2006hybrid,wallquist2009hybrid,xiang2013hybrid,kurizki2015quantum,maring2017photonic,scarlino2019coherent,clerk2020hybrid,lauk2020perspectives,awschalom2021development,kumar2023quantum}.

\subsection{Background and scope}\label{subsec:BackgroundAndScope}
The implementation of a specific edge depends on the properties of the nodes it is linking, including their underlying technological implementation, natural energy scale, physical separation, and intended function (say communication or computation).
In this work we focus on discrete (as opposed to continuous \cite{weedbrook2012gaussian}) variable quantum communication in which intermediate `flying qubits' carry quantum information along an edge realized by a guided quantum channel.
Photons are the quintessential flying qubit, and the one we consider here (though phonons could be used, e.g., in some optomechanical systems \cite{hybrid20,stannigel2012optomechanical,zivari2022chip} and other solid-state systems \cite{bienfait2019phonon}), as they can efficiently travel between nodes and they have several degrees of freedom (mode occupation number, polarization, temporal-mode, etc.) that can be used to encode quantum information. 

We focus on a guided channel (e.g., optical fiber \cite{lucamarini2018overcoming,pittaluga2021600,serafini2006distributed,brown2016co} or microwave coaxial cable \cite{kurpiers2018deterministic,axline2018demand,campagne2018deterministic,leung2019deterministic,chang2020remote,zhong2021deterministic,burkhart2021error,casariego2023propagating}) linking close-by nodes, say within a given lab or device, leaving the consideration of nonguided free-space channels to other work \cite{vallone2015experimental,yin2017satellite,chen2021integrated,de2023satellite,gonzalez2022open}. 
The main issue with distant nodes is that the fidelity of states being transferred (and likewise the degree of entanglement generated) typically decreases exponentially with the length of the connecting channel due to photon absorption and noise in the channel (the rate of this decay can be minimized by using telecom light) \cite{radnaev2010quantum,heshami2016quantum,wallucks2020quantum,arenskotter2023telecom}.
In principle, this issue can be solved using a quantum repeater, which itself needs to be able to distribute entanglement between close-by `repeater stations,' though further details are beyond the scope of this work \cite{briegel1998quantum,duan2001long,munro2015inside}. Accordingly, we will not explicitly address the difficulties of connecting distant nodes. Note that the error correction protocols we mention are still relevant for distant nodes though they will have more overhead for larger distances.

\begin{figure}[h!]  
	\includegraphics[width=\linewidth]{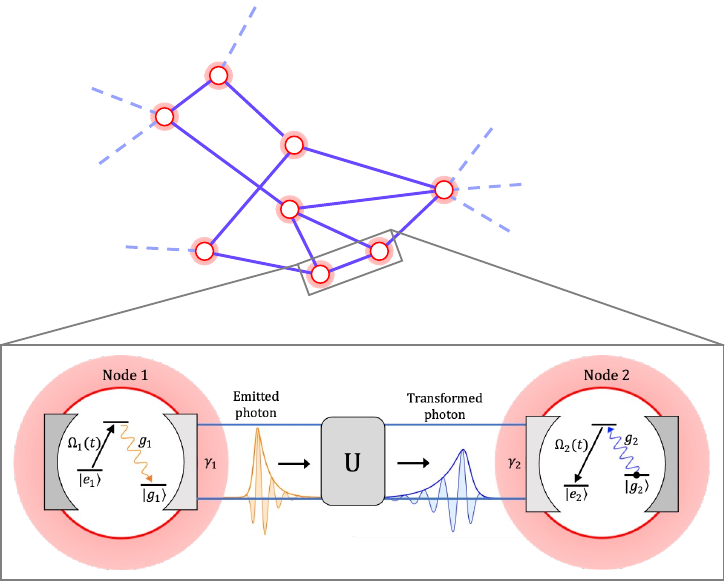}
	\caption{
		Potential subgraph representing part of a quantum network (above) with a zoomed in focus (below) on two nodes (red points) and the quantum channel (blue edges) linking them, along which our QST scheme is implemented (see the main text for details).
		The graph structure is only for illustration purposes, showing how our scheme can fit into the bigger picture of quantum networking.
		The dashed lines indicate potential continuations of the network to other nodes. The lower image is modified from figures of Refs.~\citenum{cirac1997qst} and \citenum{randles2023quantum}.
	}
	\label{fig:networkZoomedIn}
\end{figure}

In this paper we focus on the theoretical implementation of a deterministic quantum state transfer (QST) scheme capable of linking hybrid quantum nodes [in the sense of case (ii) above] via itinerant photons. In particular, we consider connecting two different types of spatially separated nodes, with potentially different spectral properties (resonance frequency and decay width). The state of a qubit prepared at the `sender' node 1 is mapped to the state of an emitted photon wave packet (facilitated by local controls) that is sent to a `receiver' node 2 via a guided channel. To optimally be absorbed at node 2, and hence to map the photonic state to a material qubit state, this wave packet must be modified, tailoring its time-frequency shape \cite{adiabaticPassageNote}.
In our previous work \cite{randles2023quantum} we showed how this can be accomplished by incorporating a unitary transformation, $U$, that time reverses (see SM~\smref{subsec:timeReversal}{A4} for some discussion of why we emphasize time reversal), frequency shifts, and stretches or compresses the intermediate photon wave packet along the quantum channel \cite{shapingNote}.
The original version of this scheme (without the unitary) was proposed in the seminal work of Cirac and co-workers \cite{cirac1997qst}. In the time domain, $U$ is given by
\be\label{eq:UTransTime}
U(t, t') = \sqrt{\xi} e^{i \omega_0 (T-t)} \delta(t' - \xi(T-t)).
\ee 
With the inclusion of this unitary transformation, we showed how to design system controls (laser pulses) that will transfer the state of a qubit at node 1 to one at node 2 even if the nodes have significantly different resonance frequencies and decay rates (provided we can implement $U$ correctly). 
This is especially important in hybrid cases, where without $U$ the likelihood of node 2 absorbing a photon emitted by node 1, even for well-designed controls, is very small. A schematic representation of how such a pair of linked nodes might fit into a larger quantum network is given in Fig.~\ref{fig:networkZoomedIn}. 

In this paper we consider and analyze realistic errors that can occur in the hybrid QST scheme of our previous work \cite{randles2023quantum}, which demonstrated how the scheme works in ideal conditions. We focus on errors in the implementation of the unitary transformation, as other common errors, such as those due to photon absorption in the transmission line, incorrect  laser controls, and cavity loss, are well-known and understood
\cite{monroe1995demonstration,wineland1998experimental,kimble1998strong,van1997ideal,van1998photonic}. This lets us show our protocol's utility in the presence of these realistic errors especially when it is  supplemented by known error correction protocols \cite{van1997ideal,van1998photonic} that can correct these common errors as well as errors in the unitary transformation using local quantum computations with auxiliary quantum emitters at each node (see Sec.~\ref{subsec:errorCorrection}). Nonetheless, error mitigation remains a crucial part of making such error correction protocols viable, specifically to keep the expected number of repetitions of a primitive transfer operation low and moreover to lessen additional overhead.

We also show that the probability of successfully transferring the state from one node to another, $\Psucc$, is determined (rather intuitively) by the overlap between the actual and ideal single-photon wave packets to be incident on node 2 denoted by $\Psi$ and $\Phi$, respectively. Specifically, we find
\be\label{eq:PsuccSynopsis}
\boxed{\Psucc = \left|\braket{\Phi}{\Psi}\right|^2
	= \left|\int_{-\infty}^\infty dt~\Phi^*(t) \Psi(t) \right|^2}
\ee
(see Sec.~\ref{subsec:UParameterErrs} for the corresponding derivation in the particular case we focus on in this work, where each node is comprised of a three-level atom in a cavity and see SM~\smref{sec:reversibleGenerality}{A} for a general argument for this result, which ultimately amounts to the  Born rule). This can be understood by interpreting the successful completion of the protocol as a photodetection event in which the receiving node can be thought of as a photodetector, that (in the case of a ``click'') would project onto a particular single-photon wave packet $\Phi$, which is determined by the parameters of node 2 and the laser driving it. 

\subsection{Outline} 
We outline the model for the interaction of the nodes and present our QST scheme in Sec.~\ref{sec:schemeAndModel}. This includes specifying the corresponding Hamiltonian and equations of motion (EOMs) for the amplitudes of various excitations as well as giving an overview of the standard errors present in such a scheme.
In Sec.~\ref{sec:UnitaryErrors} we consider errors in the unitary transformation. 
In Sec.~\ref{subsec:UParameterErrs} we show how these errors can be described in the context of quantum measurement theory in terms of positive operator-valued measures (POVMs). Then in Sec.~\ref{subsec:optTiming} we show how the ideal value of the unitary transformation’s timing parameter can be determined and in Sec.~\ref{subsec:numericalResults} we perform some further numerical analysis of the probability of successful state transfer in the presence of unitary errors. In Sec.~\ref{subsec:errorCorrection} we highlight how known error correction protocols could be used to reliably perform this state transfer procedure in the presence of realistic errors. 
In Sec.~\ref{subsec:heraldedRelevance} we consider the relevance of our work in heralded, as opposed to deterministic, schemes.
Finally, in Sec.~\ref{sec:Discussion} we discuss the utility of our scheme and provide an outlook of other contexts where it could be employed.
Additional analysis and supporting details regarding the scope of our results are provided in the \smref{SupplementalMaterial}{Supplemental Material} (SM), which is appended to the end of the main text.

\section{Scheme and model}\label{sec:schemeAndModel}
Here we summarize the model used for two linked nodes of a quantum network interacting via a quantum channel. For brevity, we leave much of the corresponding derivations and further details to Refs.~\citenum{gardiner1985input,gardiner1992wave}, the textbook \cite{gardiner2004quantum}, and our previous work \cite{randles2023quantum}.
Starting from the Gardiner-Collett model \cite{gardiner1985input}, one can show that in the quantum trajectory formalism \cite{gardiner1992wave}, the dynamics (during time intervals when no quantum jump occurs) are determined by an effective non-Hermitian Hamiltonian of the form
\be\label{eq:Heff}
H_\textrm{eff} = H_1 + H_2 + H_\textrm{tl}.
\ee
Here $H_j$ is the Hamiltonian for node $j = 1,2$ and $H_\textrm{tl}$ accounts for the nodes' interaction via the \textit{transmission line}.

\subsection{Node implementation}\label{subsec:nodeImp}
Here we consider nodes that can encode the state of a qubit in an effective two-level system with a controllable coupling to a well-defined electromagnetic field mode. 
Such a node is readily realized, at least theoretically, by a three-level $\Lambda$-type atom (or ion) in a high-$Q$ optical cavity \cite{cirac1997qst} and there has been significant recent experimental progress in such systems \cite{covey2023quantum,vogell2017deterministic,krutyanskiy2023entanglement,krutyanskiy2023telecom} (see Sec.~\ref{subsec:ExpCQED}). Accordingly, we focus on this case (both here and in our previous work \cite{randles2023quantum}) as an exemplar of the physics underlying our scheme. The overall scheme applies more generally so our use of `atom' throughout the paper can often be mapped to other material systems such as an ionic, solid-state, or superconducting qubit in analogous setups. These atomic systems along with the rest of our QST scheme are highlighted in Fig.~\ref{fig:networkZoomedIn}.
In this case, at each node the state of a qubit is encoded in the ground states of the atom, $\ket{e}$ and $\ket{g}$, which are coupled via a Raman transition through an auxiliary atomic excited state $\ket{r}$. 
We consider asymmetric cavities that preferentially couple to the transmission line, e.g., by using a partially transmitting mirror to interface  with the transmission line and a (near) perfectly reflective outer mirror.

That is, the dynamics of the first Raman qubit are controlled using a laser pulse that drives a transition from $\ket{e_1}$ to $\ket{r_1}$, which is followed by the transition $\ket{r_1}$ to $\ket{g_1}$ and the emission of a photon into cavity 1, and meanwhile leaves $\ket{g_1}$ undisturbed with no corresponding emission. Thus, if atom 1 was in the `excited' state $\ket{e_1}$, an emitted photon would leak out of cavity 1, propagate down the transmission line along which it would be transformed via $U$, before its incidence on node 2. Then atom 2, which is prepared in $\ket{g_2}$, can simply undergo the Raman process analogous to that undergone by atom 1 but in the backwards order, with the goal of inducing absorption of the photon. 
This is possible as by implementing $U$ we have effectively equalized the spectral properties of the two nodes (analogous to impedance matching).

By linearity, this process can thus be used to implement our QST scheme, where the state of atom 1, $c_g \ket{g_1} + c_e \ket{e_1}$, is transferred to the photon, $c_g \ket{0} + c_e \ket{1}$, and then to atom 2, $c_g \ket{g_2} + c_e \ket{e_2}$. 
Here we take the photonic qubit to be encoded in the occupation number degree of freedom, i.e., either the vacuum $\ket{0}$ or single-photon $\ket{1}$ state in a certain mode \cite{humphreys2018deterministic,stockill2017phase}. Other encodings could potentially be used with appropriate modifications to our scheme. There are of course tradeoffs between different encodings, e.g.,  the polarization encoding $\{\ket{H}, \ket{V}\}$ is more robust in some scenarios \cite{northup2014quantum,reiserer2015cavity} yet the occupation number encoding is necessary for the known error correction protocols we consider (see SM~\smref{subsec:otherEncodings}{A6} and \smref{sec:ECZprotocols}{D} for further discussion).

For these nodes, the effective dynamics in the single (or zero) excitation subspace are ultimately determined by the Hamiltonian ($\hbar = 1$) \cite{cirac1997qst,randles2023quantum}
\be\label{eq:HjSimplified}
H_j = i G_j(t) \p{\hc{a}_j \kb{g_j}{e_j} - a_j \kb{e_j}{g_j} }.
\ee
The operators $\hc{a}_j$ and $a_j$ are the creation and annihilation operators for cavity $j$, respectively. Likewise $\kb{e_j}{g_j}$ and $\kb{g_j}{e_j}$ are the   raising and lowering operators for the effective two-level atom $j$, respectively.
Here $G_j(t) = g_j \Omega_j(t)/2 \Delta_j$ is the Jaynes–Cummings interaction strength between the effective two-level atom and cavity at node $j$, where (as depicted in Fig.~\ref{fig:networkZoomedIn}) $g_j$ is the bare atom cavity coupling, $\Omega_j(t)$ is the user-controlled Rabi frequency envelope of the driving laser, and $\Delta_j = \om_{L j} - \om_{r j}$ is the laser detuning.
Note $G_j(t)$ is real with the implicit time-dependent laser phase, of the form $e^{i \phi_j(t)}$, factored out. 
To obtain this form for $H_j$ we selected specific laser frequencies $\om_{L j}$ to eliminate shifts to the cavity energy states and chose the laser phases $\phi_j(t)$ (the chirps specifically) to compensate for the ac Stark shift to $\ket{e_j}$ (see equation 34 of Ref.~\citenum{randles2023quantum} and the surrounding discussion for details as well as SM~\smref{subsec:generalNodeH}{B1}). Additionally, we assumed far off-resonant driving lasers (to suppress spontaneous emission \cite{brion2007adiabatic}) so that we could adiabatically eliminate the excited states $\ket{r_j}$, and we went into a rotating frame at the frequency of the laser driving each system $\om_{Lj}$ (in which the states $\ket{e}$ and $\ket{g}$ are to be interpreted).

\subsection{Interaction model}
Assuming a vacuum field input, or less strictly that no light is incident on the nodes in relevant modes, the connection of the nodes is ultimately described by \cite{randles2023quantum}
\be\label{eq:Htl}
H_\textrm{tl} = -\frac{i}{2} \p{\g_1 \hc{a}_1 a_1  + \g_2 \hc{a}_2 a_2  
	+ 2 \sqrt{\g_1 \g_2} e^{i \zeta t}\hc{a}_2 \tilde{a}_1},
\ee
which provides an effective description of the transmission line in which 
node 2 is effectively directly coupled to the transformed output of node 1. This is accounted for by the operator $\tilde{a}_1$, which is an annihilation operator for the transformed photon that exits the unitary transformation device. Note that both $a_1$ and $\tilde{a}_1$ are time-delayed operators, which implicitly account for the time delay of light propagating between the nodes.
This description highlights the role of the transmission line as an intermediary for the transfer of excitations from node 1 to node 2 (via the $\hc{a}_2 \tilde{a}_1$ term). Here $\g_j$ is the decay rate for cavity $j$ into the transmission line.
The relative phase of $\zeta = \om_{L2} - \om_{L1}$ in the excitation transfer term corresponds to oscillations at the mismatch in the frequencies of the lasers driving the two systems and is due to going into the aforementioned rotating frames for each node.

Within this model we assume the coupling to be unidirectional with photons only propagating from system 1 to 2 (there is no $a_2 \hc{a}_1$ term in $H_\textrm{tl}$), at least during the QST procedure. Ideally, this unidirectionality should be a consequence of atom 2 being prepared in a stable ground state $\ket{g_2}$, yet it can be physically imposed as necessary, e.g., by using a circulator \cite{unidirectionalNote}. Then $U$ does not affect system 1 dynamics and $a_1$ acts in the Heisenberg picture in a standard way, i.e., on $\ket{C}_1 = c_0 \ket{0}_1 + c_1 \ket{1}_1$ it acts as 
\be\label{eq:a1Action}
{}_1\bra{0}  a_1(t) \ket{C}_1 = c_1(t). 
\ee
Meanwhile, in this effective description, the second system is effectively directly coupled to the \textit{unitarily transformed} output of system 1. This is accounted for by the $\hc{a}_2 \tilde{a}_1$  term of Eq.~(\ref{eq:Htl}), where $\tilde{a}_1$ encodes the effect of $U$, acting as 
\be\label{eq:a1NewAction}
{}_1\bra{0}  \tilde{a}_1(t) \ket{C}_1 = \chi(t) c_1(f(t))
\ee 
in contrast to Eq.~(\ref{eq:a1Action}), where \cite{UPosNote} 
\begin{align}\label{eq:transFieldPrefactor}
	\chi(t) &= \begin{cases}
		0, & t_i < t < t_s \\
		\sqrt{\xi} e^{i \om_0 (T - t)}, & t_s < t < t_f \\
		1, & \textrm{elsewise} 
	\end{cases}
\end{align}
and 
\be\label{eq:fTimeArg}
f(t) = \begin{cases}
	\textrm{undefined}, & t_i < t < t_s \\
	\xi(T-t), & t_s < t < t_f \\
	t, & \textrm{elsewise}
\end{cases}.
\ee
Here $t_s$ is the time at which the transformed field starts to be produced and it is controlled via the relation $T = t_s (1 + 1/\xi)$. These functions are broken up into intervals as the part of the photon wave packet to be transformed, taken to be of duration $t_l = l/c$, must pass through the unitary transformation device (in the time interval $t_i \equiv t_s - t_l < t < t_s$), before the transformed wave packet is produced (in the interval $t_s < t < t_f \equiv t_s +t_l/\xi$). Outside of the time interval where the transformation is happening $\tilde{a}_1$ reduces to the standard $a_1$. The form of these functions for $t_s < t < t_f$ comes directly from Eq.~(\ref{eq:UTransTime}), assuming that the untransformed pulse is blocked at the transformation device during the production of the transformed field. [This assumption is not necessary, but it leads to a simpler description of how the unitary transformations effect can be encoded in the time argument of a fictitious system $\tilde{1}$ that is, effectively, directly driving system 2 (see Ref.~\citenum{randles2023quantum} for further discussion). Furthermore, the distinction of whether we block the original field for $t_s < t < t_f$ does not matter if the two Raman processes are driven by lasers with substantially different frequencies $|\zeta| = |\om_{L2} - \om_{L1}| \gg \g_{1,2}$ (see Sec.~\ref{subsec:optTiming}).] The crucial part is that the transformed part of the field (which should be the entirety of the single-photon wave packet assuming the control parameters are suitably picked and implemented) has the time-reversed and stretched argument $\xi(T-t)$.

\subsection{Model generality}\label{subsec:generalityMention} 
Importantly, our focus on a particular kind of node does not limit the scope of our work, which is meant to concern hybrid links, as the ideas behind our scheme apply to many other analogous controllable systems that could be used for either (or both) node(s). In fact, the Hamiltonian governing the system dynamics we analyze [given in Eqs.~(\ref{eq:Heff}-\ref{eq:Htl})] is quite general for deterministic QST schemes like ours, so the physics should be identical after an appropriate mapping of physical parameters. This is provided the unitary transformation $U$ can be implemented to transduce the intermediate photon between the emitting and receiving nodes energy and time scales. [Note that $U$ has a proposed implementation in the optical regime \cite{timereversal}. In other regimes, say for microwave systems or hybrid cases, such as the coupling of microwave and optical nodes, some aspects of such a transformation have been considered yet to our knowledge the entire unitary we consider has not been (see SM~\smref{subsec:UImp}{A5} for further discussion).]

In particular, we assume that we utilize nodes for which individual excitations can controllably and reversibly be transferred from the material system to a photonic mode (emission) and vice versa (absorption), often at the single-photon level, say mediated by a cavity or resonator \cite{parkins1993synthesis,cirac1997qst,reiserer2015cavity}. 
It should then be possible to recast the Hamiltonians for the nodes themselves into the standard Jaynes–Cummings form [as was done for Eq.~(\ref{eq:HjSimplified})] through the appropriate adiabatic elimination of auxiliary states, selection of unitary transformations to the node Hamiltonian, and tuning of the system and control parameters. One benefit of such coherent and reversible interactions is that you can generate different kinds of target states, e.g., tune the wave packet’s shape, amplitude, and phase (see SM~\smref{sec:wavePacketShape}{B} for further discussion). This is to be contrasted against heralded, probabilistic approaches such as those based on spontaneous emission (see Sec.~\ref{subsec:heraldedRelevance}). 
The form of the Hamiltonian describing the coupling of the nodes, Eq.~(\ref{eq:Htl}), is also common within the context of input-output theory in which one eliminates the channel from the description (via its corresponding continuous mode field operators) to obtain a simpler description where the two nodes are effectively directly coupled (see SM~\smref{subsec:modelGenerality}{A2} for further details and examples).

\subsection{System evolution}\label{subsec:systemEvolution}
Starting from Eq.~(\ref{eq:Heff}), the zero-excitation ground state evolves trivially $\ket{g g} \ket{0 0}  \ra \ket{g g} \ket{0 0}$ as 
\be\label{eq:ggTransfer}
i \frac{d}{dt} \ket{g g} \ket{0 0} = H_\textrm{eff} \ket{g g} \ket{0 0} = 0
\ee
in the interaction picture. 
Meanwhile the dynamics of a state in the single-excitation subspace 
\begin{align}\label{eq:psiSingleEx}
	\ket{\psi(t)} =\; &\alpha_1(t) \ket{e g} \ket{00} + \alpha_2(t) \ket{g e} \ket{00} \nonumber \\
	&+ \beta_1(t) \ket{g g} \ket{10} + \beta_2(t) \ket{g g} \ket{01}
\end{align}
are encoded by $\alpha_j$ and $\beta_j$, which are the state amplitudes for an excitation being in atom $j$ ($\ket{e_j}$) and cavity $j$ ($\ket{1_j}$), respectively. The corresponding amplitude EOMs are
\begin{subequations}\label{eq:coefEOMs} 
	\begin{align}
		\dot\alpha_1 & = -G_1 \beta_1, \label{eq:EOMa} \\
		\dot\beta_1 & = G_1 \alpha_1 - \frac{\gamma_1}{2} \beta_1,  \label{eq:EOMb}\\
		\dot\alpha_2 & = -G_2 \beta_2, \label{eq:EOMc} \\ 
		\dot\beta_2 & = G_2 \alpha_2 - \frac{\gamma_2}{2} \beta_2 - \sqrt{\g_2} e^{i \zeta t} 
		\Psi(t), \label{eq:EOMd}
	\end{align}
\end{subequations}
where 
\begin{align}\label{eq:transformedWP}
	\Psi(t) &:= \sqrt{\g_1} \chi(t) \beta_1(f(t)) \nonumber \\
	&\approx \sqrt{\g_1 \xi} e^{i \om_0 (T - t)} \beta_1(\xi (T - t)),
\end{align}
is the transformed wave packet (as we are working in a rotating frame, it is a slowly varying envelope function \cite{SPicPsiNote}) emitted by system 1 that is driving system 2. The approximation in Eq.~(\ref{eq:transformedWP}) is exact during the transformed field production $t_s < t < t_f$ and hence holds in the large $l$ limit. Note that by tuning the laser frequencies and phases to obtain the simple Jaynes–Cummings type node Hamiltonians of Eq.~(\ref{eq:HjSimplified}) we are fixing $\alpha_1$ and $\beta_1$ to have the same constant phase, so without loss of generality we take them both to be real. This simplified case, where we tune the control parameters so that the system 1 amplitudes are real, is of course not general, though it entirely suffices for our QST scheme and makes our analysis easier. 
It is worth mentioning that one can controllably modulate the phase of the emitted photon wave packet, $\Psi(t) = \sqrt{\g_1} \beta_1(t)$, in time by adjusting the parameters of the driving laser, namely the phase (though this can exacerbate non-Markovian effects, which need more careful treatment \cite{penas2023improving}), and then $\alpha_1$ also acquires a time-dependent phase (see SM~\smref{subsec:generalNodeH}{B1} for more details).

Note that in our setup the transfer $\ket{g_1} \ra \ket{g_2}$ happens by default, see Eq.~(\ref{eq:ggTransfer}), and so the principal goal is to transfer any excitation from the first qubit to the second, $\ket{e_1} \ra \ket{e_2}$.
Thus, in the single-excitation subspace, we want $|\alpha_1(t \ra -\infty)| = 1$ and $|\alpha_2(t \ra \infty)| = 1$; the former condition is satisfied via appropriate local state preparation while the latter requires the transmission process to be implemented correctly. Accordingly, we deem the probability of successful QST to be 
$\Psucc \equiv |\alpha_2(t \ra \infty)|^2$, with the caveat that one need also transfer the relative phase of the qubit states, which requires an interferometrically stable channel (see the discussion at the end of Sec.~\ref{subsec:UParameterErrs} for further details).

\subsection{Simplified treatment}\label{subsec:strategies}
As we have alluded to, we want to highlight the role of the unitary transformation $U$ in our QST scheme. Accordingly, we use several simplifying strategies that allow us to set aside other well known kinds of errors and quantify the impact of $U$ by the single parameter $\Psucc$. In this subsection, we list these strategies and briefly discuss how they can naturally be incorporated into our analysis (see SM~\smref{subsec:standardErrors}{D1} for further details).
These strategies include that:
\begin{itemize}
	\setlength\itemsep{0em}
	\item[1)] emission of the $\Lambda$ system into modes other than the desired cavity mode is treated post hoc,
	\item[2)] cavity field loss into modes other the relevant transmission line mode
	is treated post hoc,
	\item[3)] transmission line loss is treated post hoc,
	\item[4)] we assume the laser pulses (or other control drives) are implemented correctly,
	and
	\item[5)] we do not account for dispersion in the transmission line.
\end{itemize}
Strategies 1) and 2) are worded for the emission process from node 1, yet we likewise (due to time-reversal symmetry) apply them during absorption at node 2. 

The post hoc treatment in strategies 1),  2), and 3) is performed by multiplying our probability of success by the respective survival probabilities $P_1$, $P_2$, and $P_3$ of the photon being transferred to the desired mode during 
emission into (or absorption out of) the cavity (see SM~\smref{subsec:atomicDecay3}{B3}), 
the cavity-channel interactions,
and its propagation through the channel  \cite{vogell2017deterministic,gorshkov2007photon,morin2019deterministic}.
That is, the evolution described in Sec.~\ref{subsec:systemEvolution} is conditioned on attempts where the photon is not lost (absorbed or in an orthogonal mode).
Thus, the modified success probability of our scheme is 
\be\label{eq:tildePsuccGen}
\tilde{P}_\textrm{success} = P_1 \cdot P_2 \cdot P_3 \cdot \Psucc.
\ee
One can naturally separate the first two survival probabilities into  contributions from each of the individual nodes. Namely, letting $\mathcal{P}_\textrm{em-j}$ and $\mathcal{P}_\textrm{cav-j}$ denote the respective probabilities of the emitter and cavity at node $j$ coupling to the desired mode (which are the same for emission and absorption due to time-reversal symmetry), we have
$P_1 = \mathcal{P}_\textrm{em-1} \mathcal{P}_\textrm{em-2}$ 
and 
$P_2 =  \mathcal{P}_\textrm{cav-1} \mathcal{P}_\textrm{cav-2}$.

As an instructive example, here we consider the case where nodes 1 and 2 undergo the same amount of loss.
In particular, we take $\mathcal{P}_\textrm{em-1} = \mathcal{P}_\textrm{em-2} = C_\textrm{em}/(1 + C_\textrm{em})$
and	 
$\mathcal{P}_\textrm{cav-1} = \mathcal{P}_\textrm{cav-2} = C_\textrm{cav}/(1 + C_\textrm{cav})$. 
Here $C_\textrm{em}$ and $C_\textrm{cav}$ are cooperativity parameters that quantify how well the emitters and cavities, respectively, are able to produce photons in the desired output mode. 
Meanwhile, for strategy 3), we consider exponential transmission line loss $P_3 = e^{-x/x_\textrm{tl}}$ with attenuation distance $x_\textrm{tl}$ \cite{etaUNote}. 
(See SM~\smref{subsec:standardErrors}{D1} for further explanation and motivation of these parameters and the corresponding survival probabilities; realistic values are given in Sec.~\ref{subsec:ExpCQED}.)
In this case, we have
\be\label{eq:tildePsuccSym}
\tilde{P}_\textrm{success}
= \Psucc \cdot \p{\frac{C_\textrm{em}}{1 + C_\textrm{em}} \frac{C_\textrm{cav}}{1 + C_\textrm{cav}} }^2 e^{-x/x_\textrm{tl}}. 
\ee

Strategies 4) and 5) are used for simplicity and could be accounted for implicitly via modifications to the actual and target wave packet shapes $\Psi$ and $\Phi$ in $\Psucc$ of Eq.~(\ref{eq:PsuccSynopsis}). Strategy 4) is well-founded in that laser errors in the amplitudes $G_j$ and phases $\phi_j$ are typically negligible compared to the other errors we consider, yet such errors could easily be included in our model and numerics if necessary. 
One point of note that is obfuscated by the effective treatment of Eq.~(\ref{eq:Htl}) is the role of the relative timing of the laser pulses (which is implicitly encoded in the time-delayed operators for node 1). 
A relative timing error of $\delta t$ would result in $\Psucc$ being determined by the overlap of $\Psi(t + \delta t)$, instead of $\Psi(t)$, with the target $\Phi(t)$, degrading the excitation transfer.
Thus, in strategy 4), we additionally assume that no such relative timing errors are made, which demands care, but can be accomplished by characterizing and controlling the photon propagation time through the channel and using a common reference clock for both nodes.
In regards to strategy 5), distortion to the wave packet $\Psi$ induced by channel dispersion can be (partially) compensated for by modifying the control pulses \cite{penas2023improving} as well as the unitary transformation $U$ (see  SM~\smref{subsec:timeReversal}{A4}). However, the comprehensive modeling and treatment of such distortion effects is beyond the scope of our work. We note that channel dispersion poses less of a problem when using short channels and/or small bandwidth (large duration) photons, making strategy 5) more justified in these cases.

\subsection{Realizable parameters in cavity QED experiments}\label{subsec:ExpCQED}
We will now consider realistic values of the parameters $C_\textrm{em}, C_\textrm{cav},$ and $x_\textrm{tl}$ for atom or ion in an optical cavities type nodes. This allows us to quantify the impacts of the standard loss mechanisms that we set aside in the previous subsection.
For transmission line loss we consider two cases that are employed or sought in the literature: 
optical and telecom light, for which $x^\textrm{(opt)}_\textrm{tl} \approx 1.2-1.5$ km  and $x^\textrm{(tele)}_\textrm{tl} \approx 15-25$ km, respectively \cite{attenuationNote}.
For emitter and cavity losses, we consider several references that use nodes similar to our exemplar case and in Table \hyperlink{cooperativityTableAbridged}{\ref*{tab:cooperativityTableAbridged}} we list the corresponding cooperativities and survival probabilities they are (or would be) able to achieve.  
The average cooperativities based on this table are $C^{(\textrm{avg})}_\textrm{em} = 9.0$ and $C^{(\textrm{avg})}_\textrm{cav} = 5.9$, whereas if the maximum and minimum values are disregarded in each case, the averages become $3.6$ and $5.8$, respectively.

We note that the reported ion experiments tend to have lower emitter and cavity cooperativities than the neutral atom experiments. One main reason for this is that small mode volumes are needed to obtain large ion-light couplings $g$, yet
they also result in a stronger disturbance to the trapping field due to stray fields caused by charge build up on the dielectric mirrors in a typical Fabry-Perot type cavity setup.
This can be largely circumvented using a fiber-based Fabry-Perot cavity, in which the two cavity mirrors are each an end face of an optical fiber,
wherein the accumulated charge is distributed over the fibers' dielectric surfaces \cite{takahashi2020strong}. 
In Sec.~\ref{subsec:errorCorrection} we use these realistic $x_\textrm{tl}$ values and the tabulated cooperativity values to inform how implementable our scheme is (in consonance with an error correction protocol) in different parameter regimes (see Fig.~\ref{fig:EnVsPsucc}).

\begin{table}[htp]
	\hypertarget{cooperativityTableAbridged}{} 
	\begin{tabularx}{0.9\linewidth}{ccccccc}
		\toprule
		\multirow{2}*{\begin{tabular}{c}Emitter \\ type \end{tabular} } 
		& \multirow{2}{*}{Ref(s)} & \multirow{2}{*}{$C_\textrm{em}$} & ~$\mathcal{P}_\textrm{em}$~ & \multirow{2}{*}{$C_\textrm{cav}$} & ~$\mathcal{P}_\textrm{cav}$~ & $P_\textrm{tot}$ \\ 
		& & & (\%) & & (\%) & (\%) \\ \midrule
		\multirow{6}*{\begin{tabular}{c}Neutral \\ atom \end{tabular} } 
		& \cite{ritter2012elementary} 
		& 2.8 & 74 & \textcolor{gray}{9.0} & \textcolor{gray}{90} & 44 \\
		& \cite{reiserer2013nondestructive, reiserer2014quantum} 
		& 6.0 & 86 & 11.9 & 92 & 62 \\
		& \cite{chibani2016photon} 
		& 66.7 & 99 & 1.3  & 57 & 31 \\
		& \cite{morin2019deterministic} 
		& 2.9 & 75 & 8.0  & 89 & 44 \\
		& \cite{daiss2021quantum} A 
		& 7.7  & 89 & 11.5  & 92 & 66 \\
		& \cite{daiss2021quantum} B  
		& 6.9  & 87 & 6.0  & 86 & 56 \\
		\midrule
		\multirow{5}{*}{Ion}
		& \cite{steiner2014photon}                  
		& 0.05  & 5 & 0.5  & 32 & 0.02 \\
		& \cite{begley2016optimized,vogell2017deterministic}                
		& 0.3 & 24 & \textcolor{gray}{6.7}  & \textcolor{gray}{87} & 4 \\
		& \cite{krutyanskiy2023entanglement} A
		& 0.8  & 45 & \textcolor{gray}{0.3}  & \textcolor{gray}{20} & 0.8  \\
		& \cite{krutyanskiy2023entanglement}   B    
		& 1.9  & 66 & \textcolor{gray}{3.5}  & \textcolor{gray}{78} & 26 \\ 
		& \cite{takahashi2020strong} & 3.2 &	76 & 0.3 & 20$^*$ & 2 \\
		\bottomrule          
	\end{tabularx}
	\caption{\label{tab:cooperativityTableAbridged}
		Overview of realizable emitter and cavity parameter values based on several recent experimental analyses that consider nodes comprised of a neutral atom or ion coupled to an optical cavity (in experiments that use two such nodes, we list the values for each, labeled A and B). Here we list the emitter and cavity cooperativities ($C_\textrm{em}$ and $C_\textrm{cav}$) as well as the corresponding maximum probabilities that a photon is emitted by the atom or ion into the correct cavity mode $\mathcal{P}_\textrm{em} = C_\textrm{em}/(1 + C_\textrm{em})$
		and then from the cavity into the desired output mode (in the transmission line) 
		$\mathcal{P}_\textrm{cav} = C_\textrm{cav}/(1 + C_\textrm{cav})$. 
		In some references, $C_\textrm{cav}$ was not given and could not be calculated directly, yet an analog to $\mathcal{P}_\textrm{cav}$ is reported. In these cases we infer the values of $C_\textrm{cav}$ by backtracking (indicated via \textcolor{gray}{gray} coloring).
		Here $P_\textrm{tot} = P_1 P_2 = (\mathcal{P}_\textrm{em}  \mathcal{P}_\textrm{cav})^2$ is the combined probability of this full emission process at node 1 and symmetrically of absorption at node 2, assuming both nodes have the same cooperativities. 
		The emitter in each of the neutral atom experiments is a $^{87}$Rb atom. In the trapped ion experiments, Refs.~\citenum{steiner2014photon} and \citenum{begley2016optimized} use a single $^{174}$Yb$^+$ atom and up to 5 $^{40}$Ca$^+$ ions, respectively, while the others \cite{krutyanskiy2023entanglement,takahashi2020strong} use a single $^{40}$Ca$^+$ ion. 
		$*$: The setup of Ref.~\citenum{takahashi2020strong} is not intended to preferentially produce photons out of one mirror. Accordingly, the cavity parameter values reported here appear small (and are excluded from the reported $C^{(\textrm{avg})}_\textrm{cav}$) yet they could readily be increased by modifying the setup.
		An extended version of this table is given in SM~\smref{subsec:standardErrors}{D1}.
	}
\end{table}

\section{Unitary errors}\label{sec:UnitaryErrors}
Here we consider errors in the unitary parameter values of $\om_0$,  $\xi$, and $T$. We focus on cases where the transformation duration, $t_l$, is long enough to transform essentially all of the pulse. That is, we do not analyze errors due to the unitary not being implemented for long enough, which (at least in non-linear optical setups) could be due to not using a long enough medium for the transformation device. 
Errors such as photon absorption or distortion can be corrected for \cite{van1997ideal,van1998photonic}, though the corresponding protocols do not correct for errors due to the undesired production of a photon. Hence we assume there is no other mechanism for atom 2 to absorb an excitation, which is consistent with the assumption of a vacuum field input to system 1 needed to obtain the effective Hamiltonian of Eq.~(\ref{eq:Heff}). In particular, we assume that there are no relevant thermal excitations, which is only valid for systems at low temperatures relative to their operation frequency such that the average thermal occupation number $\overbar{n}_\textrm{th} = 1/(e^{\hbar \omega/k_B T} - 1)$ is nearly zero (it is desirable to have it be far less than 1) \cite{han2021microwave}. 
This is typically a very good approximation for optical systems though systems that operate at lower frequencies need to  be cooled. For instance, at room temperature $T = 293$ K optical light with wavelength $\lambda \approx 700$ nm will have $\overbar{n}_\textrm{th} \approx 3 \cdot 10^{-31}$, whereas microwave light with $\lambda \approx 20$ mm will have $\overbar{n}_\textrm{th} \approx 400$. Hence adequate cooling is crucial in reducing thermal noise and loss experienced by guided microwave channels such as cryogenic microwave links \cite{magnard2020microwave,casariego2023propagating}. 

The ideal values of the frequency and stretching parameters are $\om_0 = \om_{0 i} \equiv \zeta = \om_{L2} - \om_{L1}$, which compensates for the difference in frequencies of the two systems, and $\xi = \xi_i \equiv \g_2/\g_1$, which stretches the wave packet to match the receiving time scale of cavity 2 \cite{randles2023quantum}. 
Unlike for $\om_0$ and $\xi$, there is not a similarly `nice' expression for the ideal value of the timing parameter $T = t_s (1 + 1/\xi)$, which controls the starting time for the transformation $t_s$. However, the goal is simple: for a given $l$ one need simply select a $t_s$ (and hence $T$) such that the largest contributions to $\beta_1(t)$ are transformed. We can be more precise in finding the optimal value of $T$, though we defer this to Sec.~\ref{subsec:optTiming}, where the analysis will be made easier using the machinery we will develop in the following section. For now we simply note that such an optimal $T$, which we will call $T^*$, must exist.

As we are focusing on errors in the unitary parameters in this section, we assume that $G_1$ is implemented correctly such that  $\alpha_1$, $\beta_1$ are as desired (see SM~\smref{sec:wavePacketShape}{B} for a discussion of how an appropriate $\alpha_1$ or $\beta_1$ can be used to determine the corresponding $G_1$) and that \cite{G2FormNote} 
\be\label{eq:G2timeRev}
G_2(t) = \xi_i G_1(\xi_i (T_i - t))
\ee
with $T_i = T|_{\xi = \xi_i}$ 
[see strategy 4) above]. Note that with this choice, the laser phase $\phi_2(t)$ for system 2 is also time reversed and scaled relative to that for system 1, assuming they are implemented in the prescribed way.
Hence the second laser pulse is not affected by an error in the unitaries' values of $\xi$ and $T$. (In fact $T_i$ need not take on its optimal value $T_i^*$ as long as it is consistent in $U$ and $G_2$ and $l$ is long enough for the entire wave packet to be transformed.) One can show that the above choices for the unitary parameters and $G_2$'s relation to $G_1$ are an optimum, in that given solutions $\alpha_1$ and $\beta_1$, there are corresponding solutions for system 2
\be\label{eq:idealA2}
\alpha_2^i(t) = e^{i \om_{0 i} T_i} \alpha_1(\xi_i (T_i - t))
\ee
and 
\be\label{eq:idealB2}
\beta_2^i(t) = -e^{i \om_{0 i} T_i} \beta_1(\xi_i (T_i - t))
\ee 
(assuming $l$ is long enough for the entire wave packet to be transformed), which act in a time-reversed manner relative to their system 1 counterparts. Hence, with the above choices and assumptions, if atom 1 loses an excitation, $\alpha_1$ goes from one to zero, then atom 2 will absorb it, as $|\alpha_2^i|$ goes from zero to one.

\subsection{Unitary parameter errors}\label{subsec:UParameterErrs}
If an excitation is sent from atom 1, we want atom 2 to absorb it and hence to achieve $|\alpha_2(t \ra \infty)| = 1$. Thus, we analyze here the structure of $\alpha_2$'s EOM, which, after eliminating $\beta_2$ in the coupled Eqs.~(\ref{eq:EOMc}--\hyperref[eq:EOMd]{d}), can be found to be
\begin{align}
	\ddot\alpha_2 
	&=  \p{\frac{\dot{G}_2}{G_2} -  \frac{\gamma_2}{2} } \dot\alpha_2
	-G_2^2 \alpha_2 + \sqrt{\g_2} e^{i \zeta t}  G_2 \Psi(t).
\end{align}
Rearranging the terms, it follows that
\be\label{eq:grFcn}
L(t) \alpha_2(t)  =  \Psi(t)
\ee
with 
\be
L(t) := \frac{e^{-i \zeta t}}{ \sqrt{\g_2}   G_2} \br{
	\frac{d^2}{dt^2} - \p{\frac{\dot{G}_2}{G_2} -  \frac{\gamma_2}{2} } \frac{d}{dt} 
	+ G_2^2 
}
\ee
a linear operator (we \textit{assume} it is invertible) and so it has some Green's function $\Gamma^*(t, t')$ \cite{GreenFcnNote} such that 
\be
\alpha_2(t) = \int_{-\infty}^\infty dt'~\Gamma^*(t, t') \Psi(t').
\ee
Here we treat $\Psi(t)$ as some generic (possibly subnormalized)
wave packet of arbitrary shape. This is justified mathematically as, in terms of the differential equation Eq.~(\ref{eq:grFcn}), $\Psi(t)$ is just some non-homogeneous source term,  and the Green's function solution is indifferent to the origin of $\Psi$. Note we cannot design the first laser pulse $G_1$ (even if it is supplemented by our unitary) to produce arbitrary wave packets from system 1 (see SM~\smref{subsec:wavePacketShape5}{B5} for further discussion).

We will now use this Green's function to derive an expression for the limiting value of $\alpha_2$ at some \textit{end} time $t_e$ by which the amplitudes have reached steady values (in practice $t_e$ can be taken to be $+ \infty$ mathematically). Physically, we know that $|\alpha_2| \leq 1$, so with the shorthand $\Gamma(t = t_e, t') = \Gamma_e(t')$, we have
\be\label{eq:al2AmpBnd}
1 \geq |\alpha_2(t_e)|^2 = \left|
\int_{-\infty}^\infty dt'~\Gamma^*_e(t') \Psi(t')
\right|^2
= \left|
\braket{\Gamma_e}{\Psi}
\right|^2,
\ee 
where we are assuming $\Gamma_e$ is well-defined and unique (in practice, this is typically the case with appropriate boundary conditions, and our numerics in Sec.~\ref{subsec:numericalResults} substantiate the validity of the solution we find), yet we do not yet know its norm. [We focus on $\Gamma_e(t')$ because our primary goal is that the excitation is ultimately transferred (as part of the QST scheme), not to know the exact dynamics of the excitations. Accordingly, we do not attempt to compute $\Gamma(t, t')$ for all times $t$, though we do note that it must be causal so  $\Gamma(t, t' > t) = 0$.] 
This is true for arbitrary $\Psi$, which will be normalized (in the single-excitation subspace) unless there are losses, say due to photon absorption.
Thus, we can select
\be
\ket{\Psi} = \frac{\ket{\Gamma_e}}{
	\sqrt{\braket{\Gamma_e}{\Gamma_e}}
}
\ee
(up to a phase) to maximize  $|\alpha_2(t_e)|^2 = \left|\braket{\Gamma_e}{\Psi}\right|^2$ via the Cauchy–Schwarz inequality. With this selection, by Eq.~(\ref{eq:al2AmpBnd}) we have 
\be\label{eq:GammaBnd}
1 \geq |\alpha_2(t_e)|^2 
= \left|\sqrt{\braket{\Gamma_e}{\Gamma_e}}
\right|^2
= \braket{\Gamma_e}{\Gamma_e}	
\ee
so $\ket{\Gamma_e}$ is either a normalized or subnormalized quantum state. 

We know that with ideal parameters $\om_0 = \om_{0 i}$ and $\xi = \xi_i$ (again, for sufficiently long $l$ with appropriate timing $T = T_i$) our unitary will produce the ideal wave packet
\begin{align}\label{eq:PhiIdeal}
	\Phi(t) &= \Psi(t)|_{\om_0 = \om_{0 i}, \xi=\xi_i, T=T_i, l \ra \infty} \nonumber \\
	&= \sqrt{\g_2} e^{i \om_{0 i} (T_i - t)} \beta_1( \xi_i (T_i - t) ),
\end{align}
which evidently is the time reversed, stretched, and frequency shifted counterpart of the wave packet to be emitted by system 1.   
Crucially, note that for properly designed $\alpha_1$ that start at 1 at some early \textit{preparation} time $t_p$, from Eq.~(\ref{eq:idealA2}) we have $|\alpha_2^i(t = t_e)| = |e^{i \om_{0 i} T_i}| = 1$ for $ t_e \geq T_i - t_p/\xi_i$ and hence the corresponding normalized $\Phi(t)$, $\braket{\Phi}{\Phi} = 1$, would lead to perfect absorption at system 2 \cite{alpha2BFSolnNote}. Thus, again using Cauchy–Schwarz, we have that
\begin{align}
	1 &= |\alpha_2^i(t_e)|^2 
	= \left| \braket{\Gamma_e}{\Phi} \right|^2 \nonumber \\
	&\leq \braket{\Gamma_e}{\Gamma_e} 	\braket{\Phi}{\Phi} 
	= \braket{\Gamma_e}{\Gamma_e}  
	\leq 1, 
\end{align}
where we used Eq.~(\ref{eq:GammaBnd}) in the last step. It clearly follows that 
\be
\braket{\Gamma_e}{\Gamma_e}   \equiv 1
\ee 
and hence $\ket{\Gamma_e}$ corresponds to a normalized state vector, and moreover 
\be
\ket{\Gamma_e} = \ket{\Phi} 
\Longleftrightarrow 
\Gamma_e(t) = \Phi(t)
\ee
(again, up to a phase) as the states must be linearly dependent to saturate Cauchy–Schwarz. 

It thus follows that the probability of success is
\begin{align}\label{eq:PsuccDerived}
	\Psucc &\equiv |\alpha_2(t_e)|^2
	= \left|\braket{\Phi}{\Psi}\right|^2 \nonumber \\
	&= \left|
	\int_{-\infty}^\infty dt'~\Phi^*(t') \Psi(t')
	\right|^2
\end{align}
as claimed in Eq.~(\ref{eq:PsuccSynopsis}).
That is, the probability of success is given by the norm squared overlap of the incident photon wave packet $\Psi(t)$ with the ideal wave packet $\Phi(t)$. Thus, if no errors occur, the transformed wave packet $\Psi(t)$ will be equal to the normalized $\Phi(t)$ and hence $\Psucc = 1$.
It follows that system 2 can be thought of as a photodetector that in the case of a click would project on the ideal state via the POVM element
\be\label{eq:purePhiPOVM}
\hat{\Pi}_\textrm{success} =\ketbra{\Phi}{\Phi}.
\ee
Note we would only register such a click (or not) if we appended an atomic measurement of the receiving qubit state. Then, we see that for the pure input state $\rho_\textrm{in} = \ketbra{\Psi}{\Psi}$,
\be \label{eq:PsuccViaPOVM}
\Psucc = \Tr(\rho_\textrm{in} \hat{\Pi}_\textrm{success} )
= \left| \braket{\Phi}{\Psi} \right|^2,
\ee
which matches Eq.~(\ref{eq:PsuccDerived}). Note that this does not assume that $\Psi$ is normalized, it could be subnormalized due to an error at system 1, an error in the unitary, or loss during the transformation. 

More generally, such a POVM would be a weighted sum of projectors (mixed) but here it is a lone projector (pure) as our process is reversible.
For instance, if the system parameters, such as $g_j$, varied due to fluctuations in the position of the atoms or cavities, then averaging over these fluctuations would give a mixed POVM \cite{biswas2021detecting}. 
Another relevant situation would be fluctuations in the unitary transformation parameters ($\om_0$, $\xi$, and $T$), which would result in the input state $\rho_\textrm{in}$ of Eq.~(\ref{eq:PsuccViaPOVM}) being an incoherent mixture of states $\Psi$ with different parameters according to some underlying classical probability distribution. Specifically, if there is classical uncertainty or variation in the unitary transformation parameters, e.g., variation in $\om_0$ due to the finite linewidth of a control laser, then the actual probability of success would be $\Psucc(\om_0, \xi, T)$ averaged over the corresponding classical probability distribution. Note that the ideal wave packet should remain the same as it is purely determined by the parameters of node 2 and the laser driving it, $G_2$.

Note that $\Psucc$ is actually just the probability that an excitation from atom 1 is transferred, via the intermediate photonic degree of freedom, to atom 2. This assumes that no other excitations that can excite atom 2 are produced during the transmission.
We additionally assume that the phase of the initial state of atom 1 is correctly transferred to atom 2, i.e., the channel  needs to be interferometrically stable \cite{humphreys2018deterministic,covey2023quantum,stockill2017phase}. 
This can be accomplished using stable local oscillators as frequency references at each system to establish a common phase reference for both nodes. Then phase stabilization techniques can be used to maintain (and control) the phase induced by the channel, which is easier for shorter channels \cite{ball2016role,randles2023quantum,humphreys2018deterministic,stockill2017phase,valivarthi2016quantum}. We assume such considerations are taken so we can focus on the impact of $U$ rather the examination of such phase errors.
Even if such phase errors do occur, error correction protocols could be used to eliminate them (see Sec.~\ref{subsec:errorCorrection}). We will proceed under these assumptions so that $\Psucc$, as defined in Eq.~(\ref{eq:PsuccDerived}), is a good measure of the success of the entire QST scheme. Note that what `success' ultimately means is up to the particular scheme. For instance, the quantum state fidelity $\mathcal{F}$ may be a more apt measure of successful QST. Here $\mathcal{F}$ is given by the magnitude squared of the overlap of the initial (qubit) state of atom 1 with the final state of atom 2. Notably in the absence of phase errors $\mathcal{F}(|c_e|)$ is bounded below by $\Psucc = |\alpha_2(t_e)|^2$ (see SM~\smref{subsec:fidelity}{A3} for details).

\subsection{Optimal timing}\label{subsec:optTiming}
Now that we have shown this new perspective, where system 2 can be thought of as a single-photon detector, we will derive an expression that the optimal unitary timing parameter $T^*$ must satisfy. 
(Note this is new within the context of this problem; that POVMs play a key role in describing photodetection \cite{lundeen2009tomography, coldenstrodt2009proposed, natarajan2013quantum, van2017photodetector,young2018general, propp2019quantum} and measurements more generally \cite{kraus1983states,jacobs2006straightforward,nielsen2010quantum} is well known.) We compute $T^*$ assuming that the transformation is implemented correctly in all ways except that it has a limited duration $t_l$. Then the probability of success based on the Green's function argument above is 
\be
\Psucc(t_s)  =  \left|\braket{\Phi}{\Phi_l}\right|^2,
\ee
where $\Phi_l(t) = \Psi(t)|_{\om_0 = \om_{0 i}, \xi=\xi_i, T=T_i}$ is the actual transformed wave packet for some finite $l$, assuming the ideal frequency and stretching parameters as well as consistent timing; $\Phi_l(t)$ is equal to $\Phi(t)$ for $t_s < t < t_f$. Maximizing $\Psucc(t_s)$ can thus be accomplished by solving for the $t_s$ (and hence $T_i$) such that
\be\label{eq:tsStar}
\frac{d}{d t_s} \Psucc(t_s) = 0
\ee
(and verifying that the optimum is a maximum). 

Note that
\begin{align}\label{eq:PhiPhilIP} 
	\braket{\Phi}{\Phi_l} = \sqrt{\g_1 \g_2} \int_{-\infty}^\infty dt~ \Big[&e^{-i \zeta (T_i - t)}  \beta_1( \xi_i (T_i - t) ) \nonumber \\
	&\times \chi(t) \beta_1(f(t)) \Big],
\end{align}
where $\chi(t)$ and $f(t)$ characterize the different stages of the transformation and are implicitly evaluated at $\om_0 = \om_{0 i}$, $\xi=\xi_i$, and $T=T_i$ here. The parameters $T_i$, $t_i$, and $t_f$ (which are related to the different stages) all implicitly depend on $t_s$ and so Eq.~(\ref{eq:tsStar}) should be solved numerically in general. However, in typical cases the contribution to Eq.~(\ref{eq:PhiPhilIP}) due to the untransformed portion of the wave packet (i.e., outside the interval $t_s < t < t_f$) will be negligible. This is due to some combination of $l$ being long enough for the $\beta_1$ product in the integrand to be small over the relevant domain and the term being far off-resonant, with $|\om_{0 i}| \gg \g_{1,2}$, such that the phase rapidly oscillates and the integrand averages to zero.
In such a case, effectively none of the untransformed wave packet emitted from system 1 will induce a transition at system 2 and so we have 
\begin{align}
	\braket{\Phi}{\Phi_l}
	&\approx \int_{t_s}^{t_f} dt~|\Phi(t)|^2  \nonumber \\
	&= \g_2 \int_{t_s}^{t_f} dt~\beta_1^2( \xi_i (T_i - t) ) \nonumber \\
	&= \g_1 \int_{t_i}^{t_s} dt~\beta_1^2(t)
\end{align}
(the approximation gets better for large $|\om_{0 i}|$ and/or $l$). As this inner product is real, we can maximize $\Psucc(t_s)$ by finding the $t_s$ such that 
\be\label{eq:simplerIdealTCond}
0 = \frac{d}{d t_s} \int_{t_s - t_l}^{t_s} dt~\beta_1^2(t)
=  \beta_1^2(t_s) -  \beta_1^2(t_s - t_l).
\ee
This is a much simpler condition than in the general case and it can easily be solved for numerically once $G_1$ is specified and hence $\beta_1$ is determined. The corresponding solution is the ideal value of $t_s$, which we will denote by $t_s^* \equiv T_i^*/(1 + 1/\xi_i)$.

\subsection{Numerical results}\label{subsec:numericalResults} 
\begin{figure}[h!]  
	\includegraphics[width=\linewidth]{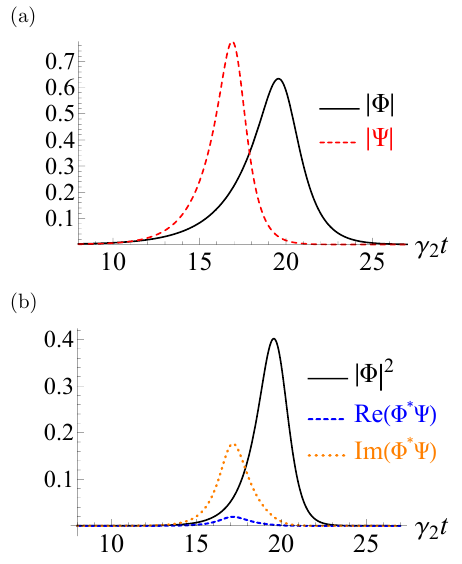}
	\caption{
		(a) Plot of the ideal wave packet modulus $|\Phi|$ for $k=2$, with the unitary parameters taking on their ideal values $\om_0 = \om_{0 i} = 50,$ $\xi=\xi_i=1/2$, $T=T_i^* = 19.7$ for  transformation duration $t_l=10$, and a non-ideal wave packet modulus $|\Psi|$ still with $\om_0 = \om_{0 i}$, but with stretching and timing errors $\xi = 0.75$ and $T = 17$. (All quantities are in units where $\g_2 = 1$.) 
		(b) Plot of the integrand in Eq.~(\ref{eq:PsuccDerived}) for computing $\Psucc$ in the ideal case (solid line) and the real and imaginary parts in the non-ideal case (dashed and dotted lines, respectively). 
		Here, for visualization purposes, we decompose Eq.~(\ref{eq:PsuccDerived}) as $\Psucc = \br{\int dt \Re\p{\Phi^* \Psi}}^2 + \br{\int dt \Im\p{\Phi^* \Psi}}^2 = 0.186$, whereas $\Psucc = \int dt |\Phi|^2 = 1$ for the ideal wave packet (which assumes the entirety of the wave packet emitted from node 1 is transformed).	
	}
	\label{fig:PhiAndPsiWavePackets}
\end{figure}

To makes plots illustrating errors in the different unitary parameters, we must specify the first laser pulse $G_1$. A natural case to consider is the $G_1$ such that the amplitude for atom 1,  $\alpha_1$, logistically decreases  from 1 to 0 
\be\label{eq:logA1}
\alpha_1(t) = \frac{1 + \tanh(-k t)}{2}.
\ee
Note, for any monotonically decreasing $\alpha_1$, one can compute the corresponding laser pulse $G_1$ that would generate it [see Eq.~(\smref{eq:G1inTermsOfAlpha1}{B7}) of  SM~\smref{subsec:wavePacketShape2}{B2}].
Then the amplitude for cavity 1 can be determined as $\beta_1 = -\dot\alpha_1/G_1$, which in turn gives the exact form of the ideal emitted photon wave packet $\Phi$ of Eq.~(\ref{eq:PhiIdeal}). (See SM~\smref{subsec:wavePacketShape2}{B2} for $\beta_1$ in this logistic $\alpha_1$ case.) We can thus compute how errors in the unitary transformation parameters, i.e., incorrect values of $\om_0$, $\xi$, and $T$, degrade the transfer resulting in a decreased $\Psucc$.
The impact of such errors on $\Psucc$ can be seen graphically as in Fig.~\ref{fig:PhiAndPsiWavePackets}, where we compare a wave packet $\Psi$ due to a unitary transformation with stretching and timing errors to the corresponding ideal wave packet $\Phi$. 

\begin{figure}[ht!]  
	\includegraphics[width=\linewidth]{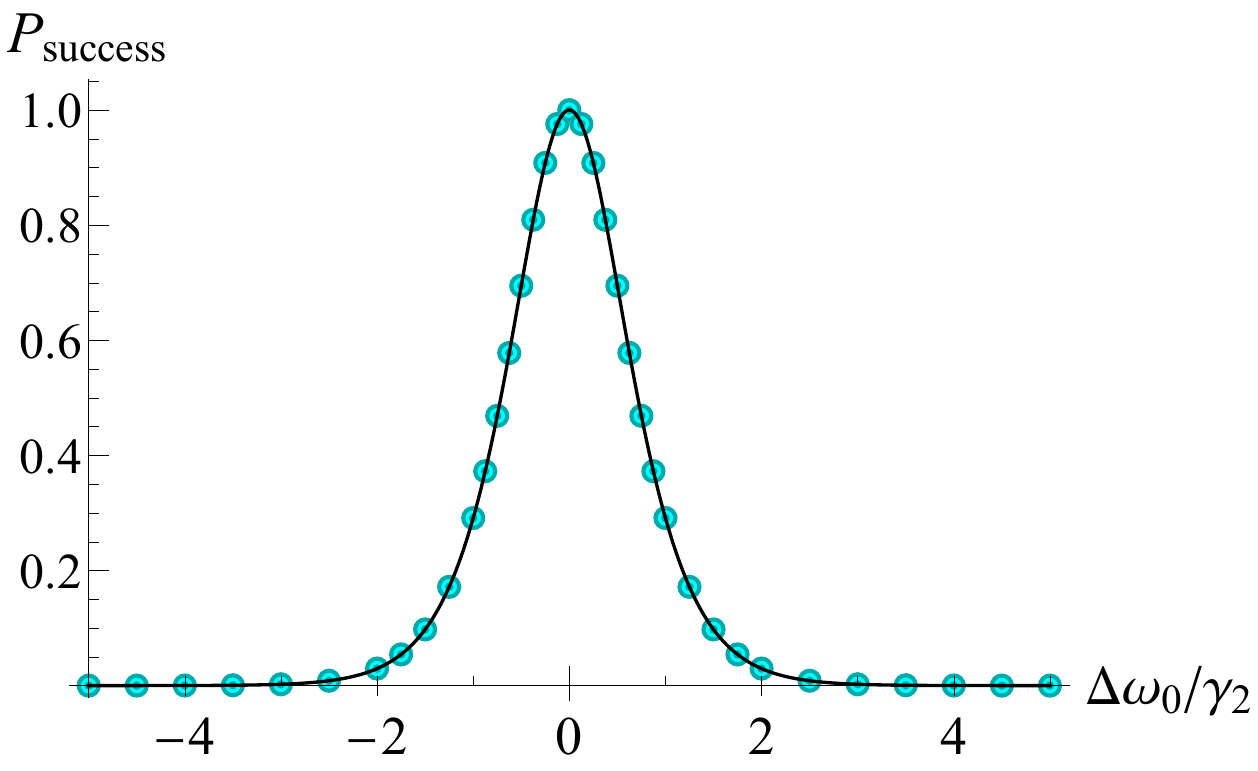}
	\caption{
		Plot of $\Psucc$ as a function of the detuning $\Delta\om_0$ assuming the other unitary parameters are ideal. We take the underlying parameter values to be the same as in Fig.~\ref{fig:PhiAndPsiWavePackets} (that is, $k=2$, $\om_{0 i} = 50$, $\xi_i = 1/2$, and $t_l=10$ in units where $\g_2 =1$). The probability of success quickly falls off away from resonance, $\Delta\om_0 = 0$. The solid line is from a state overlap calculation using Eq.~(\ref{eq:PsuccDerived}) and the overlaid points are from directly numerically solving coupled EOMs for the state amplitudes (see main text).
	}
	\label{fig:PsuccVsOm0}
\end{figure}

\begin{figure}[ht!]  
	\includegraphics[width=\linewidth]{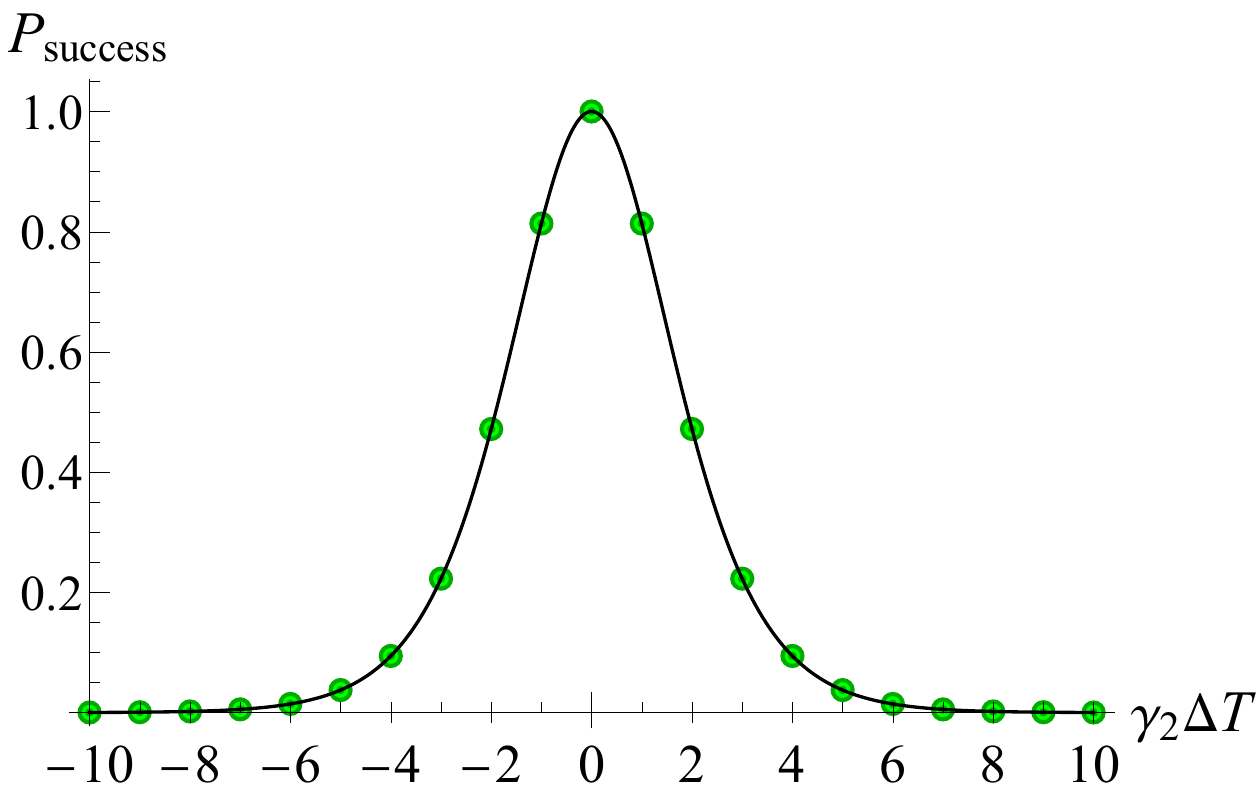}
	\caption{
		Plot of $\Psucc$ versus $\Delta T$ assuming the other unitary parameters are ideal (with the same underlying physical parameters as the previous figures). The solid line is from an overlap calculation and the overlaid points are from solving coupled amplitude EOMs (see main text).
	}
	\label{fig:PsuccVsT}
\end{figure}

\begin{figure*}[ht]  
	\includegraphics[width=0.9\linewidth]{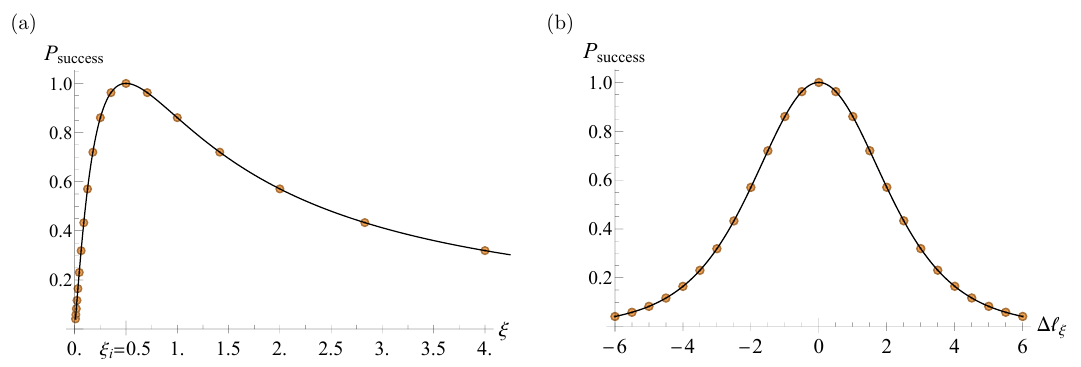}
	\caption{Plots of $\Psucc$ versus $\xi$ on (a) a linear scale and (b) a logarithmic scale with $\Delta \ell_\xi = \log_2{\xi/\xi_i}$, assuming the other unitary parameters are ideal (with the same underlying physical parameters as the previous figures). The solid line is from an overlap calculation and the overlaid points are from solving coupled amplitude EOMs (see main text).}
	\label{fig:PsuccVsXi}
\end{figure*}

Furthermore, we can compute how $\Psucc$ varies as a function of the amount of error in the various transformation parameters. For instance, we illustrate the effect of errors in the unitary parameters $\om_0$, $T$, and $\xi$ by plotting $\Psucc$ versus one of these parameters assuming the other parameters are ideal in Figs.~\ref{fig:PsuccVsOm0}, \ref{fig:PsuccVsT}, and \ref{fig:PsuccVsXi}, respectively. We work with a set of shifted `error variables' that are centered at zero: $\Delta\om_0 \equiv \om_0 - \om_{0 i}$, $\Delta \ell_\xi \equiv \log_2{\xi} - \log_2{\xi_i}$, and $\Delta T \equiv T - T_i^*$.
We consider the logarithm of $\xi$ in most of our plots as its ideal value is a ratio of two decay widths,  $\xi_i = \g_2/\g_1$, so errors in $\xi$ should scale multiplicatively, e.g., doubling $\xi$ (with respect to its ideal value) should be (about) as bad as halving it.
In each of these plots (Figs.~\ref{fig:PsuccVsOm0}-\ref{fig:PsuccVsXi}), the solid black lines are a cubic interpolation between 201 points with the independent variable's values distributed evenly over the intervals shown. The corresponding value of $\Psucc$ is given by the overlap calculation $|\braket{\Phi}{\Psi}|^2$. The overlaid colored points are calculated by numerically solving the coupled ordinary differential equations (ODEs) of Eqs.~(\ref{eq:EOMa}-\hyperref[eq:EOMd]{d}) for various values of the independent variable (with larger separations because the relevant numerics are more computationally expensive).

For concreteness we focus on the specific case of a logistic $\alpha_1$ with $k=2$ and the physical parameters $\om_{0 i} = 50$ and $\xi_i = 1/2$. Additionally, we assume a long transformation length $l=10$, for which $T^* = 19.7$, such that in the absence of errors in the unitary, effectively all of the wave packet would be transformed with  $\Psucc= 0.999995 \approx 1$.  Here all quantities are given in natural units in which $\g_2 = c =1$. Importantly, these underlying parameter values are not crucial as, at least for large $l$ as we have here, only the error variables (made dimensionless with appropriate factors of $\g_2$) and the wave packet shape (as effectively controlled by $\alpha_1$) matter (see SM~\smref{sec:sepIndex}{C}). Hence the results gleaned from this specific case apply more generally. 
The largest discrepancy in calculating $\Psucc$ between the original coupled ODEs solution method and the POVM wave packet overlap method is $1.1 \cdot 10^{-6}$, which we assume to be numerical error (in particular, we see that this specific error value goes down if we increase our error tolerance when solving the ODEs). Hence this numerical comparison serves as a strong indicator of the validity of the POVM based results we found.

The probability of success quickly falls off away from its peak value near unity as $\om_0$, $T$, and $\xi$ are shifted from their ideal values. 
The narrow peak around $\om_0 = \om_{0 i}$ in Fig.~\ref{fig:PsuccVsOm0}, which has a full width at half maximum (FWHM) of $1.4 \g_2$, illustrates the importance of frequency conversion. We note that the shape and width of the $\Psucc$ versus $\om_0$ curve do not change appreciably as the central frequency $|\om_{0 i}|$ is further increased. 
This is because, as alluded to above, only the detuning of the transformed wave packet from resonance with system 2 relative to $\g_2$, $|\Delta \om_0|/\g_2$, significantly matters during the transformation. Moreover, for large $|\om_{0 i}|$ the untransformed portions of the wave packet will not induce an excitation in atom 2. Thus, even when $|\om_{0 i}|$ is very large, e.g., $|\om_{0 i}|/\g_2 \sim 10^6$ is typical, the FWHM will be stable. Here it remains at $1.4 \g_2$ (for this logistic $\alpha_1$ with $k=2 \g_2 = \g_1$), and hence it is critical to control against frequency errors. Note that wave packets with narrower temporal shapes are less susceptible to such frequency errors, though this will be limited by emitting node parameters (see SM~\smref{subsec:formalBeta1FreqError4}{B4}).

The couplings $\gamma_j$ tend to be on the order of kHz-MHz, and are typically smaller in the microwave regime as compared to the optical. 
Hence, typical values of $\g_j$ can vary by 2 to 3 orders of magnitude between different systems and so $\xi_i = \g_2/\g_1 \sim 100 -1000$ is reasonable for hybrid interconnects (assuming $\g_2 > \g_1$, otherwise $\xi_i$ would be the reciprocal of this).
Experiments across different platforms have demonstrated the ability to shape photon wave packets, which includes stretching and compression by these orders of magnitude \cite{lavoie2013spectral,andrews2015quantum,allgaier2017highly,morin2019deterministic}. Note that the results for different $\xi_i$ can be mapped between one another because, as mentioned above, $\Psucc$ is predominantly determined by the error variables, $\Delta\ell_\xi$ in this case.
Importantly, the narrow width of $\Psucc$ as a function of $\Delta\om_0/\g_2$ does not doom us as sub-kHz-MHz level precision in frequencies (necessary to obtain $|\Delta\om_0| < \g_2$) is possible using tunable, narrow-linewidth lasers \cite{chan2004tunable,brown2006facility,tran2019ring,corato2023widely}, mid-infrared and terahertz laser sources based on difference frequency generation \cite{chen2007continuous,belkin2007terahertz}, and for many other standard lasers. For instance, in an optical setup with atoms or ions (see Ref.~\citenum{krutyanskiy2023entanglement} as a good example) 
typical laser linewidths and cavity jitters are on the order of $10-100$ kHz, meanwhile $\g_j$ are around $1-10$ MHz (see Table \hyperlink{extendedParameterTable}{ \ref*{tab:extendedParameterTable}} in the SM) so one could reasonably achieve $|\Delta\om_0|/\g_2$ of $10^{-2}$ to $10^{-3}$, for which $\Psucc > 99\%$ in Fig.~\ref{fig:PsuccVsOm0}.

We can quantify how much two of the dimensionless error variables $\{\Delta\om_0/\g_2, \Delta \ell_\xi, \g_2 \Delta T\}$ depend on one another by computing their ``index of separability'' $\mathcal{S}$ \cite{sepIndexNote}.
For an $m \times n$ matrix $A$ we define $\mathcal{S}$ in terms of its singular values $\{ \sigma_i(A) \}$ with maximum $\sigma_{\max}(A)$ as
\be\label{eq:indexofSep}
\mathcal{S}(A) := \frac{\sigma_{\max}^2( A )}{\sum_i \sigma_i^2( A )},
\ee
which is the square of the ratio of the induced 2-norm and Frobenius norm of $A$. 
This index of separability is bounded as $0 < 1/\min\{m, n\} \leq \mathcal{S}(A) \leq 1$, where the maximum $\mathcal{S}(A) = 1$ entails that $A$ is separable, i.e., can be written as an outer product of two vectors, and the minimum $\mathcal{S}(A) = 1/\min\{m, n\}$ (which is nearly zero for large matrices) entails $A$ is full rank with all equal singular values. 
However, practically the smallest observed values will be much larger than this ($\sim{0.75}$ here), which comes from comparing to random matrices (see SM~\smref{sec:sepIndex}{C}).

\begin{figure*}[ht!] 
	\includegraphics[width=\linewidth]{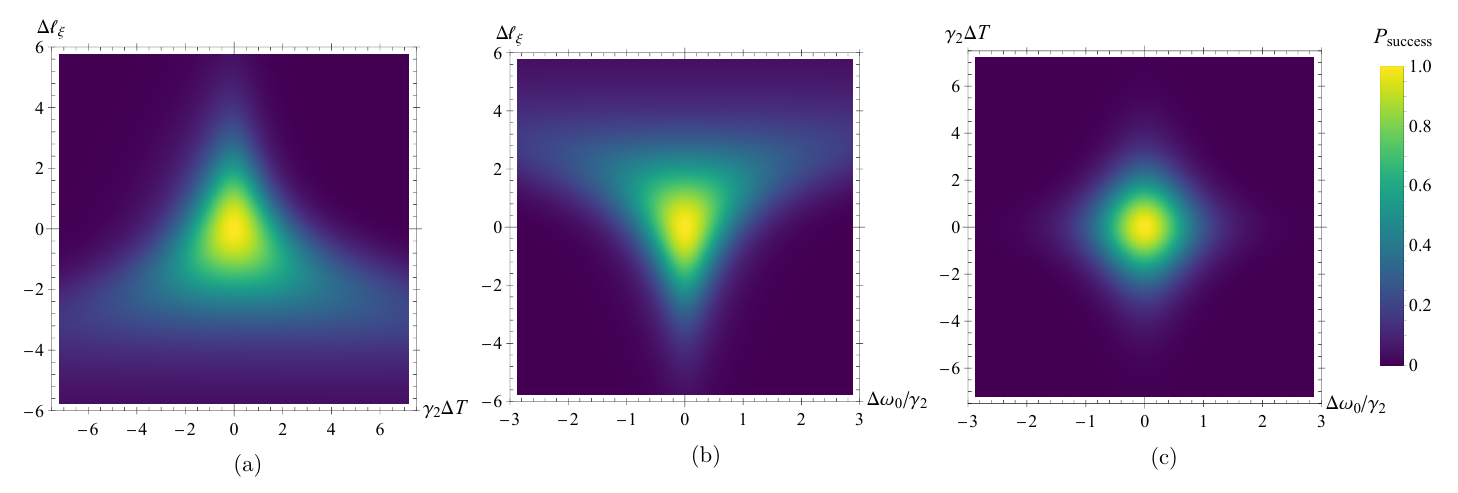}
	\caption{
		Density plot of $\Psucc$ as a function the error variable pairs for the transformation parameters (a) $T$ and $\xi$, (b) $\om_0$ and $\xi$, and (c) $\om_0$ and $T$ assuming the other unitary parameter is ideal (with the same underlying physical parameters as the previous figures). Each plot is computed as the cubic interpolation of a 61 by 61 grid of points that are evenly spaced over the region shown and is colored according to the legend. 
	}
	\label{fig:PsuccVsXiandT}
\end{figure*}

For instance, with the same underlying physical parameters as in Figs.~\ref{fig:PhiAndPsiWavePackets}--\ref{fig:PsuccVsXi}, we take the joint probability distribution for $\g_2 \Delta T$ and $\Delta\ell_\xi$  assuming $\om_0$ takes on its ideal value, $\Psucc(\Delta T, \Delta\ell_\xi)|_{\om_0 = \om_{0 i}}$, compute it on a grid of equally spaced values for $\Delta T$ and $\Delta\ell_\xi$ yielding a matrix, and then compute its index of separability to be 
\be\label{eq:STxi}
S_{T, \xi} \equiv \mathcal{S}[\Psucc(\Delta T,  \Delta \ell_\xi)|_{\om_0 = \om_{0 i}}] 
= 0.87. 
\ee
Similarly, for the other two variable pairs (keeping the third variable at its ideal value) we find
\be
S_{\om_0, \xi} \equiv \mathcal{S}[\Psucc(\Delta\om_0, \Delta \ell_\xi)|_{T = T^*}] 
= 0.87
\ee 
and
\be\label{eq:Som0T}
S_{\om_0, T} \equiv \mathcal{S}[\Psucc(\Delta\om_0, \Delta T)|_{\xi = \xi_i}] 
=  0.998. 
\ee 
[Each of these $\mathcal{S}$ values is computed on a 121 by 121 grid (matrix) with the same spacings as described in the Fig.~\ref{fig:PsuccVsXiandT} caption over a rectangular region twice as large in each direction as those depicted. 
These reported values ultimately serve as upper bounds for the separability index for large grids. Corresponding lower bounds can be calculated using the zero-mean counterparts of the $\Psucc$ matrices used here, yielding respective $\mathcal{S}$ values of $0.80$, $0.80$, and $0.97$ compared to Eqs.~(\ref{eq:STxi})--(\ref{eq:Som0T}). (See SM~\smref{sec:sepIndex}{C} for additional method details.)]

Hence  $\om_0$ and $T$ errors are largely independent of one another
(as the corresponding probability distribution is almost separable, $\mathcal{S} \approx 1$), whereas errors in $\om_0$ and $T$ are distinctly dependent on what $\xi$ error occurs. The corresponding joint probability distributions are given in  Fig.~\ref{fig:PsuccVsXiandT} and can be used to get visual intuition for the index of separability. For instance, Fig.~\ref{fig:PsuccVsXiandT} (a) illustrates that errors in $T$ and $\xi$ are dependent on each other, which can intuitively be explained as wave packet timing errors in $T$ will reduce the overlap, yet there will be relatively more overlap if one also elongates the wave packet in the time domain by choosing $\xi < \xi_i$ ($\Delta \ell_\xi < 0$).
This analysis accounts for dependencies of the errors on one another that are intrinsic to our model. We leave considerations of other error dependencies that may be due to a particular implementation of the transformation $U$ to other work.

\subsection{Error correction}\label{subsec:errorCorrection}
We have highlighted many things can go wrong in the implementation of the unitary itself, which is in addition to standard errors due to incorrect driving laser pulses, incorrect laser frequencies, fluctuating system parameters, as well as photon absorption or other losses in the transmission line, cavities, material qubits, or transformation device. With all of these possibilities, the presence of errors on a single attempt of a QST scheme like ours is inescapable, though crucially we can correct for these errors.
Accordingly, we will now discuss the applicability of a few error correction protocols to our scheme as well as their limitations. 

\textbf{ECZ protocols.} By using the occupation number encoding our scheme could naturally be improved by utilizing one of two relatively low overhead error correction protocols 	\cite{van1997ideal,van1998photonic} that capitalize on the feasibility of suppressing photon production errors $\ket{0} \ra \ket{1}$. We refer to these as the `ECZ protocols' after their authors'  last names.
These protocols can correct for photon loss and phase errors, which are recurrent in standard transmission lines, e.g., optical fibers and waveguides. They do so using auxiliary quantum emitters, again envisaged as atoms in a cavity, that can encode the state of a qubit at each node and serve as a form of redundancy. They cannot correct for errors caused by the creation of photons in the relevant mode, which is often reasonable as the production of photons can be suppressed in a well-isolated setup operating at an appropriate temperature (such that there is low thermal occupation, $\overbar{n}_\textrm{th} \ll 1$, which as previously discussed is more of a problem in microwave-based systems).

Each ECZ protocol is based on repeating a certain primitive transfer operation, which we will refer to as a trial, that includes 1 or 2 QST attempts (typically 2), local single-qubit and two-qubit entanglement gate operations, and measurements of auxiliary qubits or states. These local gate and measurement operations are assumed to be implemented perfectly, though in practice these operations will also be error prone and hence limit how close to unity the ultimate transfer fidelity can be.
In the earlier protocol \cite{van1997ideal}, measurements are performed on each trial, a successful outcome of which validates the state transfer. In the later protocol \cite{van1998photonic}, a given trial serves as a `purification' step to be iterated so that a target state is approached exponentially, at a rate that gets faster for smaller net errors in the protocol, and hence can be reached up to some error threshold in fidelity.
(See  SM~\smref{sec:ECZprotocols}{D} for further discussion of these protocols.)

Both ECZ protocols consider an effective channel in which on a single transmission attempt (which forms an integral part of both protocols) the initial state will evolve for large times (so that no excitations remain in the cavities) according to a map of the form \cite{channelMapNote}
\be\label{eq:HeffMap}
\begin{matrix}
	\ket{g_1} \ket{g_2} \\
	\ket{e_1} \ket{g_2}
\end{matrix}
\longrightarrow
\begin{matrix}
	\alpha \ket{g_1} \ket{g_2}\\ 
	\beta \ket{g_1} \ket{e_2} + \Upsilon_1 \ket{g_1} \ket{g_2} + \Upsilon_2 \ket{e_1} \ket{g_2}
\end{matrix},
\ee
where $\alpha, \beta$, and $\Upsilon_{1,2}$ are constants.
These are the form of the long time limit of the solutions to EOMs analogous to Eq.~(\ref{eq:coefEOMs}) with $|\beta| = |\alpha_2(t_e)|$ and $|\Upsilon_2| = |\alpha_1(t_e)|$, but with additional possible errors included.
Importantly for our scheme, \textit{an error in the unitary transformation falls under this class of channel} as it will contribute to photon loss. Chiefly, an incorrectly shaped incident wave packet will not be absorbed by node 2 and hence will be directed out a different spatial mode \cite{unidirectionalNote}. This is in addition to losses that are in intrinsic to the transformation device. Such errors can be accounted for via the amplitudes $\beta$ and $\Upsilon_1$.
Thus, our scheme can naturally be incorporated as an extended version of the transmission steps in the ECZ protocols, which can aptly  be utilized in our context for reliable QST between hybrid nodes.

\begin{figure}[h]  
	\includegraphics[width=\linewidth]{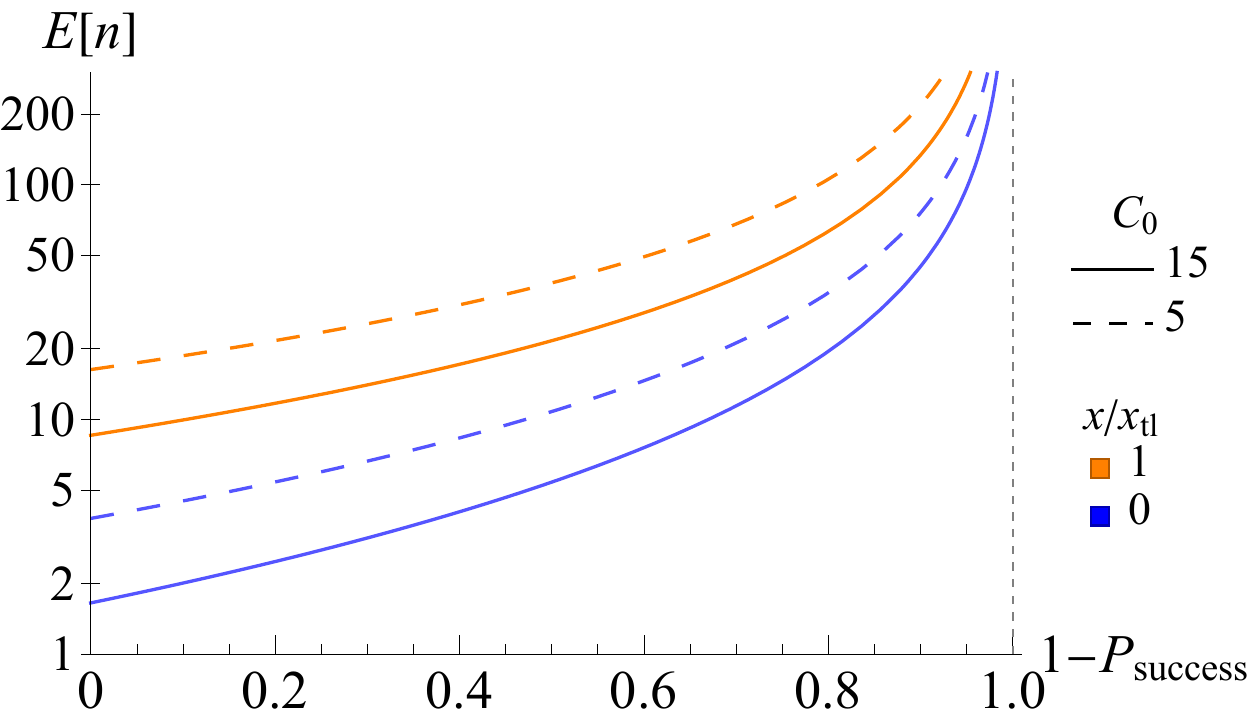}
	\caption{
		Plot of the expected number of ECZ  \cite{van1997ideal} error correction trials, $E[n]$ from Eq.~(\ref{eq:WorstEnForUnitAl}), on a log scale versus the error in the itinerant photon's temporal-frequency shape, $1 - \Psucc$. This is done for realistic cooperativities informed by the values listed in Table \protect\hyperlink{cooperativityTableAbridged}{\ref*{tab:cooperativityTableAbridged}}. Here for simplicity we take $C_\textrm{em} = C_\textrm{cav} = C_0$,
		as in both cases currently achievable cooperativity values are around $C_0 = 5$ (dashed lines), while $C_0 = 15$ (solid lines) is more optimistic yet plausible. 
		These plotted cases could equivalently represent other cooperativity pairs with the same respective combined probabilities $P_\textrm{tot} = P_1 P_2$ of $48\%$ and $78\%$.
		For each  $C_0$ value, we consider two cases: no transmission loss ($x/x_\textrm{tl} = 0$, $P_3 = 1$) and transmission loss over one attenuation distance ($x/x_\textrm{tl} = 1$, $P_3 = 1/e \approx 37\%$), which are shown in the lower (blue) and upper (orange) lines, respectively, for a given line type (solid or dashed). 
		See Sec.~\ref{subsec:ExpCQED} for standard values of $x_\textrm{tl}$ for optical and telecom light in fiber.
		As $\Psucc$ gets small, tending towards zero ($\varepsilon \ra 1$), the expected number of trials quickly grows, diverging like $1/\Psucc$ to leading order (indicated via the dashed vertical asymptote).	
		For instance, $E[n]$ reaches $10$ when $\varepsilon \approx 0.75$, which in the above cases with cooperativities of 15 and 5 corresponds to bare $\Psucc$ values of about $33\%$ and $53\%$, respectively, for $x/x_\textrm{tl} = 0$.
	}
	\label{fig:EnVsPsucc}
\end{figure}

We acknowledge that such a protocol would slow down the rate of entanglement generation due to the time it takes to repeat the primitive transfer scheme as well as to perform the requisite local computations. Importantly, however, if we can achieve reasonably small error probabilities in the necessary operations the slowdown in rate can be manageable.
We will now illustrate this by considering the expected number of repetitions $E[n]$ of the earlier ECZ protocol (until success) \cite{van1997ideal} in the case where $|\alpha| = 1$, as we have previously assumed in our scheme, yet its phase could be nontrivial due to a relative phase error
(see SM~\smref{subsec:ECZ97Summary}{D2}). Consider an error on a given transmission of $|\Upsilon_1|^2 + |\Upsilon_2|^2 = \varepsilon$ such that $|\beta|^2 = 1 - \varepsilon$ by normalization. In this model the worst case, that with the largest $E[n]$, is when we are dominated by $\Upsilon_1$ errors (i.e., $|\Upsilon_1|^2 = \varepsilon$ and  $\Upsilon_2 = 0$). In this worst case we find
\be\label{eq:WorstEnForUnitAl}
E[n] =  \frac{4}{ (1 - \varepsilon) (2 - \varepsilon)^2 },
\ee
which starts at 1 for $\varepsilon = 0$, corresponding to only needing a single trial without errors, and monotonically increases as a function of $\varepsilon \in [0, 1]$, tending towards $+ \infty$ as $\varepsilon \ra 1$ as the state transfer entirely fails for $\beta = 0$ (see SM~\smref{sec:ECZprotocols}{D} for the general derivation and further details). 
Note in fact we expect to be in this regime, $|\Upsilon_1| \gg |\Upsilon_2|$, with photon loss and unitary transformation errors being dominant such that
\be\label{eq:photonLossDomErr}
\varepsilon \approx |\Upsilon_1|^2 \approx 1 - \tilde{P}_\textrm{success},
\ee
see Eq.~(\ref{eq:tildePsuccGen}).
In Fig.~\ref{fig:EnVsPsucc} we use Eqs.~(\ref{eq:WorstEnForUnitAl}) and (\ref{eq:photonLossDomErr}) to show how $E[n]$ depends on the bare $\Psucc$ value with the inclusion of realistic standard errors (informed by the  parameter values of Table \hyperlink{cooperativityTableAbridged}{\ref*{tab:cooperativityTableAbridged}}).
Knowing the dependence of $E[n]$ on $\Psucc$ for experimentally achievable parameter regimes can inform how much unitary transformation loss or error (quantified by $\Psucc$) can be tolerated in a given implementation.
Importantly, we find that if small unitary transformation error probabilities, $1 - \Psucc$, can be achieved, then few trials (protocol repetitions) will typically be needed when linking nodes with good (yet realizable) cooperativities linked by relatively short channels (e.g., with lengths around an order of magnitude smaller than the attenuation distance $x_\textrm{tl}$ or less).

\textbf{Alternate protocols and their scope.} 
Depending on the achievable $E[n]$, the magnitude of errors in the local operations needed for the ECZ protocols, and other implementation details,
the employment of another correction protocol or even a nondeterministic heralded approach may be appropriate (see Sec.~\ref{subsec:heraldedRelevance}).
There are several alternate error correction protocols that can be used for deterministic quantum communication via photons. 
For consistency, here we focus on protocols that use the mode occupation number encoding (see SM~\smref{subsec:otherEncodings}{A6} for a discussion of other encodings). Such protocols can be realized using multi-photon states as a type of redundancy, e.g., using so-called  binomial  \cite{michael2016new} or cat \cite{burkhart2021error} codes, or using photonic graph states (such as cluster states) generated from a single emitter \cite{lindner2009proposal,zhan2020deterministic,thomas2022efficient}.
Each of these alternate protocols comes with their own technical challenges, especially in hybrid interfacing contexts like ours where we would need to transform all the itinerant photons in a logical state preserving way. Moreover, both the sender and receiver would need to be able to interface with these multiphoton logical states. For instance, such binomial codes are realizable with superconducting circuit technology \cite{axline2018demand} but are more challenging for other implementations with less control of the requisite multiphoton states. Similarly, only a few emitters, including quantum dots and atoms in cavities, have demonstrated viability in the generation of such photonic graph states. Accordingly, one advantage of the ECZ protocols, especially for hybrid interconnects, is their relative simplicity as the generation of single photon states is a standard, feasible task for a wide variety of node implementations.

To make such protocols (including those of ECZ) applicable in the NISQ (Noisy Intermediate-Scale Quantum \cite{preskill2018quantum}) era with error prone operations on small numbers of qubits one has to compromise between the potential noise reduction of a given error correction protocol versus the realistic `cost' of its implementation in a given deterministic QST proposal or experiment. This cost includes additional resources such as local operations and controls as well as auxiliary emitters that can simultaneously be coupled at each node, which all come with their own potential errors as well as slower rates. 
For instance, when adapted to our hybrid case, the earlier ECZ protocol \cite{van1997ideal} is contingent on the errors in the unitary parameters, laser pulses, and system parameters [specifically $\alpha$ and $\beta$ in Eq.~(\ref{eq:HeffMap})] being consistent between subsequent transmissions in a given trial. If this systematic condition breaks down, then the later ECZ protocol \cite{van1998photonic} should potentially be used as it can iteratively purify more general random errors in the channel even with potentially non-Markovian decoherence. However, this more comprehensive nature of the later protocol is at the expense of additional overhead, in the form of additional local operations, especially for large errors, so a compromise must be made. 

\subsection{Relevance in heralded schemes}\label{subsec:heraldedRelevance}
At this point, it is useful to contrast our deterministic scheme against  commonly employed heralded schemes for QST and/or remote entanglement generation.
Here we give the basic idea of such heralded schemes and discuss how the ideas underlying our work can be applied to them.
Such a heralded scheme may be appropriate if the ECZ or alternate error correction protocols discussed above for a deterministic scheme are not realizable (see SM~\smref{subsec:heraldedProtocols}{A7} for some discussion of the pros and cons of deterministic versus heralded schemes).
In many such schemes  \cite{craddock2019quantum,krutyanskiy2023entanglement,dyckovsky2012analysis,pompili2021realization,stockill2017phase,humphreys2018deterministic}, two nodes each emit a photon that encodes the state of a qubit in one of its degrees of freedom (typically using a polarization encoding). Said photonic qubits states are (ideally) maximally entangled to their emitter's qubit state. Then a photonic Bell-state measurement is implemented in which the two photons interfere at a 50:50 beam splitter and a subsequent coincident detection of photons with orthogonal qubit states ideally heralds the creation of a remote entangled state between the matter qubits at each node, otherwise the procedure is repeated until success. 

The quality of the generated remote entanglement is ultimately determined by how indistinguishable the two photons are, whose states we denote by $\ket{\psi_A}$ and $\ket{\psi_B}$.
Namely, the fidelity of the remote entangled state relative to a target maximally entangled state is determined by the mode overlap of the two photons $\mathcal{C} \equiv \braket{\psi_A}{\psi_B}$ as \cite{craddock2019quantum} 
\be
\mathcal{F} =  \frac{1 + |\mathcal{C}|^2}{2}
\ee
(note this ignores detector background counts and imperfect emitter-photon entanglement, which will further degrade $\mathcal{F}$ \cite{krutyanskiy2023entanglement}).
The key physics behind this is that for indistinguishable photons, $|\mathcal{C}| = 1$, the beam splitter removes which-path information leading to the entanglement swapping  from the emitter-photon pairs to the emitters with unit fidelity (after an appropriate heralded measurement).
Meanwhile, the interference of distinguishable photons at the beam splitter results in the addition of a classical mixture to the otherwise entangled (heralded) joint state of the emitter qubits. This combination preserves the joint state populations yet reduces the coherences, thereby lowering the state fidelity (i.e., it is a two-qubit dephasing channel). 

Thus, having nearly indistinguishable photons as inputs to the photonic Bell-state measurement of such a heralded scheme is crucial for obtaining high fidelity remote entanglement. 
This can be connected to our work in that a transformation like our $U$ (but without time reversal) could be used to make the photons emitted by heterogeneous nodes nearly indistinguishable and thus to achieve heralded hybrid remote entanglement generation (or QST via an additional quantum teleportation step).
Moreover, the analysis of the errors in such a transformation, to be employed in a heralded scheme, carries over almost directly from our work with $|\mathcal{C}|^2$ effectively replacing $\Psucc$ as a figure of merit for the success of the scheme. 

\section{Discussion}\label{sec:Discussion} 
In this paper we have demonstrated theoretically how unitary transformations to the temporal-spectral mode of a photon serving as a flying qubit between hybrid quantum nodes can be used to drastically improve the probability of successfully transferring quantum information between the nodes. We showed that the probability of transferring an excitation from one node to another, which is a good measure of successful quantum state transfer, is given by the modulus squared of the overlap of the spectral shapes of the actual and ideal photon wave packets.
Doing so makes the role of the unitary transformation apparent: it should transform the emitted photon to one with this ideal spectral shape (which can be calculated). Importantly, our scheme applies quite generally to any nodes in which controlled quantum light-matter interaction can be realized to reversibly transfer the state of a material qubit to a photonic degree of freedom and back, not just for the three-level atom in a cavity type nodes we focus on.

We analyzed the impact of errors in the implementation of the unitary transformation. This includes showing how the success of the protocol depends on the deviations from the ideal parameters as well as quantifying how much these parameter errors depend on one another. This analysis, along with our considerations of more standard errors, can be used to determine what kinds of errors dominate in a given physical setup, say to form an error budget. Furthermore, we discussed how our scheme can naturally be incorporated with and aid known error correction protocols to significantly suppress or potentially eliminate unavoidable errors in deterministic quantum state transfer, even between hybrid systems. Such an error-corrected adaptation of our scheme could be used to distribute entanglement in a quantum network or for distributed quantum computing.
As a next step, our methods could be applied to other material systems and the unitary transformation's physical implementation could be further considered, especially for cases outside of the optical regime.

\bibliographystyle{unsrt}
\bibliography{nextPaper_parameterErrorAnalysis}

\begin{thebibliography}{100}
\bibitem{hybridNote}
We limit our usage of `hybrid quantum' to these cases, though the phrase is
  used more broadly, e.g., to indicate hybrid classical-quantum technologies or
  hybrid continuous-discrete variable systems. Alternatively, our use of
  `hybrid' in discussing differing types of systems and nodes of a quantum
  network could be replaced by `heterogeneous' to be contrasted with
  homogeneous implementations connecting like systems.
\bibitem{etaUNote}
The probability of photon loss due to the unitary transformation itself, e.g.,
  due to absorption or scattering at $U$ or if $U$ is intrinsically
  probabilistic, is implicitly accounted for by the potential subnormalization
  of the wave packet incident on node 2, $\Psi$. For a given setup this can be
  made more explicit if an underlying transformation device efficiency $0 \leq
  \eta_U \leq 1$ can be characterized, then one can simply map $P_3 \ra \eta_U
  P_3$.
\bibitem{GreenFcnNote}
We choose to define the complex conjugated Green's function, $\Gamma^*(t, t')$,
  so that Eq.~(\ref{eq:al2AmpBnd}) takes a more standard form for a complex
  inner product.
\bibitem{alpha2BFSolnNote}
This whole argument applies more generally as long as the EOM for the analog of
  $\alpha_2$ is linear and there is an ideal solution with $|\alpha_2^i(t_e)| =
  1$.
\end{thebibliography}

\newpage\null\newpage
\title{ Supplemental Material for ``Success probabilities in time-reversal based hybrid quantum state transfer''} 
\maketitle
\phantomsection\label{SupplementalMaterial}

\setcounter{section}{0}
\setcounter{footnote}{0}
\renewcommand{\thepage}{SM~\arabic{page}}
\setcounter{page}{1}
\subsection*{Contents}
\noindent \hyperref[sec:reversibleGenerality]{A.~Scope of results} \hfill \pageref{sec:reversibleGenerality} \\
\hspace*{1em} \hyperref[subsec:PsuccGenerality]{1.~Generality of $P_\textrm{success}$} \hfill \pageref{subsec:PsuccGenerality} \\
\hspace*{1em} \hyperref[subsec:modelGenerality]{2.~Generality of model Hamiltonian} \hfill \pageref{subsec:modelGenerality} \\
\hspace*{1em} \hyperref[subsec:fidelity]{3.~Fidelity as a measure of success} \hfill \pageref{subsec:fidelity} \\
\hspace*{1em} \hyperref[subsec:timeReversal]{4.~Utility of time reversal} \hfill \pageref{subsec:timeReversal} \\
\hspace*{1em} \hyperref[subsec:UImp]{5.~Unitary implementations} \hfill \pageref{subsec:UImp} \\
\hspace*{1em} \hyperref[subsec:otherEncodings]{6.~Photonic qubit encodings} \hfill \pageref{subsec:otherEncodings} \\
\hspace*{1em} \hyperref[subsec:heraldedProtocols]{7.~Heralded protocol comparison} \hfill \pageref{subsec:heraldedProtocols} \\

\vfill\break

\noindent \hyperref[sec:wavePacketShape]{B.~Determining the wave packet} \hfill \pageref{sec:wavePacketShape} \\
\hspace*{1em} \hyperref[subsec:generalNodeH]{1.~Node Hamiltonians} \hfill \pageref{subsec:generalNodeH} \\
\hspace*{1em} \hyperref[subsec:wavePacketShape2]{2.~Wave packet for a given $\alpha_1$} \hfill \pageref{subsec:wavePacketShape2} \\
\hspace*{1em} \hyperref[subsec:atomicDecay3]{3.~Impact of spontaneous decay} \hfill \pageref{subsec:atomicDecay3} \\
\hspace*{1em} \hyperref[subsec:formalBeta1FreqError4]{4.~Formal $\beta_1$  and frequency errors} \hfill \pageref{subsec:formalBeta1FreqError4} \\
\hspace*{1em} \hyperref[subsec:wavePacketShape5]{5.~Producible wave packets} \hfill \pageref{subsec:wavePacketShape5}  

\vspace*{1mm} 

\noindent \hyperref[sec:sepIndex]{C.~Index of Separability} \hfill \pageref{sec:sepIndex} 

\vspace*{1mm} 

\noindent \hyperref[sec:ECZprotocols]{D.~Error analysis and correction} \hfill \pageref{sec:ECZprotocols} \\
\hspace*{1em} \hyperref[subsec:standardErrors]{1.~Accounting for standard errors} \hfill \pageref{subsec:standardErrors} \\
\hspace*{1em} \hyperref[subsec:ECZ97Summary]{2.~First ECZ protocol summary} \hfill \pageref{subsec:ECZ97Summary} \\
\hspace*{1em} \hyperref[subsec:errorCorrectionScope]{3.~Scope of ECZ protocols} \hfill \pageref{subsec:errorCorrectionScope} 

\onecolumngrid

\vspace{0.5em}
\noindent\makebox[\linewidth]{\rule{0.8\paperwidth}{0.4pt}}
\vspace{0.5em}

Here we provide additional discussion, calculations, and background to supplement the main text. We label the equations presented here by prepending their section number (A--D), thus distinguishing them from equations in the main text, which we will refer to directly. We do not change the figure nor table numbering from the main text, nor do we restart their numbering.

\setcounter{equation}{0}
\renewcommand{\theequation}{A\thesection\arabic{equation}}
\renewcommand{\theHequation}{A\thesection\arabic{equation}} 
\phantomsection
\section*{A.~Scope of results}\label{sec:reversibleGenerality}
The atom in cavity type node we focus on in the main text is meant to be an elucidating example, which should be contrasted against other implementations of controlled light-matter interaction of individual quanta. As mentioned in Sec.~\mainref{subsec:generalityMention}, our general theory aims to capture many of such implementations, which we will further substantiate here. 
In this section we discuss the applicability of our work beyond such atom in a cavity type nodes, the impact of various choices we made in the QST scheme, and relevant alternatives to said choices.
This includes a more general argument for the probability of successful excitation transfer formula derived in the main text, $\Psucc$, in Sec.~\hyperref[subsec:PsuccGenerality]{A1} and moreover a discussion of why our model Hamiltonians themselves are quite general in Sec.~\hyperref[subsec:modelGenerality]{A2}.
In Sec.~\hyperref[subsec:fidelity]{A3} we discuss the quantum fidelity of the initial and final qubit states as a different measure of a successful QST scheme than $\Psucc$.
In Sec.~\hyperref[subsec:timeReversal]{A4} we discuss the impact of including time reversal as part of the unitary transformation to the itinerant single photons.
In Sec.~\hyperref[subsec:UImp]{A5} we highlight relevant implementations of the proposed unitary transformation.
In Sec.~\hyperref[subsec:otherEncodings]{A6} we discuss the impact of using a mode occupation number encoding for the photonic qubits and contrast it to other commonly used encodings, especially polarization. Finally, in Sec.~\hyperref[subsec:heraldedProtocols]{A7} we contrast our deterministic scheme to heralded implementations, showing that heralded schemes, especially hybrid ones, could benefit from (or be made possible by) a unitary photon transformation similar to the one we consider.

\phantomsection
\subsection*{1.~Generality of $\boldsymbol{P_\textrm{success}}$}\label{subsec:PsuccGenerality}
To illustrate the wide-reaching applicability of our scheme for quantum networking and communications we will now show how and why Eq.~(\mainref{eq:PsuccSynopsis}) of the main text applies quite generally, under modest assumptions, for nodes implemented using other systems including those just mentioned. 
To see this we interpret the receiving node to be a binary-outcome photodetector (the simplest possibility), which either ``clicks'' if the incident light is in a certain small fraction of all possible modes or does not click, corresponding to a null response where there was either no light or the incident light was in a mode the detector was not sensitive to. 
[Note that this is just a thought experiment as in our actual QST scheme it is crucial that we do not actually perform a measurement (which would cause the state to collapse out of the desired superposition state). In particular, we do not check the measurement outcome, which could be done by appending an atomic measurement of the state of qubit at node 2 (doing so would turn our QST scheme into an actual measurement).]

As is contemporarily standard (see the clarifying note in Sec.~\mainref{subsec:optTiming}), such a quantum measurement can be mathematically represented as a positive operator-valued measure (POVM), which is a set $\{ \hat{\Pi}_n \}$, where one can associate a POVM element $\hat{\Pi}_n$ (a positive Hermitian measurement operator) with each of the possible measurement outcomes labeled by $n$. The fact that upon measurement, one of the possible outcomes must occur is encapsulated by the completeness relation $\sum_n \hat{\Pi}_n = \mathds{1}$, where the sum runs over all $n$ and $\mathds{1}$ is the identity operator on the Hilbert space associated with the system being measured.	
Here the outcomes are ``click'' or ``no click'' with corresponding POVM elements 
\be
\hat{\Pi}_\textrm{click} = \sum_i w_i \ket{\phi_i}\bra{\phi_i}
\ee
(this form results from diagonalizing the POVM element, which is possible as it is positive and Hermitian),
where the sum runs over an orthonormal set of modes $\{ \ket{\phi_i} \}$ with weights $0 \leq w_i \leq 1$,
and 
$\hat{\Pi}_\textrm{no-click} = \mathds{1} - \hat{\Pi}_\textrm{click}$, respectively.
The probability of a click can then be expressed as
\be\label{eq:Pclick}
P_\textrm{click} = \Tr(\rho_\textrm{in} \hat{\Pi}_\textrm{click}). 
\ee 

In our case of single-photon detection, time reversibility guarantees that (under ideal conditions\footnote{ 
	These ideal conditions include that, as previously assumed, no 
	thermal or other unexpected excitations will interact with the receiving node, that the system parameters are fixed (do not fluctuate), that there are no dark counts, which would be represented by an additional term in the click POVM element 
	that projects onto the vacuum, 
	and moreover we do not consider multiple incident photons in the detection time.
	These are realistic assumptions in situations like ours, where the single photon excitation is mapped onto a material qubit. This is to be contrasted against standard photodetectors that are destructive and hence irreversible. Accordingly, we do not consider them in this work.
}) 
our effective photodetector (the receiving node) will project onto one single-photon wave packet. That is, the POVM element (and corresponding ideal wave packet) is pure (rank one).
This is simply because the absorption process is reversible, it is time reversed and complex conjugated relative to the pure (under the aforementioned ideal conditions) emission process \cite{cirac1997qst,gorshkov2007photon}.
Thus, here with the deterministic (and reversible) production of single-photon wave packets, we have the pure POVM element
\be
\hat{\Pi}_{\textrm{click}} = \ket{\Phi}\bra{\Phi},
\ee
where $\ket{\Phi}$ is one of the orthornomal basis vectors.
Inserting this into Eq.~(\ref{eq:Pclick}) for a pure input state $\rho_\textrm{in} = \ketbra{\Psi}{\Psi}$ we obtain Eq.~(\mainref{eq:PsuccSynopsis}).
Note that this click outcome corresponds to the successful transfer of an excitation, which is why we call it $\Psucc$ in the main text, though additional considerations (e.g., the use of local oscillators as phase references) are needed to ensure that the phase is transferred. 

We can define a probability of success for a premature detection time $t_p \leq t_d \leq t_e$ as
\be
\Psucc(t_d) := 
\left|\int_{t_p}^{t_d} dt~ \Phi^*(t) \Psi(t) \right|^2,
\ee
which is zero at the preparation time $t= t_p$ and approaches the probability of success $\Psucc(t_e)$ as $t_d$ increases towards $t_e$. 
This can be thought of as the overlap of our state $\Psi$ with the ideal state up until time $t_d$, that is, with the subnormalized state $\Phi_{t_d}(t') := \Phi(t') \Theta(t_d - t')$ with $\Theta$ the Heaviside step function. Thus we can consider the corresponding (unnormalized) POVM element
\be\label{eq:PiSucct}
\hat{\Pi}_\textrm{success}(t_d) = \ketbra{\Phi_{t_d}}{\Phi_{t_d}}.
\ee
Equivalently, we can factor out the subnormalization and write the above POVM element as
\be\label{eq:PiSucct2}
\hat{\Pi}_\textrm{success}(t_d) = W(t_d) \ketbra{\tilde\Phi_{t_d}}{\tilde\Phi_{t_d}}
\ee
with
\be
W(t) := \int_{t_p}^t dt'~ |\Phi(t')|^2 = \int_{t_p}^\infty dt'~ |\Phi(t') \Theta(t - t')|^2 = \braket{\Phi_t}{\Phi_t} 
\ee 
interpreted as an efficiency $0 \leq W \leq 1$ and $\ket{\tilde\Phi_t} = W^{-1/2}(t) \ket{\Phi_t}$ a normalized single-photon state \cite{propp2020project}. 
Importantly, no matter how we write the success POVM element describing a click, the effective photodetector is (and in fact must be\footnote{
	Just as a POVM element of a general measurement never depends on the input state on which the measurement is performed.
}) independent of the actual incident wave packet $\Psi$ and only depends on the detection time $t_d$ and on the measurement parameters (those of the receiving system and the controls used to manipulate it), which determine the ideal wave packet $\Phi$.
However, we cannot specify the form of the ideal wave packet $\Phi(t)$ without further system details, namely, the particular form of the effective Hamiltonian describing our scheme, which can be obtained by focusing on specific kinds of nodes. In this work we focus on nodes comprised of a three-level atom in an optical cavity, though, as we will next discuss, the corresponding Jaynes–Cummings interaction Hamiltonian is quite general.  
This allows us to explicitly determine the ideal wave packet $\ket{\Phi}$ in the temporal mode $\Phi(t)$ as in Sec.~\mainref{subsec:UParameterErrs}.

\phantomsection
\subsection*{2.~Generality of model Hamiltonian}\label{subsec:modelGenerality}
As we have discussed, the broad scope of Eq.~(\mainref{eq:PsuccSynopsis}) is quite intuitive. Potentially
more surprising (at least at first thought) is that our Hamiltonian itself $H_\textrm{eff}$, Eq.~(\mainref{eq:Heff}), is quite general. 
To obtain $H_\textrm{eff}$ one starts by considering two spatially separated systems (nodes) interacting via an intermediate channel that hosts a quasi-one-dimensional electromagnetic field \cite{gardiner1992wave} for which the only propagating degree of freedom of light in the channel is the longitudinal mode (the polarization and two transverse spatial modes are fixed, 
though this can straightforwardly be extended). Then, by employing the standard rotating-wave approximation and a Markov approximation (detailed in \cite{gardiner1992wave}) that includes assuming a flat coupling\footnote{
	\label{symbolNotationFootnote1}Our use of the symbols $\g$ and $\kappa$ here and in Ref.~\citenum{randles2023quantum} follows the notation of Refs.~\citenum{gardiner1992wave} and \citenum{gardiner1993driving}.
} $\kappa_j(\om) = \sqrt{\g_j/2\pi}$ within a narrow bandwidth of positive frequencies, one can eliminate the continuum mode bosonic bath operators of the channel, 
as is standard in input-output theory \cite{gardiner1992wave,gardiner1993driving,carmichael1993quantum}, to obtain the Heisenberg EOMs for system operators. 
The corresponding dynamics of the state of both systems in the Schr\"{o}dinger picture can then be described by a Lindblad master equation or equivalently in the quantum trajectory formalism \cite{dalibard1992wave,gardiner1992wave,molmer93} 
in which a given pure state trajectory of the effectively directly coupled systems is induced by an effective Hamiltonian.
This Hamiltonian will be of the same three part form as $H_\textrm{eff}$ of Eq.~(\mainref{eq:Heff}) (see Appendix I of Ref.~\citenum{randles2023quantum}), where $H_{1,2}$ are the node Hamiltonians and $H_\textrm{tl}$ accounts for their interaction [it takes on the specific form of Eq.~(\mainref{eq:Htl}) assuming a unidirectional coupling \cite{unidirectionalNote} and a vacuum field input to system 1; the phase $\zeta$ depends on what rotating frames we go into for each node]. 

In our QST scheme and often in quantum networking, it is desirable to use nodes in which some internal state of a material encoding a qubit can be reversibly converted into a quantum excitation (often a single photon). Hence, almost by construction, in the appropriate parameter regime(s), each node must be describable by a standard Jaynes–Cummings-like interaction, i.e., via a Hamiltonian of the general form 
\be\label{eq:genericNodeHam}
H_{\textrm{a}\sigma} = g(t) \hc{a} \sigma^- +  g^*(t) a \sigma^+
\ee  
in the interaction picture, under the rotating wave approximation (as was already assumed to obtain $H_\textrm{tl}$ via input-output theory). Here $a$ and $\sigma^-$ 
are annihilation operators ($\hc{a}$ and $\sigma^+$ are their Hermitian conjugate creation operators) for a photonic and material excitation (the cavity and atom in our case), respectively, and $g(t)$ is a time-dependent coupling between them.
This Hamiltonian describes a system in which excitations with similar energies are transferred between light and matter. When an excitation is removed from the material, a photon is produced, $\hc{a} \sigma^-$, (though other excitations of light could potentially be used, say in a two-photon transition) or vice versa during absorption, $a \sigma^+$.
This form assumes we are in the single-excitation regime as we focus on material dynamics in the two-level subspace encoding the state of the qubit, $\{\ket{g}, \ket{e} \}$ here.
As emphasized by Ref.~\citenum{kimble2008quantum}, for quantum networking applications it is desirable for the light-matter coupling $g(t)$ to be ‘user controlled’ so that one can tune its amplitude and phase (say using lasers driving the system) so as to deterministically induce emission and absorption of photon wave packets of particular shapes.
Our effective node Hamiltonians Eq.~(\mainref{eq:HjSimplified}), obtained in the case where each node is comprised of an atom in cavity, are of this same form with $\sigma^- = \kb{g}{e}$ and $g(t) = i G(t)$, though the phase can be tuned more generally, see Sec.~\hyperref[subsec:generalNodeH]{B1}.

The generality of the Jaynes–Cummings (JC)  model beyond systems of atoms or molecules in a cavity (our setting here) was not obvious from its conception \cite{jaynes1963comparison}.
In fact it took decades for the strong coupling regime, which is necessary to realize JC physics, to be reached in cavity QED experiments \cite{meschede1985one,kimble1998strong}. 
Since then, additional groundbreaking cavity QED experiments have demonstrated the ability to engineer highly-controllable interactions of individual photons with a small number of quantum emitters (single atoms interacting with single photons in our case) with the corresponding weak light-matter interaction enhanced by a high-$Q$ cavity. Such advances are relevant for our scheme, which is a modified and extended version of a seminal cavity QED based QST protocol \cite{cirac1997qst}, and for quantum networking more generally. See Refs.~\citenum{reiserer2015cavity} and \citenum{reiserer2022colloquium} for relevant reviews (we point to Refs.~\citenum{boozer2007reversible,ritter2012elementary,welte2018photon,samutpraphoot2020strong} for some innovative demonstrations).
Meanwhile, other technologies have emerged that can similarly facilitate such highly-controllable quantum light-matter interactions to implement fundamental primitives like QST and entanglement generation (which in some cases are heralded as opposed to deterministic).
These include experimental implementations (and accompanying theory) using trapped ion qubits  \cite{stute2013quantum,stephenson2020high,schupp2021interface,krutyanskiy2023telecom},
collective atomic excitations \cite{hammerer2010quantum},
various forms of superconducting qubits \cite{pechal2014microwave,kurpiers2018deterministic,campagne2018deterministic,axline2018demand,zhong2021deterministic,burkhart2021error},  
and electronic spin qubits [nitrogen-vacancy (NV) centers in diamond are a common realization] \cite{pompili2021realization}.
These are not limited to photonic flying qubits; notably see Ref.~\citenum{neuman2021phononic} which uses an intermediate phonon for hybrid quantum state transfer between a superconducting microwave qubit and a solid-state spin qubit. In their main text they do this using a direct coupling, though in their Supplementary Note 1 they also consider a `pitch-and-catch' protocol with the phonon serving as a flying qubit that could propagate between distant versions of such nodes connected by a waveguide.
These many implementations make the generality of JC physics manifest. 
See Ref.~\citenum{larson2022jaynes} for further exposition of the JC model including its applicability in many experimental settings, including those just mentioned, as well as its history, theoretical details, and extensions. 

The precise means of obtaining a Hamiltonian of this form depend on the nodes' physical implementation, though some common steps (as mentioned in the main text) include adiabatically eliminating other states, selecting an appropriate rotating frame to go into (i.e., selecting which unitary transformations to perform on  the node Hamiltonians), and tuning the system parameters to eliminate undesired terms. The physical interpretation of the operators in $H_{\textrm{a}\sigma}$ and the methods of controlling the coupling $g(t)$ likewise depend on the node.
However, as long as such a Hamiltonian is realizable with a given node implementation (as it often is for deterministic QST schemes), then the analysis presented here and in Ref.~\citenum{randles2023quantum} should be applicable. In particular, the equations we derive, such as the excitation amplitude EOMs Eq.~(\mainref{eq:coefEOMs}) and other subsequently derived expressions (or similar variations thereof), should apply more generally to these other types of nodes.
Thus, even though we emphasize connecting two nodes that are each comprised of an atom in a cavity, our results are relevant to hybrid cases where two different types of nodes are connected.

\phantomsection
\subsection*{3.~Fidelity as a measure of success}\label{subsec:fidelity}
The probability of successfully transmitting an excitation between the systems $\Psucc = |\alpha_2(t_e)|^2$ [given in Eq.~(\mainref{eq:PsuccDerived}) of the main text] is a useful measure of the success of a state transfer attempt. 
This can be contrasted to the quantum state fidelity
\be
\mathcal{F} = |\braket{\psi_i}{\psi_f}|^2
\ee
(some authors 
refer to the square root of this quantity as `the fidelity'), which accounts for impact of phase errors in addition to excitation loss.
Here the initial state (of atom 1) is $\ket{\psi_i} =  c_g \ket{g} + c_e \ket{e}$	(using the same notation for the coefficients as Sec.~\mainref{subsec:nodeImp} of the main text) and the final state (of atom 2) is of the form
\be
\ket{\psi_f} = e^{i \theta_f} \sqrt{1- |c_e|^2|\alpha_2(t_e)|^2} \ket{g} + c_e |\alpha_2(t_e)| \ket{e},
\ee
where $\theta_f$ encodes any relative phase errors between $\ket{g}$ and $\ket{e}$ (we already factored out any phase error in the term $\alpha_2(t_e) e^{i \phi_2(t_e)}$ 
and included it in $\theta_f$, up to an unphysical global phase).\footnote{
	Note that it does not matter if the unitary induces a relative phase $\phi$ during the transmission from node 1 to 2, as long as a  $-\phi$ phase is induced during the reverse transmission from node 2 to 1. For instance, say $(a \ket{0}_1 + b \ket{1}_1) \ket{0}_2 \ra \ket{0}_1 (a \ket{0} _2+ b e^{i \phi} \ket{1}_2)$ under a QST procedure, then the basis $\{\ket{0}_2, \ket{1'}_2 =e^{i \phi} \ket{1}_2\}$ can be used for system 2, effectively relabeling the excited state to account for this relative phase.
	This is ok provided this phase error is systematic and reversible. That is, if when the state is transferred back to system 1, it obtains the opposite phase $-\phi$, e.g.,
	$\ket{0}_1 (a \ket{0} _2+ b \ket{1'}_2) \ra (a \ket{0} _1+ b e^{-i \phi} \ket{1'}_1) \ket{0}_2 = (a \ket{0} _1+ b \ket{1}_1) \ket{0}_2$, which is the original state. Otherwise, this is a genuine phase error (which is assumed to be the case for $\delta\theta$ in this section).
}
Note there is no phase error if $\theta_f = \arg(c_g)$ so using the relation $|c_g|^2 + |c_e|^2 = 1$ along with shorthand\footnote{
	Note that within given SM sections we occasionally represent a new variable with a symbol that has already been used. For instance, here we define the excited state population $x$, which is not to be confused with the transmission line propagation distance. The quantity under consideration should be clear from context.
	\label{note:repSymbol}
} 
$x = |c_e|^2$, $a = |\alpha_2(t_e)|$, and $\delta\theta = \arg(c_g^* e^{i \theta_f})$ we have
\be\label{eq:generalFidelity}
\mathcal{F} = \left| e^{i \delta\theta} \sqrt{(1-x)(1- a^2 x)} + a x \right|^2
= a^2 x^2 + (1-x)(1- a^2 x) + 2 \cos(\delta\theta) \sqrt{(1-x)(1- a^2 x)} a x. 
\ee

\begin{figure*}[ht] 
	\includegraphics[width=\linewidth]{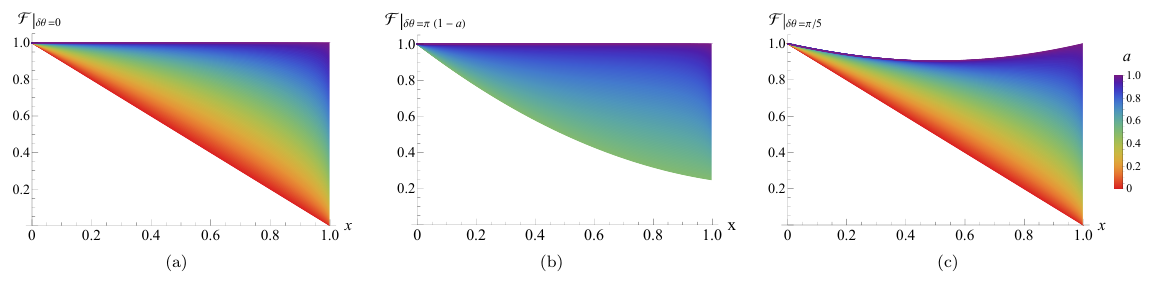} 
	\caption{
		Plots of the fidelity $\mathcal{F}$ as a function of $x$ for various $a$ as indicated by the legend. In each of the subfigures we consider a different case of the phase error: (a) $\delta\theta = 0$, (b) $\delta\theta = \pi(1-a)$, and (c) $\delta\theta = \pi/5$. 
		In each of case $\mathcal{F}(x)$ increases monotonically as a function of $a$ pointwise for each $x$ over the shown values of $a$. To obtain this behavior in case (b), and hence to avoid crossings (which would make this type of figure futile), we only includes curves for $a \geq 1/2$.   
	}
	\label{fig:fidelityPlots}
\end{figure*}

Note this dramatically simplifies for $|c_e| = 1 \implies x =1$ to $\mathcal{F}|_{x=1} = a^2 = |\alpha_2(t_e)|^2$, which is the probability of successfully transferring an excitation $\Psucc$ considered above.
The more general behavior of $\mathcal{F}(x, a)$ depends on what the phase error $\delta\theta$ is, three possibilities of which are illustrated in Fig.~\ref{fig:fidelityPlots}.
In the ideal case without phase errors $\delta\theta = 0$, achieved by phase matching for a given $\alpha_2$, the fidelity $\mathcal{F}$ monotonically decreases from 1 to $a^2$ as $x$ goes from 0 to 1, see Fig.~\ref{fig:fidelityPlots} (a).
Hence in this ideal phase matched case the considered $\Psucc$ is a lower bound for $\mathcal{F}$.
This fidelity is clearly minimized if $\delta\theta = \pi$, yet one should be able to tailor the scheme so that, at least for large $a$ near $1$, the phase error is negligibly small. The simplest possible realization of this is the phase error model $\delta\theta = \theta_0 (1 - a)$, where assuming the phase error gets smaller as $a$ increases then $ 0 \leq \theta_0 \leq \pi$. 
Even in the worst such case with $\theta_0 = \pi$, corresponding to a phase flip error for $a=0$, the average value of $\mathcal{F}$ as a function of $x$ is always well above $a^2$, 
and moreover for $a \geq 1/2$ in the model we again find (as in the ideal case) that $\mathcal{F}$ monotonically decreases from 1 to $a^2$, see Fig.~\ref{fig:fidelityPlots} (b).
Thus $\Psucc = a^2$ again serves as a lower bound provided the transfer satisfies the modest condition $a \geq 1/2$. 
This lower bound restriction on $a$ of $1/2$ gets smaller as $\theta_0$ decreases, approaching $a \geq 0$ for the ideal case $\theta_0 = 0$ considered above.
This may be overly simplistic, though the development of a general phase error model is beyond of the scope of this work and would likely depend on many system-specific figures of merit.
We note that  similar results seem to hold for other phase error models provided $\lim_{a \ra 1} \delta\theta = 0$.
If this does not happen, i.e., if $\lim_{a \ra 1} \delta\theta = \delta\theta_1 \neq 0$, then even for perfect excitation transfer $a=1$, the fidelity will drop below one $\forall~0 < x < 1$ (which just excludes the end points) and will reach a minimum value of 
\be\label{eq:minA1Fidelity}
\min_x \mathcal{F}(x, a=1) = \cos^2\p{ \frac{\delta\theta_1}{2} }
= 1 - \frac{\delta\theta_1^2}{4} + \mathcal{O}\p{ \delta\theta_1^4 }
\ee
at $x=1/2$ (assuming $\delta\theta$ is independent of $x$). See Fig.~\ref{fig:fidelityPlots} (c) for an example of this behavior in the case of a constant error of $ \delta\theta = \pi/5$.

\begin{figure*}[ht] 
	\includegraphics[width=0.65\linewidth]{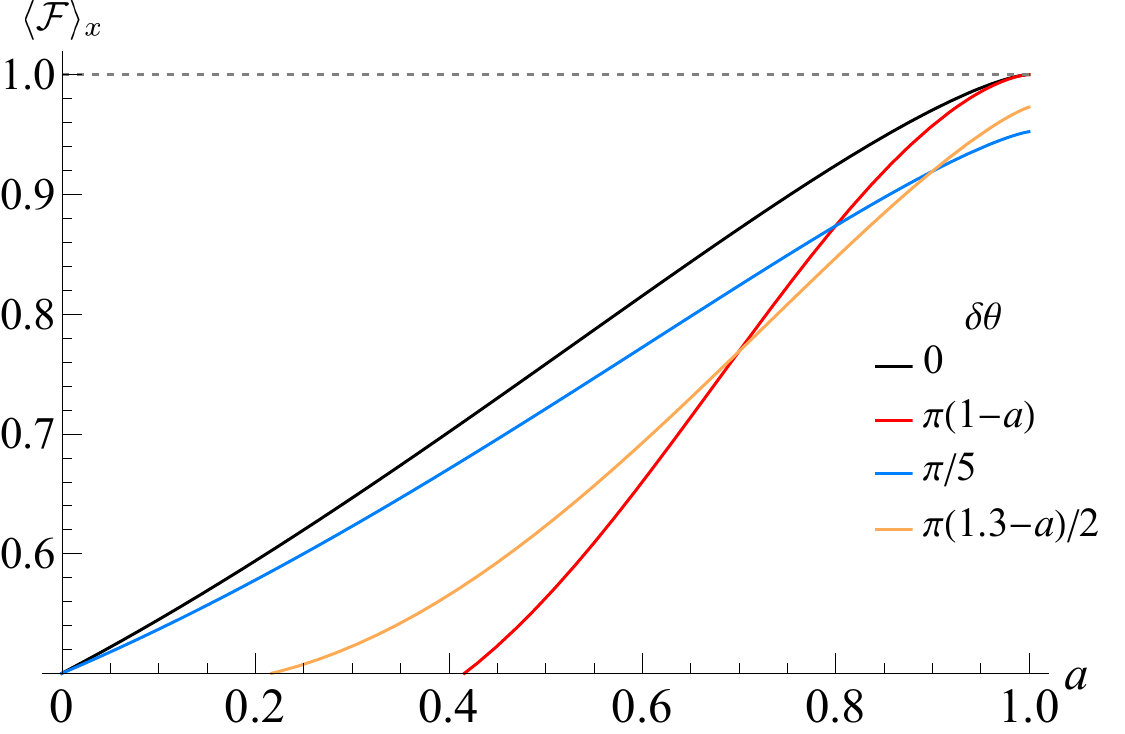} 
	\caption{
			Plots of the average fidelity $\langle \mathcal{F} \rangle_x$ for the three error models for $\delta\theta$ considered in Fig.~\ref{fig:fidelityPlots} (a-c) as well as an additional model that combines the simple phase error model with $\theta_0 = \pi/2$ (see SM text) with a small residual offset:
			$\delta\theta = \pi(1 - a)/2 + 3\pi/20$ (these respective cases are shown as black, red, blue, and orange lines). 
	}
	\label{fig:avgFidelityPlot}
\end{figure*}

Due to the presence of $\ket{e_1} \ket{g_2} \ra \ket{g_1} \ket{g_2}$ (photon loss) errors, the fidelity tends to be much lower when the excited state population $x$ is large 
(at least for small $a$, see Fig.~\ref{fig:fidelityPlots}).
One way to quantify the typical fidelity of an arbitrary initial state $\ket{\psi_i}$ is by averaging over the $x$ parameter. In particular, we express the initial state in the Bloch sphere representation as $\ket{\psi_i} = \cos(\theta_{Bs}/2) \ket{g} + e^{i \phi_{Bs}} \sin(\theta_{Bs}/2) \ket{e}$
and average $\mathcal{F}$ over the polar angle $\theta_{Bs}$. Doing so using Eq.~(\ref{eq:generalFidelity}), wherein $x = \sin^2(\theta_{Bs}/2)$, and again assuming $\delta\theta$ is independent of $x$, we find 
\be
\langle \mathcal{F} \rangle_x \equiv \int_0^\pi \frac{d \theta_{Bs}}{\pi}  \mathcal{F}[x(\theta_{Bs})]
=
\frac{1}{2} + \frac{a^2}{4} + \frac{ \cos(\delta\theta) \br{a \sqrt{1-a^2} (2 a^2 -1) + \sin^{-1}(a) } }{ 2\pi a^2 },
\ee
which is plotted in Fig.~\ref{fig:avgFidelityPlot} for several phase error models. As $a \ra 1$, corresponding to $\Psucc = a^2$ near unity, this averaged fidelity approaches $\lim_{a \ra 1} \langle \mathcal{F} \rangle_x = 1 - \delta\theta_1^2/8 + \mathcal{O}(\delta\theta_1^4)$, which is near one for small residual phase offsets $\delta\theta_1$, similar to Eq.~(\ref{eq:minA1Fidelity}).

\phantomsection
\subsection*{4.~Utility of time reversal}\label{subsec:timeReversal}
We emphasize the time reversal aspect of our protocol, and of the transformation specifically, both because it naturally fits within the quantum networking goal of being able to realize controlled reversible light-matter interactions to transfer the state of a qubit to and from a material system and a photon
and  
because it is not typically emphasized by other researchers considering similar QST schemes to ours.
For the former point, we note that the incorporation of a time-reversal operation as part of the unitary transformation in our scheme 
is motivated by the notion that the absorption of a flying qubit is the time-inversed complex conjugate of the emission process \cite{gorshkov2007photon}, and moreover, that this still holds for nodes with different resonance frequencies and decay rates, provided we implement a transformation to effectively match up these quantities. 
Note that including the time-reversal operation to the intermediate photon is redundant in the often considered special case where the produced photon \cite{ritter2012elementary,pechal2014microwave,maring2017photonic,kurpiers2018deterministic,campagne2018deterministic,axline2018demand,reuer2022realization} or phonon \cite{bienfait2019phonon,neuman2021phononic}
wave packet envelope is time-symmetric (as in seminal proposal by Ref.~\citenum{cirac1997qst}).
However, in some cases, an asymmetric wave packet may be preferable, such as if it is easier to produce the corresponding drives, e.g., laser pulses, (note that errors in the drives 
will generally lead to an asymmetric shape even when the target shape is time symmetric \cite{khanbekyan2017cavity,morin2019deterministic})
or if a certain wave packet shape is less susceptible to errors [for instance,  wave packets with longer durations (e.g., due to slow driving) can be emitted with higher efficiency (see Sec.~\hyperref[subsec:atomicDecay3]{B3}) yet they take longer to produce (potentially reducing the QST rate) and are more susceptible to frequency errors (see Sec.~\hyperref[subsec:formalBeta1FreqError4]{B4})].
Even in cases where the intended wave packet is symmetric, a time-reversal transformation may correct for any error-induced time asymmetry, assuming the error is systematic in that it effects the receiving and sending nodes in the same way. For instance, dispersion (or other distortion) during propagation in the channel can make an initially time-symmetric wave packet asymmetric 
and it can induce a nontrivial relative phase to the photonic qubit state.
Hence, the unitary transformation we propose could be used in the middle of the channel to cancel out such asymmetric distortions and phase errors, analogous to a spin-echo. 
If only time reversal was implemented, this would typically still result in a wave packet that is broadened and frequency shifted 
due to dispersion. However, assuming this happens in a systematic and characterizable way, these net shifts could be compensated for using our stretching and frequency shift parameters of $\xi$ and $\om_0$, respectively.
Note that such distortion can alternatively be taken into account by modifying the drives themselves.
For instance, the emission inducing drive $G_1(t)$ could be designed to produce 
wave packets that are less-susceptible to 
reshaping and the absorption inducing drive $G_2(t)$ could be modified to better receive the distorted wave packet \cite{penas2023improving}. 

Accordingly, to not restrict ourselves to this special time-symmetric case, we include time reversal in our unitary transformation. This is nice theoretically, as then our QST scheme is applicable  more generally for asymmetric photon wave packets, and in fact is indifferent 
to the photon wave packet shape (provided we know the temporal window it occupies). Furthermore, it may be natural to incorporate the time-reversal transformation as part of the photon manipulation scheme along with frequency conversion and mode stretching or compressing, as is the case in Ref.~\citenum{timereversal}.
Importantly, much of our analysis would still apply if a time-reversal transformation was not implemented. There would just be a corresponding error due to how asymmetric the wave packet is. Moreover, such analysis would still hold if a different transformation was implemented or no transformation was implemented at all (which is relevant when small frequency shifts, small time asymmetries of the photon, and compression or stretching factors of near unity are needed). This is because the result for $\Psucc$, Eq.~(\mainref{eq:PsuccSynopsis}), does not depend on how or whether the incident wave packet $\Psi(t)$ has been transformed. 

Returning to the latter point of a lack of emphasis, we note that recent work on hybrid quantum interconnects and networking often emphasizes the need for frequency conversion and wave packet shaping, in the form of bandwidth conversion, of the intermediate photon wave packet with little explicit mention of correcting for time asymmetry \cite{fisher2016frequency,maring2017photonic,morin2019deterministic,awschalom2021development,maruf2023widely}.
Some QST experiments produce photons with manifestly time-asymmetric temporal shapes, often producing wave packets that are smoothed-out versions of an exponentially decaying function 
\cite{stute2013quantum,schupp2021interface,huie2021multiplexed,krutyanskiy2023telecom}.
This may suffice for establishing photonic entanglement (see SM~\hyperref[subsec:heraldedProtocols]{A7} for such a case) but would lead to issues  in achieving state transfer or entanglement generation across material systems realized using absorption unless the corresponding photons' shapes are modified or a time-reversal transformation (or more general wave packet shaping) is employed. 
Perhaps this lack of emphasis is because the intermediate photon can often be engineered to have a symmetric (or nearly symmetric) temporal shape or because it is well-known and currently a less pressing issue than frequency and decay rate matching (especially in the current era where hybrid frequency conversion can only be achieved with low probability).

To illustrate the impact of temporal shape -- and hence how pressing errors in it may be -- we consider the highly asymmetric, though typical \cite{kielpinski2011quantum}, case of a normalized wave packet with an exponentially decaying envelope $\Psi(t) = \sqrt{\g} \Theta(t) e^{-\g t/2}$ being sent between two identical nodes.
Assuming the absorption process is the reverse of the emission process that generated this wave packet (including any time-dependent system controls used), the ideal wave packet for absorption will be the time inverse of $\Psi(t)$ up to a delay $T_d$, i.e., $\Phi(t) = \Psi(T_d - t)$. Then the  probability of success as determined by Eq.~(\mainref{eq:PsuccSynopsis}) is
$\Psucc = \g^2 T_d^2 \Theta(T_d) e^{-\g T_d}$,
which has a maximum value of $4/e^2 \approx 0.54$ for $T_d = 2/\gamma$. This indicates that while time-reversal transformations (or the use of time-symmetric photon wave packets) is necessary for optimal transfer, it is only responsible for relatively small inefficiencies in a QST scheme like ours as compared to the impact of suboptimal frequency, stretching, and timing.
The practical utility of time-reversal should thus be deferred to a case-by-case basis depending on experimental details including how naturally a  time-reversal operation can be included (say as part of an already needed transformation) and what success probabilities are desired (namely whether a potential increase by a factor of $\lesssim 2$ is worth the effort of including time reversal).

\phantomsection
\subsection*{5.~Unitary implementations}\label{subsec:UImp}
The scheme we propose is contingent on being able to implement the unitary transformation $U$ of Eq.~(\mainref{eq:UTransTime}), especially for the hybrid interconnection of heterogeneous nodes where one needs to transduce the intermediate photon between the emitting and receiving nodes disparate energy and time scales.
The form of the unitary transformation we consider was inspired by Ref.~\citenum{timereversal}, which proposed an optical implementation of a transformation that amounts to $U$, including time reversal, and is based on a three or four wave-mixing process that can be achieved in realistic nonlinear media.
In our scheme we need to be able to control both laser frequencies $\om_{L j}$ and convert the frequency of the emitted wave packet by their difference $\zeta = \om_{L2} - \om_{L1}$ all at a precision less than $\g_2$, which is typically on the order of kHz-MHz. If a wave-mixing unitary implementation, like that of Ref.~\citenum{timereversal}, is used one must thus be able generate a pump pulse at the difference frequency $|\zeta|$.
For atomic systems, as considered here, standard optical lasers could typically be used. However, when doing intraband frequency conversion, e.g.,~microwave-optical transduction, lower frequency lasers or alternate drives may be necessary, which may be limiting. 

Considering systems in the optical to near-infrared domain is a convenient starting point as $U$ has a proposed implementation in that regime \cite{timereversal} and several aspects of it have been demonstrated experimentally including implementations of optical and telecom frequency conversion \cite{allgaier2017highly,krutyanskiy2017polarisation,ikuta2018polarization,bock2018high,radnaev2010quantum,kaiser2019quantum,arenskotter2023telecom}
and shaping (typically bandwidth conversion).\footnote{
We already mentioned relevant optical photon wave packet shaping experiments in Sec.~\mainref{subsec:numericalResults} of the main text, namely see Refs.~\cite{lavoie2013spectral,andrews2015quantum,allgaier2017highly,morin2019deterministic}
}
Additionally, such systems can reasonably be connected via optical fiber (over modest distances) \cite{heshami2016quantum}. 
Hence direct connection is not detrimental for short distances though relatively small frequency conversion to the telecom band (by a factor of $\sim$$2$), that can be achieved using nonlinear optical methods, can still be valuable if employed.
One such experiment worth mentioning is Ref.~\citenum{maring2017photonic}, which demonstrated the heralded hybrid QST of a spin-wave excitation of a laser cooled atomic ensemble to a collective optical excitation of a receiving rare-earth doped crystal. They did so by employing quantum frequency conversion at both nodes 
to transduce from the atomic ensemble's natural optical frequency 
to the telecom 
for transmission along a 
fiber followed by conversion back to the optical 
for interfacing with the crystal. Their method is notably different from our scheme as it heralded as opposed to deterministic and hence they do not have to be as concerned with the precise control pulses and the corresponding wave packet temporal shape.

The development of methods to control the wave packet shape of microwave photons is an active area of research \cite{pechal2014microwave,andrews2015quantum,forn2017demand}
as are microwave channels (see Sec.~\mainref{subsec:BackgroundAndScope}).
For instance, with some adaptations, the microwave temporal and spectral mode converter of Ref.~\citenum{andrews2015quantum} could potentially be used to implement the considered unitary in the microwave regime. Their mode converter uses the vibration of the aluminum drumhead to catch, modify (which includes shaping and small frequency shifting), and then release the quantum microwave signal.
The extent of frequency conversion needed to link optical-optical systems, microwave-microwave systems, or even optical-telecom is modest compared to the order $\sim$$10^5$ frequency discrepancy that needs to be bridged for microwave to optical (or telecom) transduction. Even with the associated difficulties, microwave-optical transduction is rapidly developing area of research in which there have been many recent developments in various approaches \cite{andrews2014bidirectional,sahu2022quantum,casariego2023propagating}
(see Ref.~\citenum{han2021microwave} for a review)
and in applications such as the hybrid interfacing of microwave and optical nodes and for optically linking microwave nodes \cite{mirhosseini2020superconducting,delaney2022superconducting,kyle2023optically}.
Note these approaches emphasize frequency conversion, over shaping operations like time reversal and stretching or compression, because naturally it is the largest obstacle in microwave-optical interfacing. 
An implementation of our scheme in such a microwave-optical case could include cascaded transformations, somewhat like Ref.~\citenum{maring2017photonic}, say microwave to optical frequency conversion followed by optical tailoring of the photon wave packet, though significant advancement in current methods or new methods would be needed for this to be viable.

\phantomsection
\subsection*{6.~Photonic qubit encodings} \label{subsec:otherEncodings} 
In this work we focus on using an occupation number encoding for the photonic qubit state in which there is either a photon (in a particular singled-out mode), represented by $\ket{1}$, or there is not, $\ket{0}$. We do so for simplicity and so that the ECZ error correction protocols could be utilized, which would allow us to maintain a deterministic scheme even with errors and loss. However, this is not the only encoding that could be used for our scheme. Accordingly, in this section we compare the utility of different ways of encoding the state of a qubit in the degrees of freedom of a photon with a focus on occupation number versus polarization encodings.

\textbf{Polarization encoding.}
We will start by considering the polarization encoding in which the photonic qubit is encoded in an orthogonal polarization basis, we will refer to the linearly polarized horizontal and vertical basis $\{\ket{H}, \ket{V}\}$ for concreteness though others, such as right and left circularly polarized basis, could equally well be used. Note that one can easily change polarization basis, e.g., change between linear and circular polarized light using a quarter-wave plate. Such a polarization qubit can be mapped to and from many quantum emitters that have polarization dependent transitions. This can be achieved by appropriately selecting which electronic or spin states of the material are used to encode the qubit, e.g., considering selection rules, as well as using polarized drives.
One realization of this, which is a straightforward modification of our scheme, uses two 
well-chosen states (typically ground states, though other long-lived states may be appropriate), such as two hyperfine states of an ion split by a magnetic field \cite{northup2014quantum,krutyanskiy2023entanglement}, that are both coupled to an auxiliary level, which we will denote $\ket{r}$ here, via their own Raman transition through an intermediate upper level $\ket{i}$. The material qubit state is encoded in these states, which we will denote by $\ket{g}, \ket{e}$ for consistency. 
Thus, if the emitter was in state $\ket{g}$ we could apply a properly polarized pulse $\Omega_g(t)$ to induce a transition to $\ket{r}$ which would be accompanied by the production of a photon with polarization state $\ket{H}$ say.
Otherwise, if the emitter was in state $\ket{e}$ we would apply a different pulse $\Omega_e(t)$ with different polarization, again inducing a transition to $\ket{r}$, except a photon with a relatively orthogonal polarization, $\ket{V}$, would be produced. (This uses the level structure of the `ECZ atoms' shown later in Fig.~\ref{fig:atomConfigs} with $\Omega = \Omega_e$ and an additional Raman field $\Omega_g$ applied to the $\ket{g} \leftrightarrow \ket{r}$ transition. The dynamics of both cases are the same as for three-level $\Lambda$-type atoms we have detailed.) Hence by applying both pulses, i.e., a bichromatic Raman field, a single photon should always be generated but its polarization will depend on the emitter state such that for a generic superposition we would achieve QST from the material qubit to the photonic polarization qubit:  $c_0 \ket{g} + c_1 \ket{e} \ra c_0 \ket{H} + c_1 \ket{V}$.
Note that in practice, this transfer is difficult to implement (see the discussion of photon production in heralded setups below in Sec.~\hyperref[subsec:heraldedProtocols]{A7}). 	

Now returning to our context of QST between hybrid material nodes, we would first need to deterministically generate such a photon in a superposition of two polarization states. The photon would then propagate down a transmission line, along which it should be transformed so as to be absorbed by the receiving node.
One notable difference here for the polarization, as opposed to the occupation number, encoding is that one would typically need two different transformations, one for each polarization, as different unitary parameters $\om_0$, $\xi$, and $T$ would generally be needed to `impedance match' both Raman transitions,
e.g., due to different relative energy splittings of the states of both nodes
as well as different phase matching conditions in the implementation of  $U$ for each polarization. 
One way of achieving this would be by using a polarizing beamsplitter to direct the two different polarization components into different transmission lines along which the two separate unitary transformations are performed, after which the lines are recombined, say using another polarizing beamsplitter. 
Note that this splitting of modes into different fibers is already necessary for the transmission of a polarization encoded state as in a single fiber, even if it is `polarization maintaining,' the relative phase between two orthogonal modes in superposition, say $\ket{H}$ and $\ket{V}$, will rapidly drift.
However, by using a polarizing beamsplitter one can ensure that the initial polarizations in each of the split channels are linear and orthogonal with respect to each other. By using a polarization maintaining fibers for these channels, with the now linearly-polarized light properly aligned to either the fast or slow axis of the fiber, the states $\ket{H}$ and $\ket{V}$ will be stable and their relative phase can be made to be preserved or to be systematic at least. 
Finally, the time reversed and stretched counterpart of the original bichromatic Raman field would be applied to the receiving node, initialized in $\ket{r}$,  to induce absorption and hence QST between the two nodes.

\textbf{Tradeoffs.} 
We will now contrast the advantages and disadvantages of the occupation number versus polarization encodings in our scheme.
In the occupation number encoding, photon production $\ket{0} \ra \ket{1}$ errors can be highly suppressed (note microwave systems must be cooled down for this to be the case) but photon loss errors $\ket{1} \ra \ket{0}$ are common. This entails that one cannot distinguish the zero photon state $\ket{0}$ from a lost photon, $\ket{1} \ra \ket{0}$, which is a major downside in a single QST trial. However, this can be rectified using error correction such as the ECZ protocols \cite{van1997ideal,van1998photonic}.
Alternatively, in the polarization encoding different basis states, say horizontal $\ket{H}$ and vertical $\ket{V}$ polarizations, can mix 
while propagating in a fiber or when interacting with systems. 
That is, an initial state $\ket{\psi} = h \ket{H} + v \ket{V}$ can be mapped onto a new state 
by a nonidentity unitary (neglecting absorption) matrix $\ket{\psi} \ra U_\textrm{mix} \ket{\psi}$ so both $\ket{H} \ra \ket{V}$ and $\ket{V} \ra \ket{H}$ errors can occur. Importantly this can be highly suppressed through careful mode-matching to the transmission line for each polarization and/or appropriate use of polarization maintaining fibers as described above.
Nonetheless error correction protocols like those of ECZ, that capitalize on the ability to suppress one of the encoding states turning into the other, 
cannot be employed for the polarization encoding because of this mixing.

Therefore, in realizations where the polarization encoding has been deemed appropriate, alternative heralded schemes are likely preferable compared to deterministic schemes supplemented by costly full-fledged error correction (that can correct for both photonic bit flip errors).
For instance, one can post select (a means of heralding) on cases where the photon has not been lost, say by detecting if the receiver remains in the auxiliary state $\ket{r}$ (in the above example)  \cite{reiserer2015cavity}. 
This allows one to detect photon loss errors but not amplitude nor phase errors in the transferred state. Accordingly, if one can suppress bit-flip (mixing) errors, e.g., $\ket{H} \leftrightarrow \ket{V}$, and ensure the phase stability of the channel (similar to the occupation number encoding), then this type of heralding for a pitch-and-catch scheme using the polarization encoding (or using another dual-rail code such as the time-bin encoding) is a promising, currently implementable, alternative to the occupation number encoding.
One additional prerequisite for such a scheme is that both polarizations need to undergo nearly identical amplitude damping (photon loss) as elsewise, if $\ket{r}$ is measured and not occupied, the relative populations in $\ket{e}$ and $\ket{g}$ will be incorrect.
Such a scheme is more applicable for the QST of half of a relatively easy to generate Bell pair, i.e., remote entanglement generation, which could then be used for quantum teleportation, as opposed to the QST of an arbitrary qubit whose preparation is likely more demanding so one would not want to lose its state.  

Another notable difference between these encodings is in the control and stability of the transferred qubits phase, which is crucial for QST as discussed at the end of Sec.~\mainref{subsec:UParameterErrs} and Sec.~\hyperref[subsec:fidelity]{A3} (which shows how phase errors degrades the fidelity of the received state). As we have noted, in the occupation number encoding one needs to track and stabilize the relative phase between $\ket{0}$ and $\ket{1}$ using local oscillators and active phase stabilization, which is additional work but the methods to do so are well established (see the discussion and references at the end of Sec.~\mainref{subsec:UParameterErrs}).
This can be contrasted with polarization, which is interferometrically stable so one does not need be as concerned with active phase stabilization (e.g., maintaining the length of the transmission line) though one still needs well-defined clocks at both nodes to serve as phase references. 
Moreover, in the polarization encoding, one needs to be careful to not induce phase errors during transmission in the two split channels or by the separate unitary transformations discussed above (when applicable).
Notably, many of the aforementioned optical to optical or telecom frequency conversion experiments (in Sec.~\hyperref[subsec:UImp]{A5}) are polarization preserving and thus the polarization encoding can be made robust to phase errors even with a unitary transformation \cite{krutyanskiy2017polarisation,ikuta2018polarization,kaiser2019quantum,arenskotter2023telecom}. 

The polarization basis is commonly used in QST experiments, especially in the optical regime, where standard devices that are useful for manipulating and maintaining polarization can be employed \cite{blinov2004observation,ritter2012elementary,stute2013quantum,reiserer2015cavity}. 
Ref.~\citenum{boozer2007reversible} is a notable exception that explicitly demonstrated a reversible  cavity-based mapping of a coherent superposition between an optical field and an atom using the occupation number encoding. 
In the microwave regime it is more common to use the occupation number encoding  \cite{pechal2014microwave,kurpiers2018deterministic,axline2018demand,campagne2018deterministic,reuer2022realization},
due at least in part to a lack of microwave analogs of standard optical elements for manipulating polarization.
This discrepancy in the suitability of different photonic encodings from heterogeneous nodes is one reason we focus on the occupation number encoding (as discussed in Sec.~\mainref{subsec:errorCorrection}).

\textbf{Alternative encodings.}
Some other possibilities are encoding in temporal modes (wave packets with orthogonal time-frequency shapes), in arrival time (as done by Ref.~\citenum{maring2017photonic}), in space (path taken), or in frequency.
Using temporal modes is less viable in our scheme as their generation is typically achieved using parametric down-conversion as controlled by an applied pump field \cite{brecht2015photon}, which is quite different than the controlled emission (and absorption) processes we are concerned with. Additionally, the unitary transformation and distortion in the fiber may disrupt the temporal modes used so that they are no longer orthogonal leading to mixing errors as mentioned for polarization. Such state mixing can likewise be a problem for time- or frequency-bin encoding if there is too much overlap between the basis states or if the spacing of these states changes (say due to propagation in the channel or the unitary) in their respective domain (time or frequency).
The task of getting nodes to emit (and absorb) one of the specific basis states in a time or frequency bin encoding, dependent on the state of a material qubit, 
is quite system specific. In the frequency encoding transitions with different frequencies could be utilized. Different transitions could also be utilized to distinguish arrival-time encoded flying qubit basis states, or local gates could be used to perform SWAP-like operations, systematically rearranging the amplitudes of given material states.
Similar to polarization, using these encodings, would typically require the use of two unitary transformations one for each of the basis states. This need not be the case when encoding in arrival time if a suitably long transformation is employed, provided one correspondingly changes the timing of the controls at the receiving node.  
Note that it can be highly nontrivial to map between different photonic encodings in an efficient quantum state preserving manner.
For instance going between the occupation number and polarization encodings is nontrivial, whereas polarization and path can be converted between using a polarizing beamsplitter.

\phantomsection
\subsection*{7.~Heralded protocol comparison} \label{subsec:heraldedProtocols}
\begin{figure*}[ht] 
\includegraphics[width=0.7\linewidth]{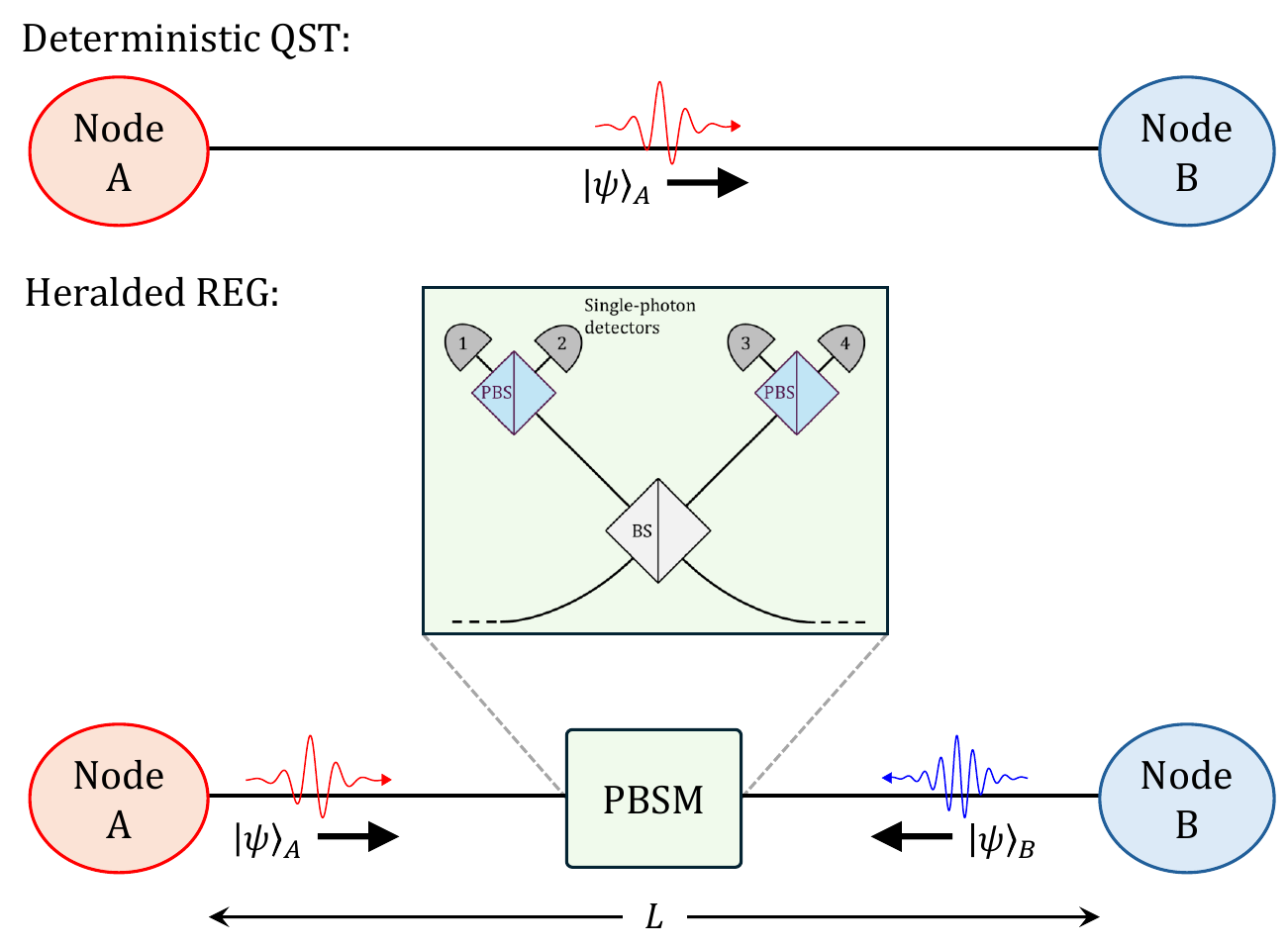} 
\caption{
		Schematic overview of a deterministic QST protocol versus a heralded REG protocol linking two nodes, A and B, separated by a distance $L$ in both cases. The unitary transformation is omitted in both diagrams for simplicity yet would generally be needed when interfacing heterogeneous nodes. See Sec.~\mainref{sec:schemeAndModel} of the main text for an overview of the deterministic case. In the prototypical heralded case, a photonic Bell-state measurement (PBSM) is employed, wherein two photons, one emitted from each node, are interfered at a balanced beam splitter (BS) and output photonic state is measured (in the photonic qubit encoding basis) using single-photon detectors. The coincident detection of orthogonal photonic qubit states heralds a remote entangled state of the qubits at nodes A and B. Here we illustrate the PBSM for a linear polarization photonic encoding, $\{ |H \rangle, |V \rangle \}$, for which the BS outputs are each further split using a polarizing beam splitter (PBS) so that the photons' polarizations can be measured.
}
\label{fig:detVsHer}
\end{figure*}

Here we contrast deterministic QST and remote entanglement generation (REG) protocols against commonly employed heralded (non-pitch-and-catch) protocols with the same aims. In Sec.~\mainref{subsec:heraldedRelevance} of the main text we introduced the general setup of such heralded protocols and described the relevance of our work to them. Here we contrast (mostly qualitatively) the merits and downsides of deterministic versus heralded protocols. 
For concreteness, we focus on comparing deterministic QST protocols with the occupation number encoding, like the one we analyzed in this work, against heralded REG protocols using the polarization encoding such as that employed by Ref.~\citenum{krutyanskiy2023entanglement} (see Fig.~\ref{fig:detVsHer} for a schematic comparison). Recall that QST and REG protocols can be mapped onto each other by transferring half of a local Bell-pair (QST $\ra$ REG) and quantum teleportation (REG $\ra$ QST).
We will make this comparison across several categories: resource overhead, photon production and absorption, photon detection, and photon transmission, which are displayed in a list with each category being followed by a comparison of corresponding details relevant to deterministic and heralded protocols individually (labeled via D and H, respectively) or both.

\begin{itemize}
\item \textbf{Resource overhead.} 
\\
D: A single deterministic QST attempt has the glaring problem of being highly susceptible to photon loss errors. Especially when using the occupation number encoding, one cannot distinguish between a photon loss error $\ket{1} \ra \ket{0}$ and the transmission of the ground state. 
Error correction, such as the ECZ protocols, can be used to address this; however, it requires an appreciable overhead, e.g., in the ability to perform local quantum operations at the nodes being connected (for ECZ) or for generating more exotic photonic states to encode the state of the qubit being transferred (see Sec.~\mainref{subsec:errorCorrection}). 
\\
H: This additional overhead is avoided in heralded REG protocols, as although photon loss will lower the entanglement generation rate, it does not itself degrade the fidelity of a successfully heralded entangled state.
Note that there is an intrinsic maximum heralding probability of $50\%$ due to the photonic Bell-state measurement (as identical photonic qubit states do not lead to a remote entangled state).
\item \textbf{Photon production and absorption.} 
\\
Both: In each case there is loss due to emitter-cavity and cavity-channel couplings as quantified by the probabilities $\mathcal{P}_\textrm{em}$ and $\mathcal{P}_\textrm{cav}$, respectively, at each nodes. Because these probabilities are the same for emission and absorption at a given node, they will equally affect both deterministic QST (emission from one node followed by absorption at the other) and heralded REG (emission from both nodes), assuming each node has the same cooperativity in both cases. \\
H: One potential drawback of heralded REG schemes is that they require the challenging controlled production of a photon whose qubit state is in a particular entangled state with its emitter's qubit, e.g., in the polarization encoding this can be accomplished by simultaneously driving two polarization-dependent transitions (as discussed in the previous subsection), which often experience some cross-coupling. 
For instance, Ref.~\citenum{krutyanskiy2023entanglement} were able to achieve ion-photon fidelities (relative to a target maximally entangled state with best-fit relative phase) of $92.9\%$ and $95.5\%$ for their nodes A and B, respectively.
\item \textbf{Photon detection.} \\
Both: Controlling the presence of undesired photons due to spontaneous emission of the nodes or environmental noise (especially in microwave systems) is crucial. \\
D: If such an undesired photon is produced due to spontaneous emission from the upper level in the Raman process it will lead to some combination of amplitude damping, photon loss, and phase errors, i.e., $\beta$ and $\Upsilon_1$ errors in Eq.~(\mainref{eq:HeffMap}). Notably, such an error can potentially be corrected for using the ECZ protocols.
Otherwise, if the error is caused by some other spurious photon causing the unwanted transition $\ket{g_2} \ra \ket{e_2}$ for the component of the first qubit in the zero-excitation subspace, then it cannot be corrected for the ECZ protocols. 
Fortunately, the receiving node is quite selective as a photodetector, in that only input photons with an appreciable overlap with the target shape $\Phi(t)$, as quantified in Eq.~(\mainref{eq:PsuccDerived}), will cause this ruinous $\ket{g_2} \ra \ket{e_2}$ error, which would require the incident light to have a component with a specific timing, frequency, and wave packet shape.
This is likewise the case for the dual-rail code (the heralded pitch-and-catch scheme) mentioned in the previous subsection \hyperref[subsec:otherEncodings]{A6}, in which case spontaneous emission will likely lead to a (detectable) photon loss error as opposed to a qubit error.
\\
H: In heralded schemes that leverage conventional single-photon detectors, such a spontaneously emitted photon is likely to lead to a detection event, yet it is distinguishable from the desired photon so it gives which-path information and thus ultimately lowers the remote state fidelity. 
Accordingly, one must compromise between entanglement quality and generation rate, i.e., using smaller coincidence windows to lessen the impact of spontaneous emission \cite{krutyanskiy2023entanglement}.
This is likewise an issue for any environmental photons at a frequency the photodetectors are sensitive too (this can be partially alleviated by filtered out light at frequencies away from the photon's central frequency).
Moreover, the success of heralded schemes heavily relies on how good the single-photon detectors used are, which is not an issue faced by our deterministic QST scheme. Namely, detector imperfections, including their bare inefficiency, timing jitter, and dark counts, reduce the achievable rates and can contribute to apparent (false) coincidence events.
\item \textbf{Photon transmission.} 
\\
Both: The role of fiber-cavity alignment (mode matching) is similar in both cases and must be done at each node.
Transmission-line loss typically occurs exponentially with propagation distance. Thus, as discussed in Sec.~\mainref{subsec:strategies}, the probability to propagate a distance $x$ in a given transmission line with attenuation distance $x_\textrm{tl}$ is $e^{-x/x_\textrm{tl}}$ (referred to as $P_3$ in the main text).
Thus, when linking nodes over a set distance $L$, as in Fig.~\ref{fig:detVsHer}, the probability of retaining a single photon over its full-propagation in a deterministic scheme is $e^{-L/x_\textrm{tl}}$, which is equal to the probability of retaining both photons in a heralded scheme no matter where the heralded measurement is performed $0 < x_\textrm{meas} < L$ (with $0$ and $L$ the respective positions of nodes 1 and 2) as $e^{-x_\textrm{meas}/x_\textrm{tl}} e^{-(L - x_\textrm{meas})/x_\textrm{tl}} = e^{-L/x_\textrm{tl}}$ (ignoring additional fiber length needed for the photonic Bell-state measurement itself).
(The situation is slightly different for heralded protocols implemented using the occupation number encoding. For instance, Ref.~\citenum{stockill2017phase} and \citenum{humphreys2018deterministic} only want one of two potential photons to be detected, as two-photon contributions lead to an error in their cases.)
\\
H: Dispersion can be made less of a problem in heralded protocols if it is undergone symmetrically for each of the photons emitted by the separate nodes,
as the particular shape of the interfered photons is not as important as them being nearly indistinguishable. This would require the Bell-state measurement to take place near the center of both locations via the transmission line.
\end{itemize}

\textbf{Outlook.}
As more complex quantum networking nodes with multiple interacting qubits, each constituting a small quantum computer, become more commonplace, error corrected deterministic QST and REG using a photon-number encoding become more viable as an alternative to analogous heralded methods.
Moreover, we note that if one can controllably produce polarization (or time- or frequency-bin) encoded photonic qubits in a given heralded REG experiment 
(which is requisite for producing indistinguishable photons and obtaining high-fidelity remote entangled states), they should be able to controllably produce a single-photon (in a particular well-chosen mode) for occupation number encoding (likely with higher success rate).
Therefore, we recommend the further consideration of deterministic quantum state transfer using the photon number encoding in such setups and more generally as a potential means of progressing distributed and hybrid quantum computing technologies. 

\setcounter{equation}{0}
\renewcommand{\theequation}{B\thesection\arabic{equation}}
\renewcommand{\theHequation}{B\thesection\arabic{equation}} 
\phantomsection
\section*{B.~Determining the wave packet}\label{sec:wavePacketShape} 
Here we consider how to design the laser pulse $G_1(t)$ to produce a specific amplitude $\alpha_1(t)$ or $\beta_1(t)$. Without the unitary transformation, the wave packet emitted by system 1 is $\Psi(t) = \sqrt{\g_1} \beta_1(t)$ \cite{randles2023quantum}, whereas with the transformation the wave packet is $\Psi(t) = \sqrt{\g_1 \xi} e^{i \om_0 (T - t)} \beta_1(\xi (T - t))$, assuming $l$ is large enough so that the entire wave packet is transformed.
We start by discussing the Hamiltonians for the nodes and the corresponding amplitude EOMs as well as their limitations in Sec.~\hyperref[subsec:generalNodeH]{B1}. We then show how $G_1$ and the corresponding $\beta_1$ can be determined from $\alpha_1$ in a `nice' case where we select certain laser frequency and phase in Sec.~\hyperref[subsec:wavePacketShape2]{B2}. 
Next, in Sec.~\hyperref[subsec:atomicDecay3]{B3} we analyze the impact of spontaneous emission during the Raman processes employed at both nodes, ultimately giving credence to our post hoc treatment of it [strategy 1)]. 
In Sec.~\hyperref[subsec:formalBeta1FreqError4]{B4} we provide a formal solution to $\beta_1$ that makes its relation to the driving pulse $G_1$ intuitive and lets us analyze frequency errors.
Ultimately, in Sec.~\hyperref[subsec:wavePacketShape5]{B5} we show 
that even with the unitary transformation we cannot produce a wave packet $\Psi(t)$ with arbitrary shape, though there is a large class of wave packets we can produce that are useful for QST.

\phantomsection
\subsection*{1.~Node Hamiltonians}\label{subsec:generalNodeH} 
In the main text we made several choices to put the Hamiltonians describing the dynamics of the atom in a cavity type nodes into the simple JC form of Eq.~(\mainref{eq:HjSimplified}).
Doing so allowed us to emphasize the crucial aspects of our QST scheme.  However, we can obtain more control of the photon wave packet, namely its phase, if we do not fix the laser frequency and phase in the previously mandated way. 
Hence, here we consider a more general node Hamiltonian in which we have not yet selected the phase and frequency of the lasers driving each system, though we have already adiabatically eliminated the excited state and gone into an appropriate rotating frame. Then $H_j$ is given by equation 34 of Ref.~\citenum{randles2023quantum}, which in the single- (or zero-) excitation subspace simplifies slightly to
\be\label{eq:HjExtended}
H _j = d_j \hc{a}_j a_j + \delta\om_j(t) \kb{e_j}{e_j}  - i G_j(t) \p{e^{i \phi_j(t)} a_j \kb{e_j}{g_j} - \Hc},
\ee
with $d_j \equiv g_j^2/\Delta_j - \delta_j$, $\delta \om_j(t) := \Omega^2_j(t)/4 \Delta_j$, and $G_j(t) := g_j \Omega_j(t)/2\Delta_j$, where  $\delta_j = \om_{Lj} - \om_{cj}$ is the Raman detuning between the $j$th laser and cavity frequency and the other parameters are defined in the main text. We see that there is still a Jaynes–Cummings like atom-cavity coupling term, which relative to the generic node JC Hamiltonian of Eq.~(\ref{eq:genericNodeHam}) has $a$ and $\sigma^- = \kb{g}{e}$ as cavity and atomic lowering operators, respectively, and $g(t) = i G(t) e^{-i \phi(t)}$; however, there are additional energy shifts. 
With these system Hamiltonians as well as Eqs.~(\mainref{eq:Heff}) and (\mainref{eq:Htl})\footnote{
Note that in the same frame as Eq.~(\ref{eq:HjExtended}), $H_\textrm{tl}$ is given by Eq.~(\mainref{eq:Htl}) precisely as written, corresponding to equation 36 of Ref.~\citenum{randles2023quantum}.
\label{note:Htl}	
}
for the state 
\be
\ket{\psi(t)} = \alpha_1(t) e^{i \phi_1(t)} \ket{e g} \ket{00} + \alpha_2(t) e^{i \phi_2(t)} \ket{g e} \ket{00} 
+ \beta_1(t) \ket{g g} \ket{10} + \beta_2(t) \ket{g g} \ket{01},
\ee
now with the laser phases explicitly factored out, we obtain the following single-excitation dynamics:
\begin{subequations}\label{eq:generalAmpEOMs} 
\begin{align} 
	\dot\alpha_j &= -G_j \beta_j - i \p{\dot\phi_j + \delta\om_j} \alpha_j \qquad (j=1,2), \label{eq:genEOMa} \\
	\dot\beta_1 &= G_1 \alpha_1 - \p{ \frac{\gamma_1}{2}  + i d_1 } \beta_1, \label{eq:genEOMb} \\
	\dot\beta_2 &= G_2 \alpha_2 - \p{ \frac{\gamma_2}{2}  + i d_2} \beta_2 - \sqrt{\g_2} e^{i \zeta t} 
	\Psi(t). \label{eq:genEOMc}
\end{align}
\end{subequations}
Note that $\alpha_j$ are slowly varying complex amplitude envelopes as we factored out the laser phases $e^{i \phi_j(t)}$. 
We see that our simpler dynamics given in Eq.~(\mainref{eq:coefEOMs}) result from this if we select `nice' laser phases such that $\dot\phi_j + \delta\om_j \equiv 0$ and laser frequencies such that $d_j = 0$.

Alternatively, we can obtain these simpler dynamics of the main text by first going into a rotating frame via the unitary transformation $U = U_1 U_2$ with 
\be
U_j = \exp\left\{i \br{ d_j t \hc{a_j} a_j + \int dt~\delta\om_j(t) \ketbra{e_j}{e_j} } \right\}
\ee
for $j = 1,2$. In this frame the system Hamiltonians of Eq.~(\ref{eq:HjExtended}) become
\be
H_j = - i G_j(t) \p{ e^{i \theta_j(t) } a_j \ketbra{e_j}{g_j} - \Hc }
\ee
with $\theta_j(t) := \phi_j(t) + \int dt~\delta\om_j(t) - d_j t$. Note that $H_\textrm{tl}$ is also modified in this new frame, in particular $\zeta \ra \zeta' \equiv \zeta +d_2 - d_1$ in Eq.~(\mainref{eq:Htl}).
Now by selecting the `nice' laser frequencies and phases for each node, such that $\theta_j \equiv 0$,\footnote{
Technically we just have that $\theta_j$ are constants, as with $\dot\phi_j + \delta\om_j \equiv 0$ and $d_j = 0$ we have $\theta_j(t) = \phi_j - \int dt~\dot\phi_j = \textrm{constant}$. However, we can fix their value by redefining the phase of $\ket{e_j}$ or equivalently fixing an appropriate lower bound of the improper integrals.
}
we obtain the effective node Hamiltonians of Eq.~(\mainref{eq:HjSimplified}).
Note that as $d_j = 0$ in this case, $H_\textrm{tl}$ is ultimately unchanged in this frame, with $\zeta' = \zeta$, so we still can still report Eq.~(\mainref{eq:Htl}) as the transmission line interaction Hamiltonian in this frame (considering the specification of footnote \ref{note:Htl}).
Note we must also account for the modification to the state $\ket{\psi}$ in this new frame as
\be
\ket{\psi'} = U \ket{\psi} 
= \alpha_1 e^{i \p{\phi_1 + \int dt~\delta\om_1}} \ket{e g} \ket{00} + \alpha_2 e^{i \p{\phi_2 + \int dt~\delta\om_2}} \ket{g e} \ket{00} 
+ \beta_1 e^{i d_1 t}  \ket{g g} \ket{10} + \beta_2 e^{i d_2 t}  \ket{g g} \ket{01}
\ee
such that for the selected laser parameters the additional phases cancel out and we simply obtain Eq.~(\mainref{eq:psiSingleEx}).

We will work with this simpler case in the following two subsections as the extra freedom in possible amplitudes we could have by making other choices is not necessary for conveying the points we want to make.
We will return to this more general case without specifying a specific laser frequency or phase in Sec.~\hyperref[subsec:wavePacketShape5]{B5} where we consider the class of $\beta_1$ and hence wave packets that can possibly be produced. This includes analysis of the possible dynamical phases of $\beta_1$, which have been constant thus far as we have assumed the `nice' laser frequency and phase are used.
We note that several other groups have conducted similar analysis of the controlled generation of single-photon wave packet shapes from quantum emitters in the context of cavity QED  \cite{gorshkov2007photon,specht2009phase,muller2017spectral,morin2019deterministic}.
Also see Ref.~\citenum{penas2023improving} for analysis showing how dynamically tuning the system controls [$G_j(t)$ and $\phi_j(t)$ in our work] can be used to compensate for errors caused by distortion of photons in the quantum channel [related to strategy 5) in Sec.~\mainref{subsec:strategies}] as well as non-Markovian effects induced by quickly modulating the coupling of a quantum emitter to a corresponding photonic mode. 
To this end, we note that one needs to be careful when considering wave packets with more complicated phases, especially if they are fast varying, as they may be incompatible with the Markov approximation used to obtain the tractable amplitude EOMs considered here, Eq.~(\ref{eq:generalAmpEOMs}). This Markov approximation is standard in input-output theory, as mentioned in Sec.~\hyperref[subsec:modelGenerality]{A2}, and effectively assumes that the produced photon wave packet only experiences small variations about a central carrier frequency.
However, this will not be the case when using large frequency bandwidth photons, which can be useful in achieving faster rates [additionally, in our protocol they are more robust to frequency errors, see Eq.~(\ref{eq:PsuccFreqError})], as then corrections beyond the Markov approximation are needed to obtain analogous EOMs that take into account non-trivial memory kernels and can be used to engineer improved system controls \cite{penas2023improving}. 
A similar photon bandwidth dependent tradeoff exists between protocol rates and material dispersion. Namely, larger frequency bandwidth photons are more susceptible to dispersion effects while small bandwidth photons are temporally long so the drives must be comparably slow, ultimately leading to a decreased rate of QST or entanglement generation. 

\phantomsection
\subsection*{2.~Wave packet for a given $\boldsymbol{\alpha_1}$} \label{subsec:wavePacketShape2} 
In Ref.~\citenum{randles2023quantum} we found that in the special case mentioned above, where we select the laser phases and frequencies such that $\dot\phi_j + \delta\om_j \equiv 0$ and $d_j = 0$, by specifying $\alpha_1$ (which can be taken to be real in this case without loss of generality as discussed in Sec.~\mainref{subsec:systemEvolution}) the corresponding laser pulse (if it exists) is
\be\label{eq:G1inTermsOfAlpha1}
G_1(t) =   \frac{-\dot\alpha_1 e^{\gamma_1 t/2} }{\sqrt{e^{ \gamma_1 t_p} \beta _1^2(t_p) - 2\int_{t_p}^t dt' e^{\gamma_1 t'} \dot\alpha_1(t') \alpha_1(t')}}.
\ee
Note that this formula has been derived in the large atomic cooperativity limit. Similar formulae that explicitly account for cooperativity can also be derived, though they often rely on different approximations \cite{gorshkov2007photon,morin2019deterministic}. 
Importantly, in Sec.~\hyperref[subsec:atomicDecay3]{B3} we find that the losses due to not accounting for the presence of spontaneous decay in the drives $G_j$ themselves are minimal and so the above formula can still be used for nodes with moderate cooperativities.
The overall minus sign here encodes that of Eq.~(\ref{eq:genEOMa}) starting from the initial condition at $t = t_p$. However, the amplitudes EOMs are preserved (remain true) if precisely two of the functions $\left\{ G_j, \alpha_j, \beta_j \right\}$ are negated for both $j=1,2$. Here $t_p$ is some early time by which the first atomic state is \textit{prepared} in the desired superposition state, $c_g \ket{g_1} + c_e \ket{e_1}$, so $\beta _1(t_p) = \sqrt{1- \alpha_1^2(t_p)}$ should be zero. The corresponding wave packet is then determined by $\beta_1(t) = -\dot\alpha_1(t)/G_1(t)$ as given by Eq.~(\mainref{eq:EOMa}).
From Eq.~(\ref{eq:G1inTermsOfAlpha1}) we see that only some amplitude functions $\alpha_1(t)$ have a corresponding pulse $G_1(t)$. Specifically, we have the consistency condition that $G_1$ is real though it need not be positive (this is by construction as the corresponding phase has already been taken out and used to cancel the ac Stark shift of the qubit) so that the argument of the square root in Eq.~(\ref{eq:G1inTermsOfAlpha1}) must be positive and hence we have the condition on $\alpha_1$:
\be\label{eq:alpha1Ineq}
1- \alpha_1^2(t_p) - 2\int_{t_p}^t dt' e^{\g_1 (t' - t_p)} \dot\alpha_1(t') \alpha_1(t') \geq 0.
\ee
As an intuitive example, we cannot sustain Rabi-like oscillations $\alpha_1(t) = \cos(\Omega t)$ unless the cavity does not let any excitations decay out of it, i.e., if $\g_1 = 0$. Note, however, that any $\alpha_1 \geq 0$ that monotonically decreases, as is appropriate for QST, can be produced (at least in principle, in practice this is limited due to a finite emitter cooperativity). Qualitatively, this restriction dictates that we cannot make $|\alpha_1|$ increase too much relative to the natural transmission rate out of the first cavity.

For the logistic $\alpha_1$ of Eq.~(\mainref{eq:logA1}), we can use Eq.~(\ref{eq:G1inTermsOfAlpha1}) to solve for the corresponding pulse $G_1$, and then find the first cavity amplitude 
\be\label{eq:cavity1AmpLogAl1}
\beta_1(t) = z \sqrt{
\frac{1}{(1+z^2)^2} + (1-r) \p{\frac{1}{1+z^2} - r \Phi_L(-z^2, 1, 1+r) }
}
\ee
with $z(t) := e^{k t}$ and $r \equiv \g_1/2 k$. Here $\Phi_L$ is the analytically continued Lerch transcendent (for general parameter values a special function is needed to express the exact solution). 
For $r=1/2$ (which aligns with the case used for most plots within the main text with  $k=\g_1 = 2$) we have 
\be\label{eq:beta1rOfOneHalf}
\beta_1(t)|_{r= 1/2} = \sqrt{ \frac{z^2 - 1}{2 (z^2 + 1)} + \frac{\arctan(z)}{2 z} }
= \frac{1}{2} e^{-k t/2} \sqrt{ 2 \arctan(e^{k t}) + \sech(k t) \tanh(k t) },
\ee
which is not symmetric in time $t$ [see Figure 3 (a) of Ref.~\citenum{randles2023quantum}], whereas for $r=1$ (so $k = \g_1/2$) we have
\be
\beta_1(t)|_{r= 1}  = \frac{z}{1+ z^2} = \frac{\sech(k t)}{2},
\ee
which is symmetric in time. In the special limit $r \ra 0$ ($k \ra \infty$ with $\g_1 > 0$ constant)
we find that 
\be\label{eq:a1Largek}
\alpha_1(t) \overset{r \ra 0}{\longrightarrow}  \Theta(-t),
\ee
with $\Theta(t)$ the Heaviside step function. Meanwhile the leading-order asymptotics  for $\beta_1$ and $G_1$ are  \cite{ferreira2004asymptotic}
\be\label{eq:b1Smallr}%
\beta_1(t) \overset{r \ll 1}{\approx} 
\begin{cases}
\sqrt{1 - \frac{1}{ \p{1+ e^{2 k t}}^2 }}, & t < 0 \\
\sqrt{e^{- \g_1 t} - \frac{1}{ \p{1+ e^{2 k t}}^2 }}, & t \geq 0
\end{cases} \overset{r \ra 0}{\longrightarrow} 
\Theta(t) e^{-\g_1 t/2}
\ee
and
\be
G_1(t)  \overset{r \ll 1}{\approx} 
\frac{k \sech(k t)}{\sqrt{2 +e^{2 k t} }}
\overset{r \ra 0}{\longrightarrow} \frac{\pi}{2} \delta(t).
\ee 
The normalization of the delta function laser pulse, $\underset{r \ra 0}{\lim}~G_1$, comes from computing the area of $G_1(t) \forall t$ for small $r$, yielding $\pi/2 + \mathcal{O}(r)$.\footnote{
Note that the given $r \ll 0$ ($k \gg \g_1$) asymptotics of the functions $\beta_1$ and $G_1$ converge pointwise, whereas the given $r \ra 0$ limits do not converge at $t=0$, where the product $k t$ is undefined.
This can lead to confusion when trying to verify that these limits are consistent with Eqs.~(\mainref{eq:EOMa}-\hyperref[eq:EOMb]{b}) 
due to indeterminate limiting values at $t=0$. 
For instance, even though $G_1$ approaches an impulse response with $\pi/2$ normalization as $k \ra \infty$, and $\alpha_1(t=0)=1/2 \forall k$, consideration must be taken when considering their product $G = G_1 \alpha_1$ as
$\underset{k \ra \infty}{\lim} G(t)= \delta(t) \neq \underset{k \ra \infty}{\lim} G_1(t) \underset{k \ra \infty}{\lim} \alpha_1(t) = \frac{\pi}{2} \delta(t) \Theta(-t) = \frac{\pi}{4} \delta(t)$
because more carefully $G_1(t) \overset{r \ra 0}{\longrightarrow} \frac{\pi}{2} \delta(t + \mathcal{O}(r))$.
Importantly, there are only seeming discrepancies in the unphysical limit where $k \ra \infty$. For large but finite values of $k$, the functions $\alpha_1$, $\beta_1$, and $G_1$ are smoothed out and we avoid any pathologies in dealing with $\Theta(t)$ and $\delta(t)$.
}

\phantomsection
\subsection*{3.~Impact of spontaneous decay}\label{subsec:atomicDecay3} 
Here we further justify why, under the limits we consider in this work (of large detuning $\Delta_j$ and cooperativity $C_{j}$ for each node),
the impact of spontaneous decay of the material qubit, with rate $\Gamma_j$, can be set aside [see strategy 1)]. We start by explicitly incorporating spontaneous decay into our model effective Hamiltonian and computing the resulting dynamics in the single-excitation subspace. We then show how our dynamics are retrieved in the  aforementioned limits. Next, we numerically investigate the impact of spontaneous decay and find that for moderate cooperativities and large (yet achievable) detunings, we obtain a survival probability for emission (or absorption) of just under $\mathcal{P}_\textrm{em-max} = C_\textrm{em}/(1 + C_\textrm{em})$ with $C_{\textrm{em}} \ra C_j = 4 g_j^2/(\g_j \Gamma_j)$ for node $j$, validating it as a nearly achievable $C_\textrm{em}$-dependent maximum.
Moreover, we show that, without even modifying our drives $G_j$, the emitted wave packet shape does not substantially change for nonzero $\Gamma_j$ (again, under the above limits) as compared to the the ideal case with no spontaneous emission. 
In particular, once photon loss has been accounted for via $\mathcal{P}_\textrm{em}$, 
for moderate $C_j$ one can obtain a near unity state overlap (fidelity) of the wave packet emitted in the presence of spontaneous emission with the target wave packet. 
As an example, for the achievable value of $C_1 = 5$, these overlaps are around $99 \%$ across each of several instances we consider below (see Fig.~\ref{fig:overlapsVsC1}).
Thus, the ensuing analysis serves as a partial validation of the post hoc treatment of strategy 1) (see Sec.~\mainref{subsec:strategies}
of the main text and Sec.~\hyperref[subsec:standardErrors]{D1}) 
and further, it demonstrates why, in the rest of our analysis, we do not need to consider errors caused by the drives $G_j$ not being tuned for specific decay rates $\Gamma_j$ (such errors do occur yet they are significantly less relevant than $\mathcal{P}_\textrm{em}$). 
Our numerical treatment is meant to be indicative rather than fully comprehensive; see Refs.~\citenum{gorshkov2007photon,vogell2017deterministic,morin2019deterministic} for some related analysis, e.g., Ref.~\citenum{gorshkov2007photon} are able to explicitly show the existence of the aforementioned upper bound $\mathcal{P}_\textrm{em-max}$ for emission and/or absorption, albeit in slightly different parameter regimes.

When employing the quantum jump (stochastic wavefunction) method we use in our work, the effective system dynamics are governed by a non-Hermitian Hamiltonian, $H_\textrm{eff}$ of Eq.~(\mainref{eq:Heff}) in our case. In this context, the role of spontaneous emission from the upper level in the Raman process can be accounted for using a complex detuning parameter $\Delta_j \rightarrow \Delta_j + i \Gamma_j/2$ in $H_\textrm{eff}$ \cite{cirac1997qst}. 
Here $\Gamma_j$ is the spontaneous decay rate of the upper level $\ket{r_j}$ (elsewhere in our work it is referred to as $\Gamma_{sd}$ for a generic node). 
As the transmission line Hamiltonian does not depend on $\Delta_j$, we need only analyze the impact of spontaneous decay on the system Hamiltonians $H_j$ and the resulting amplitudes EOMs.
Thus, we apply the map $\Delta_j \rightarrow \Delta_j + i \Gamma_j/2$ to Eq.~(\ref{eq:HjExtended}) and as $\Delta_j$ only appears in the denominator of part of $d_j$, $\delta\om_j$, and $G_j$ we consider the map (for each $j=1,2$)
\be
\frac{1}{\Delta} \ra \frac{1}{\Delta + i \Gamma/2} 
= \frac{1}{\Delta} \frac{1}{1 + i \Gamma_r}
\approx \frac{1}{\Delta} \p{ 1 - i \Gamma_r + \mathcal{O}(\Gamma_r^2) },
\ee
where we have defined the rescaled decay rate $\Gamma_r = \Gamma/(2\Delta)$.
Therefore, in the the system Hamiltonians, we have the replacements:
$d \ra \frac{g^2}{\Delta (1 + i \Gamma_r)} - \delta$,
$\delta\om \ra \frac{\delta\om}{1 + i \Gamma_r}$,
and
$G \ra \frac{G}{1 + i \Gamma_r}$
for each $j$. Thus, the ensuing amplitude EOMs, previously Eqs.~(\ref{eq:generalAmpEOMs}a-c), likewise  undergo this map and become
\begin{subequations}
\begin{align} 
\dot\alpha_j &= \frac{-G_j}{1 + i \Gamma_{r,j}} \beta_j - i \p{\dot\phi_j + \frac{\delta\om_j}{1 + i \Gamma_{r,j}} } \alpha_j \qquad (j=1,2), \\
\dot\beta_1 &= \frac{G_1}{1 + i \Gamma_{r,1}} \alpha_1 - \br{ \frac{\gamma_1}{2}  + i \p{\frac{g_1^2}{\Delta_1 (1 + i \Gamma_{r,1})} - \delta_1} } \beta_1, \\
\dot\beta_2 &=  \frac{G_2}{1 + i \Gamma_{r,2}} \alpha_2 - \br{ \frac{\gamma_2}{2}  + i \p{\frac{g_2^2}{\Delta_2 (1 + i \Gamma_{r,2})} - \delta_2} } \beta_2 - \sqrt{\g_2} e^{i \zeta t} 
\Psi(t),
\end{align}
\end{subequations}
where the $G_j$ and $\delta\om_j$ shown here are the original functions with the $\Gamma_j$ dependence already factored out.

To make the resulting EOMs more tractable (and ultimately more useful), we will again select the `nice' laser frequency and phase satisfying $\delta = g^2/\Delta$ and $\dot\phi  =  -\delta\om$ for each node. 
Recalling $\Gamma_{r,j} = \Gamma_j/(2\Delta_j)$, noting that $\delta\om_j = \Delta_j G_j^2/g_j^2$, defining the cooperativity $C_j \equiv 4 g_j^2/(\g_j \Gamma_j)$, and doing some algebra we find
\begin{subequations}\label{eq:ampEOMsWithDecay}
\begin{align} 
\dot\alpha_j &= \frac{-G_j}{1 + i \Gamma_{r,j}} \beta_j - \frac{2 G_j^2}{\g_j C_j (1 + i \Gamma_{r,j})}  \alpha_j \qquad (j=1,2), \label{eq:sdEOMa} \\
\dot\beta_1 &= \frac{G_1}{1 + i \Gamma_{r,1}} \alpha_1 - \frac{\g_1}{2} \p{ 1  + \frac{\Gamma_{r,1}^2 C_1}{1 + i \Gamma_{r,1}} } \beta_1, \label{eq:sdEOMb} \\
\dot\beta_2 &=  \frac{G_2}{1 + i \Gamma_{r,2}} \alpha_2 - \frac{\g_2}{2} \p{ 1  + \frac{\Gamma_{r,2}^2 C_2}{1 + i \Gamma_{r,2}} } \beta_2 - \sqrt{\g_2} e^{i \zeta t} 
\Psi(t). \label{eq:sdEOMc}
\end{align}
\end{subequations}
Note that if we take the large cooperativity $C_j \ra \infty$ and large detuning $|\Delta_j| \ra \infty$ limits (namely, $\Gamma_{r,j} \ra 0$ here)\footnote{
For these simultaneous limits to be defined, specifically in the $\dot\beta_j$ equations, we consider the single $\Gamma_j \ra 0$ limit with the other parameters constant such that $\Gamma_{r,j}^2 C_j \ra 0$.
}
we obtain the original EOMs Eqs.~(\mainref{eq:coefEOMs}a-d).
If we only take the large detuning limit (as is consistent with the adiabatic elimination of the upper level already performed), the $\beta_j$ EOMs are the same as in the main text, yet the atomic excitation amplitude EOMs acquire a new term: 
\be\label{eq:sdEOMaLargeDetuning}
\dot\alpha_j = -G_j \beta_j - \frac{\tilde{\Gamma}_j}{2} \alpha_j \qquad (j=1,2), 
\ee
with $\tilde{\Gamma}_j(t) := \frac{4 G_j^2(t)}{\g_j C_j} = \p{  \frac{ \Omega_j(t)}{2 \Delta_j} }^2 \Gamma_j $,
which is the same as equation (41) of Ref.~\citenum{vogell2017deterministic}. 

Now we want to solve these amplitude EOMs and determine how well an excitation is transferred from the first emitter to its cavity at node 1 and then from cavity 2 to the emitter at node 2.
For concreteness, we will focus on emission from node 1 as absorption at node 2 is the effective time-reversed process so the impact of spontaneous decay can be gleaned from either. In particular, we aim to analyze the impact of spontaneous decay (nonzero $\Gamma_j$) on the produced wave packet $\Psi(t) = \sqrt{\g_1} \beta_1(t)$, i.e., how $\Psi$'s normalization and shape compare to the target wave packet $\Psi_0(t) := \Psi(t)|_{\Gamma_1 \ra 0}$. The same kinds of degradations will likewise occur during and affect absorption (with potentially different parameters).
As even the node 1 EOMs for $\alpha_1$ and $\beta_1$, which are uncoupled from node 2, are analytically quite unwieldy, 
we will solve them numerically focusing on the case considered in Sec.~\mainref{subsec:numericalResults} of the main text with a target logistically decreasing $\alpha_1$, see Eq.~(\mainref{eq:logA1}). 
The corresponding applied drive $G_1$ is given by Eq.~(\ref{eq:G1inTermsOfAlpha1}) such that in the ideal, target case (with $\Gamma_1 \ra 0$) the cavity amplitude is given by Eq.~(\ref{eq:cavity1AmpLogAl1}), which here we deem $\beta_1^{\textrm{(target)}}(t)$.
Based on this target case, we will work in terms of the dimensionless time $\tau = k t$ (e.g., $1/k$ is the natural unit of time), often making a change of variables to $z = e^\tau$ for ease of our numerics.

Thus, our current task is to analyze the wave packets $\Psi(\tau) = \sqrt{2 r} \beta_1(\tau)$ produced when applying the unaltered $G_1(\tau)$ (calculated based on the idealized target case) in the presence of spontaneous decay, i.e., with nonzero $\Gamma_1$ and hence finite $C_1$, 
and compare them to the target wave packet $\Psi_0(\tau) = \sqrt{2 r} \beta_1^{\textrm{(target)}}(\tau)$. 
Note that here $r=\g_1/2k$, as in the previous subsection, and the prefactor of $\beta_1$ has changed to maintain wave packet normalization after the change of variable: $\int_{-\infty}^\infty d\tau |\Psi_0(\tau)|^2 =1$.
We make this comparison across various cooperativity values, $C_1$, and under two spontaneous decay models for node $j=1$: \\ 
\indent
\begin{minipage}{\textwidth}
i) in the large detuning $\Gamma_{r,1} \ra 0$ limit given in Eq.~(\ref{eq:sdEOMaLargeDetuning}) for $\alpha_1$ and Eq.~(\mainref{eq:EOMb}) for $\beta_1$
and \\
ii) for finite $\Delta_1$ using Eq.~(\ref{eq:ampEOMsWithDecay}), where as a reasonable instance we will consider cases with $\Gamma_1 = \g_1$ (as is often nearly the case, see Table \ref{tab:extendedParameterTable}) and take the modest $\Delta_1 = 5 \max\{ g_1, 2\g_1\}$, which will typically be on the order of $\sim\!100$ MHz for optical nodes (we do not want it so large that we cannot distinguish this case from model i), such that $1/\Gamma_{r,1} = 5 \max\{\sqrt{C_1}, 4 \}$.
\end{minipage}
We denote the respective wave packets produced under these the models as $\sqrt{2 r} \beta_1^{(m)}(\tau)$ with $m=$ i, ii.

\begin{figure}[htp]
\centering
\hspace{-2em}
\begin{minipage}{0.35\textwidth}
(a) $r=1/4$
\includegraphics[height=18em]{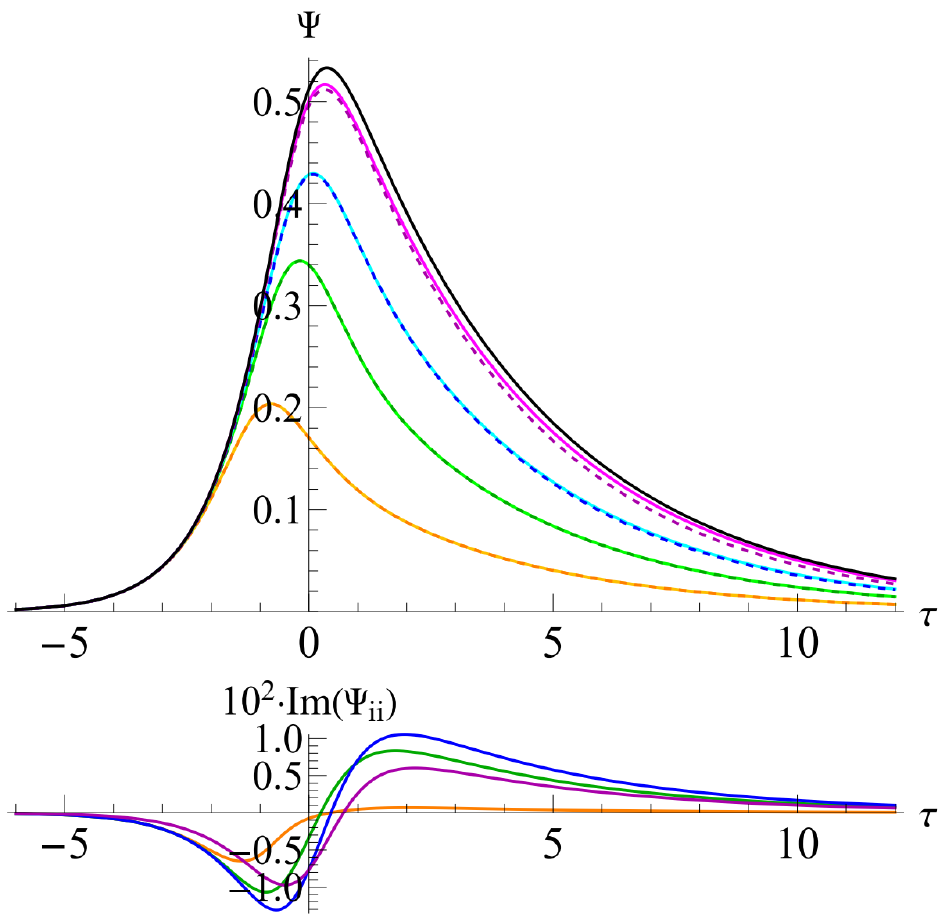}
\end{minipage}\hfill
\begin{minipage}{0.35\textwidth}
(b) $r=5$
\includegraphics[height=18em]{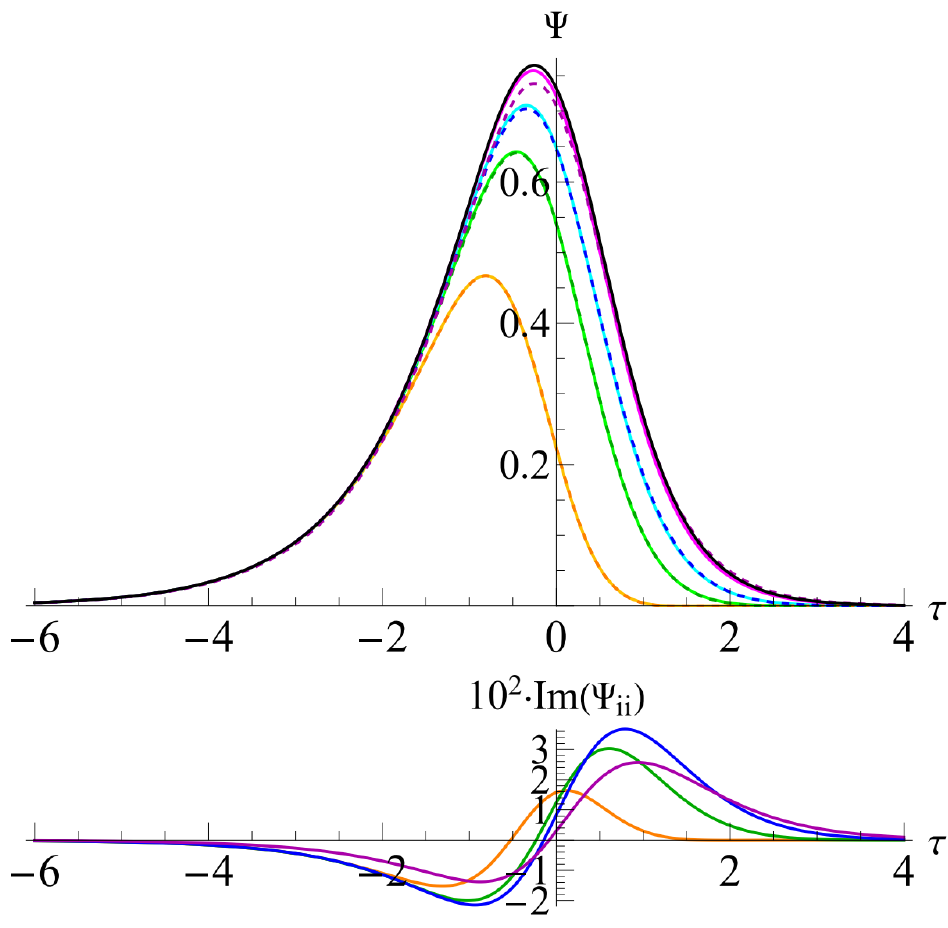}
\end{minipage}\hfill
\hspace{-2em}
\begin{minipage}{0.18\textwidth}
\vspace{1em}
\includegraphics[width=\textwidth]{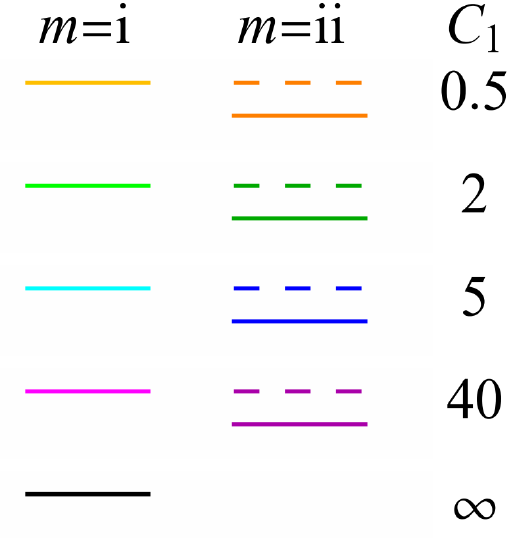}
\end{minipage}
\caption{
Plots of $\Psi$ as a function of $\tau$ for each model $m=$ i, ii, with several $C_1$ values including the ideal $C_1\ra \infty$ case, and for two disparate $r$ values. 
In the model ii, a nontrivial phase is acquired so in the upper panel we are actually just plotting the real component (dashed dark lines) with the small imaginary component (solid dark lines) being plotted in the lower panel (note the scaling).
This does not affect model i (solid light lines), wherein $\Psi$ is real.
(a) Faster driving case with $r=1/4$, which closely resembles Eq.~(\ref{eq:b1Smallr}).
(b) Slower driving case with $r=5$, which under model ii acquires a larger relative imaginary component as compared to the small $r$ case in (a) (which encourages the use of a larger detuning).
For each $r$ and under each model, the wave packets quickly approach the target shape as $C_1$ is increased as quantified in Figs.~\ref{fig:etaVsC1} and \ref{fig:overlapsVsC1}.
}
\label{fig:wavepacketsWithDecay}
\end{figure}

In Fig.~\ref{fig:wavepacketsWithDecay} we plot $\Psi(\tau)$ for the cavity amplitudes $\beta_1^{\textrm{(target)}}(\tau)$, $\beta_1^{(\textrm{i})}(\tau)$, and $\beta_1^{(\textrm{ii})}(\tau)$ (broken up into real and imaginary components) for several $C_1$ values and two disparate $r$ values. 
Under both models, for large $C_1$, the wave packets produced closely match the target shape. 
We can quantify the extent to which the wave packets match by considering two key parameters for each model:  
the probability of an excitation being transferred from the emitter to the desired cavity mode 
\be
\mathcal{P}_\textrm{em} = \int_{\tau_0}^{\tau_f} d\tau \big|\Psi(\tau) \big|^2
\leftrightarrow 2 r  \int_{\tau_0}^{\tau_f} d\tau \big|\beta_1^{(m)}(\tau) \big|^2
\equiv \eta_m^2,
\ee 
where $\eta_m$ is the subnormalization of the wave packet,
and 
the overlap (fidelity) of the renormalized wave packet with the 
target wave packet
\be
\mathfrak{o}_m \equiv \left| 2 r  \int_{\tau_0}^{\tau_f} d\tau \p{  \frac{1}{\eta_m}\beta_1^{(m)}(\tau)}^* \beta_1^{\textrm{(target)}}(\tau) \right|^2.
\ee
By using these two parameters we separate the impacts of spontaneous decay on photon loss ($\mathcal{P}_\textrm{em}$) and photon shape ($\mathfrak{o}_m$).
The integrals bounds should extend over all times in principle, yet in practice the $\beta_1$ functions we consider here decay rapidly away from $\tau = 0$ so the bounds we use here of $\tau_0 = -15$ and $\tau_f = 15$ more than suffice for our purposes.\footnote{
Note that over this interval the target wave packets are slightly subnormalized. Namely, for the considered $r$ values of $\{ 1/4, 1/2, 1, 2, 5 \}$ we find $1 -  \int_{\tau_0}^{\tau_f} d\tau |\Psi_0(\tau)|^2$ of $\{ 4.6 \cdot 10^{-4}, 2.3 \cdot 10^{-7}, 1.9 \cdot 10^{-3}, 2.7 \cdot 10^{-7}, 6.5 \cdot 10^{-6} \}$, respectively. For comparisons between $r$ values, we account for these slight subnormalizations by dividing the corresponding $\beta_1$'s by $\br{\int_{\tau_0}^{\tau_f} d\tau |\Psi_0(\tau)|^2 }^{1/2}$.
Note these deviations are small enough that this step could suitably be omitted as each target wave packet is essentially normalized.
}

\begin{figure}[htp]
\centering
\begin{minipage}{0.48\textwidth}
(a) \hspace{22em} \textcolor{white}{ }
\includegraphics[width=\textwidth]{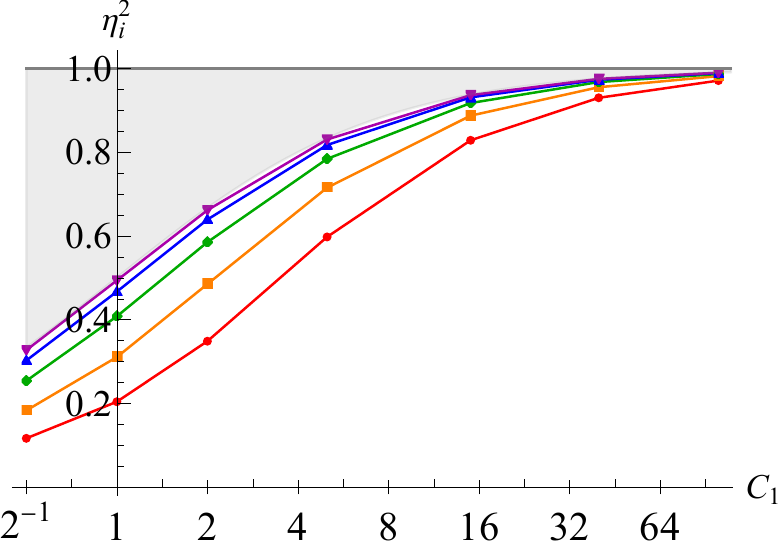}
\end{minipage}\hfill
\begin{minipage}{0.48\textwidth}
(b) \hspace{22em} \textcolor{white}{ }
\includegraphics[width=\textwidth]{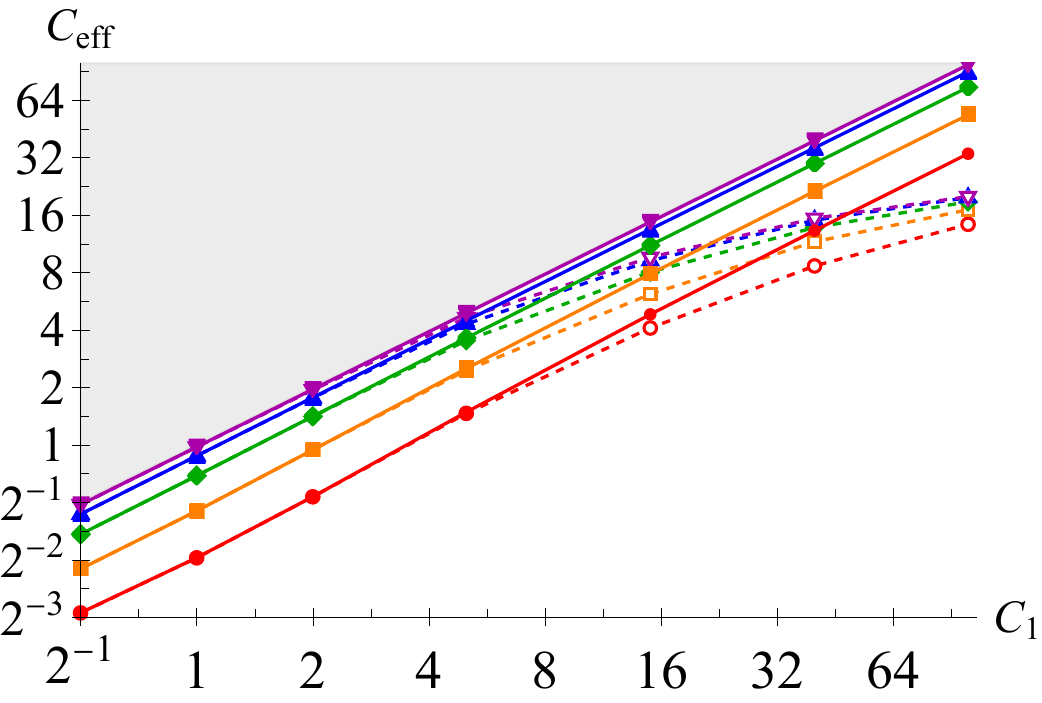}
\end{minipage}\par
\vskip\floatsep
\begin{minipage}{0.52\textwidth}
(c) \hspace{24em} \textcolor{white}{ }
\includegraphics[width=0.98\textwidth]{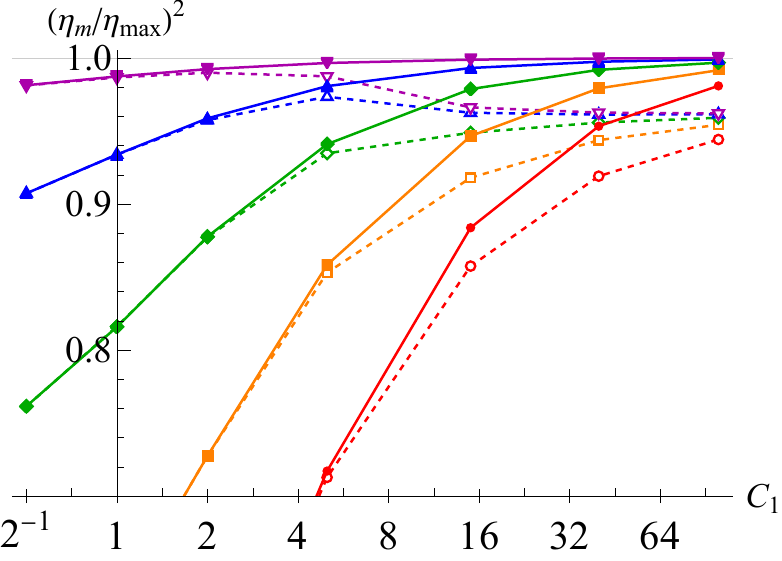}
\end{minipage}
\begin{minipage}{0.32\textwidth}
\includegraphics[width=0.8\textwidth]{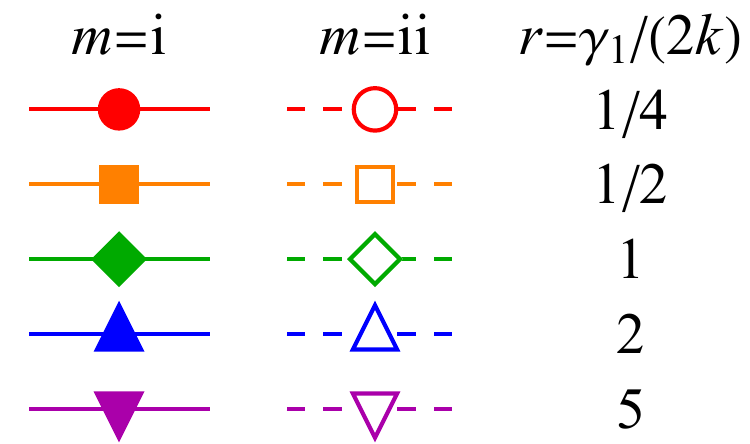}
\vspace{5em}
\end{minipage}
\caption{
Plots quantifying wave packet subnormalization $\eta_m$ due to photon loss caused by spontaneous decay during the Raman process 
for various $C_1$ and $r$ values under each model $m$ (see the legend).
(a) Plot of $\mathcal{P}_\textrm{em} = \eta_\textrm{i}^2$ versus $C_1$ for several $r$ values. 
The model ii curves are omitted in this case to reduce visual clutter yet are included in (b) and (c).	
(b) Plot of the effective cooperativity $C_\textrm{eff}$, 
which satisfies $\mathcal{P}_\textrm{em} = C_\textrm{eff}/(1 + C_\textrm{eff})$, versus $C_1$, which shows how fast driving lowers the achieved cooperativity relative to the ideal value. 
(c) The relative deviations of each $\eta_m^2$ from $\eta_\textrm{max}^2 \equiv \mathcal{P}_\textrm{em-max} = C_1/(1+C_1)$ as a function of $C_1$.
For large $r$ these curves approach the $C_1/(1 + C_1)$ bound, which is indicated via the light gray shading in (a) and (b) and the 1 line in (c).
In (b) and (c) we see that under model ii we end up doing significantly worse than model i for large $C_1$ as the values of $C_\textrm{eff}$ and $(\eta_m/\eta_\textrm{max})^2$ seem to converge to constant sub optimal values for each $r$. Note this does occur and is a result of how we choose the detuning $\Delta_1$. This behavior is ameliorated if one chooses even larger detunings, say scaling as $\sim C_1 \g_1$ for large $C_1$.
}
\label{fig:etaVsC1}
\end{figure}

\begin{figure}[htp]
\centering
\hspace{-2em}
\begin{minipage}{0.48\textwidth}
(a) 
\includegraphics[width=\textwidth]{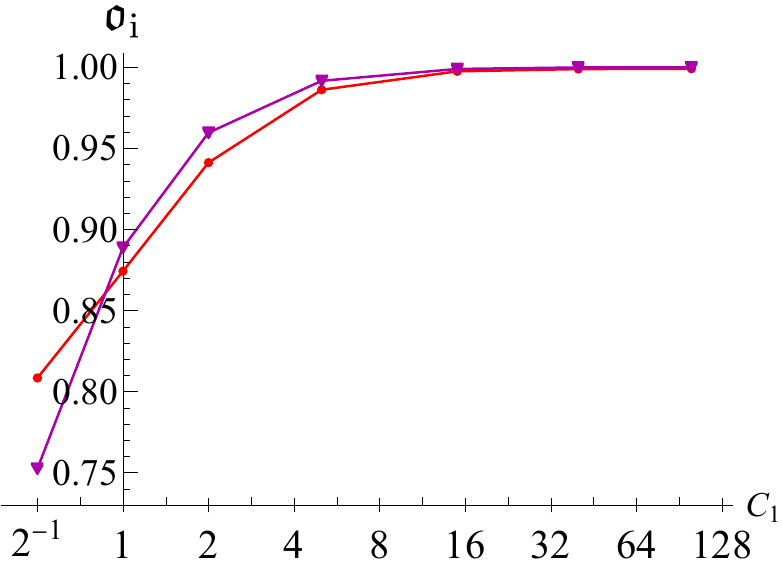}
\end{minipage}\hfill
\begin{minipage}{0.48\textwidth}
(b) 
\includegraphics[width=\textwidth]{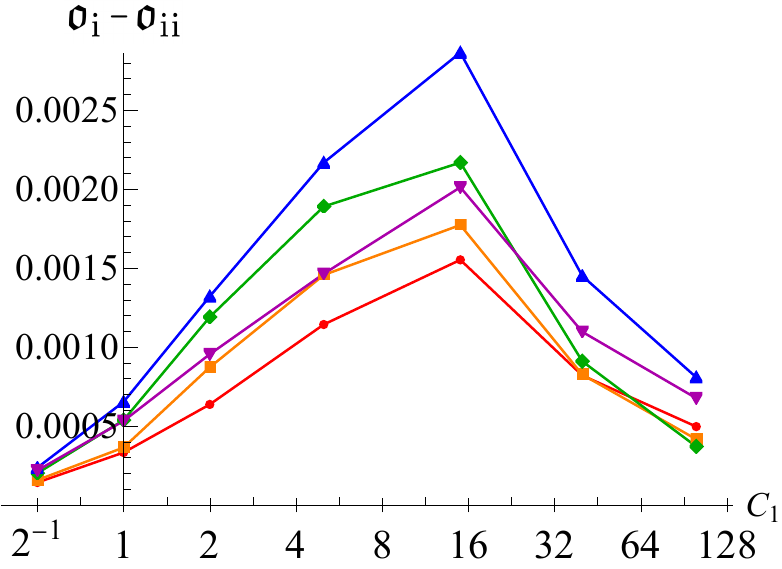}
\end{minipage}
\caption{
(a) Plot of the overlap $\mathfrak{o}_i$ versus $C_1$ for $r=1/4$ (red, circles) and $r=5$ (purple, triangles), other $r$ values are omitted to reduce visual clutter.
(b) Plot of the overlap differences between each model $\mathfrak{o}_\textrm{i} - \mathfrak{o}_\textrm{ii}$ versus $C_1$ for each of the $r$ values considered in Fig.~\ref{fig:etaVsC1}, using the same coloring at point marker scheme.
We see that $\mathfrak{o}_\textrm{i}$ quickly approaches 1 as $C_1$ increases.
For instance, for $C_1 = 5$, the smallest overlap $\mathfrak{o}_\textrm{i}$ we find is $98.6\%$ for $r=1/4$ and the overlap slightly increases as $r$ increase up to $99.2\%$ for $r=5$. Under model ii the overlaps only shift down slightly by $\sim 0.1\%$ as can be seen in (b).
}
\label{fig:overlapsVsC1}
\end{figure}

In Fig.~\ref{fig:etaVsC1} we plot $\mathcal{P}_\textrm{em} =\eta_m^2$ (in several forms) as a function of $C_1$ for several $r$ values. We find that $\mathcal{P}_\textrm{em}$ approaches 1 as $C_1$ tends to infinity for each $r$ and this approach happens faster for larger $r$, which correspond to a slower driving with small $k$ relative to $\g_1$. Accordingly, there is a trade off between photon generation rate (where one wants want large $k$) and successful emission. 
Notably, $\eta_\textrm{i}^2$ nearly saturates the $\mathcal{P}_\textrm{em-max} = C_1/(1 + C_1)$ bound for relatively large $r \gtrsim 5$, as is consistent with Refs.~\citenum{gorshkov2007photon,vogell2017deterministic,morin2019deterministic} and our strategy 1). This near saturation also happens in model ii for suitably large detunings.
Beyond subnormalization, the shape of the wave packets produced for finite $C_1$ will differ from target shape as is quantified by the overlap $\mathfrak{o}_m$. 
If this shape difference is considerable, then additional considerations would be required in our analysis to quantify this spontaneous decay induced shape mismatch as it would implicitly degrade the wave packet overlap in $\Psucc$. 
Accordingly, in Fig.~\ref{fig:overlapsVsC1} (a) we plot $\mathfrak{o}_\textrm{i}$ versus $C_1$ for the extremal values of $r$ we consider and in Fig.~\ref{fig:overlapsVsC1} (b) we show the overlap deviations between the models i and ii. We find that 
the overlaps quickly approach 1 as $C_1$ increases so the shape is largely left unchanged.
Moreover, if nodes 1 and 2 have similar cooperativities, the impact of this shape mismatch will be further suppressed. For instance, in model i, simulating emission and absorption between nodes with identical cooperativities $C$ (and assuming the unitary transformation is implemented perfectly with no other errors) one finds that $|\alpha_2(t_f)| = \mathcal{P}_\textrm{em}$ exactly, 
so the subnormalization fully accounts for excitation loss in the whole process. 
[We have performed some similar analysis of $\eta_m$ and $\mathfrak{o}_m$ for other drives (including a smoothed out square wave, a $\sech$ shape, and a $\sech$ shape with an additional small constant drive for $t>0$) and the corresponding wave packets, finding the same general behaviors as shown for this example.]
Thus, we do not need to modify the drives themselves as wave packet shape mismatch errors due to spontaneous decay are essentially negligible (for large $\Delta_j$ and moderate $C_j$). 
Accordingly, the impact of spontaneous decay 
can justifiably be treated post hoc via strategy 1) for large detuning and $\mathcal{P}_\textrm{em}$ nearly saturates its maximum value for relatively slow drives $G_j$.

\phantomsection
\subsection*{4.~Formal $\boldsymbol{\beta_1}$ and frequency errors}\label{subsec:formalBeta1FreqError4} 
Note that Eq.~(\mainref{eq:EOMb}) has the formal solution
\be
\beta_1(t) = \int_{t_p}^t dt' e^{-\g_1 (t - t')/2} G_1(t') \alpha_1(t'),
\ee
which is for the case where one selects the `nice' laser frequency and phases. Note that, as above, $t_p$ is a preparation time at which any excitation is solely in atom 1, it can mathematically be taken to be $-\infty$; doing so and defining $ H(t) := \Theta(t)  e^{-\g_1 t/2} $ and $G(t) := G_1(t) \alpha_1(t)$ it follows that
\be\label{eq:beta1AsConvolution}
\beta_1(t) = \int_{-\infty}^\infty dt' G(t') H(t-t')  = (G*H)(t).
\ee
We thus see that $G_1$ and $\alpha_1$, through their product $G$, act to ``smear out'' $H$ via the convolution of Eq.~(\ref{eq:beta1AsConvolution}).
This guides the intuition that $\beta_1$ should be narrowest for\footnote{
The unit weight follows from $G$ preserving the normalization of the wave packet with respect to the convolved function $H$. 
To be clear, we are not claiming that any $G(t)$ with this property can be produced; only certain $\alpha_1$ and hence $G$ are possible.
}
$G(t) = \delta(t)$ such that $\beta_1(t) = H(t) = \Theta(t) e^{-\g_1 t/2} $, which is exactly the $\beta_1$ we found in the $k \ra \infty$ limit above.

This was a slight degression, but the point we want to make is that the decay rate $\g_1$, which is fixed by the cavity-transmission-line interaction, ultimately limits the possible wave packets that can be produced. 
This is good to know because narrower wave packets are less susceptible to frequency errors. To see this, note that for ideal values of $\xi$ and $T$, assuming $l$ is large enough to transform the entire wave packet, we have 
\be\label{eq:PsuccFreqError}
\Psucc 
= \left| \int_{-\infty}^\infty d \mathcal{T}~
e^{i \frac{\Delta\om_0}{\g_2} \mathcal{T}}
\beta_1^2(\mathcal{T}/\g_1)
\right|^2
=  \left| 
\mathcal{F}\br{\beta_1^2}\p{ \frac{\Delta\om_0}{\g_2} }
\right|^2,
\ee
where $\mathcal{T} = \g_2 (T^* - t)$ is a dimensionless time variable with Fourier conjugate frequency variable $\Delta\om_0/\g_2$ with $\mathcal{F}[\cdot]$ denoting the Fourier transform.
Thus, if the oscillations in the phase $e^{i\frac{\Delta\om_0}{\g_2} \mathcal{T} }$ occur over short time scales relative to the temporal duration of $\beta_1$, then the integrand will tend to average out to zero. We can protect against this by reducing $\beta_1$'s temporal duration 
(broadening $\mathcal{F}\br{\beta_1^2}$) as then the phase will be more stable in the window where $\beta_1$ is appreciable, leading to a larger probability of success. 
As an example of this, in the limit of a impulse response laser pulse $G_1$, i.e.,~as $k \ra \infty$ in the logistic $\alpha_1$ case, we have that $\beta_1$ is given by  Eq.~(\ref{eq:b1Smallr}) and hence 
\be\label{eq:PsuccFreqErr}
\Psucc 
= \left| \int_{0}^\infty d \mathcal{T}~
e^{i \frac{\Delta\om_0}{\g_2} \mathcal{T}}  e^{-\mathcal{T}}
\right|^2
= \frac{1}{1 + (\Delta\om_0/\g_2)^2},
\ee
which is Lorentzian function of $\Delta\om_0$ with peak value of unity and FWHM of $2 \g_2$ [which is wider than the FWHM of $1.4 \g_2$ for the $k=2 \g_2$ case considered in the main text as $\beta_1(t)$ is narrower here]. 

This can be contrasted against the case where no unitary transformation is implemented. Using the same pulses $G_j$,
\be
\Psucc = \left| \int_{-\infty}^\infty dt'~\Phi^*(t') \Psi(t') \right|^2
=  \g_1 \g_2 \left|
\int_{-\infty}^\infty dt'~ e^{i \om_{0 i} t'} \beta_1(\xi_i (T_i-t')) \beta_1(t')
\right|^2,
\ee
which will clearly tend to average out to zero due to fast-oscillating phase $e^{i \om_{0 i} t}$ unless both $|\Psi(t)|$ and $|\Phi(t)|$ are narrowly peaked around the same time.
We can evaluate this integral exactly in the $k \ra \infty$ limit considered above, where $\beta_1(t) = \Theta(t) e^{-\g_1 t/2}$ so with $ g =(\g_1-\g_2)/2$ as intermediate shorthand
\be\label{eq:PsuccNoU}
\Psucc(T_i > 0) 
= \g_1 \g_2 e^{-\g_2 T_i} \left|
\frac{e^{(i \om_{0 i} -g) T_i} - 1}{i \om_{0 i} -g}
\right|^2 
\leq \frac{4 \g_1 \g_2 e^{-\g_2 T_i}}{\om_{0 i}^2 + g^2}
\leq \frac{4 \g_1 \g_2  }{\om_{0 i}^2 + (\g_1-\g_2)^2/4}
\leq \frac{4 \g_1 \g_2  }{\om_{0 i}^2}
\ee
while $\Psucc(T_i \leq 0) = 0$. 
This rightmost bound in Eq.~(\ref{eq:PsuccNoU}) is clearly relavent for large $|\om_{0 i}|$, in which case it naturally takes on a form similar to Eq.~(\ref{eq:PsuccFreqErr}) in the case of large $|\Delta\om_0|$. Clearly we see that without a unitary transformation, $\Psucc$ rapidly decreases as $|\om_{0 i}|$ gets large relative to $\g_{1,2}$ (as one would expect). 

\phantomsection
\subsection*{5.~Producible wave packets}\label{subsec:wavePacketShape5} 
Similar to how we specified $G_1$ in terms of $\alpha_1$ in Eq.~(\ref{eq:G1inTermsOfAlpha1}), we can alternatively specify $G_1$ in terms of $\beta_1$. Thus, just as we limited possible $\alpha_1$ in Sec.~\hyperref[subsec:wavePacketShape2]{B2}, here we will show that not all $\beta_1$  can be produced (and hence that system 1 cannot produce wave packets with entirely arbitrary temporal shapes).
From the EOMs for system 1, Eqs.~(\ref{eq:genEOMa}-\hyperref[eq:genEOMb]{b}), only can derive that  
\be\label{eq:genG1fromBeta1}
G_1^2(t) = \frac{e^{2i \theta_0(t)} \br{\dot\beta_1(t) + \p{i d_1 + \g_1/2} \beta_1(t)}^2 }{ 
\alpha_1^2(t_p) e^{2i \theta_0(t_p)} - 2 \int_{t_p}^t dt' e^{2i \theta_0(t')} \br{\dot\beta_1(t')  + \p{i d_1 + \g_1/2} \beta_1(t') } \beta_1(t') 
},
\ee
which is valid for different choices of laser frequency and phase determined by the parameters $\dot\theta_0 := \dot\phi_1 + \delta\om_1 = \dot\theta_1|_{d_1 = 0}$ and $d_1 = g_1^2/\Delta_1 - \delta_1$ as defined above.
As $G_1(t)$ is the laser pulse magnitude, which we picked to be positive without loss of generality by factoring out the phase $e^{i \phi_1(t)}$ in Eq.~(\ref{eq:HjExtended}), as a consistency condition we must choose $\phi_1(t)$ or equivalently $\theta_0(t)$ to cancel off the phase induced by the $\beta_1$ terms in the other parts of the expression.
In principle we could work with the condition that the whole expression for $G_1[\beta_1]$ must be real in order to determine the possible functions $\theta_0(t)$ and hence $\phi(t)$. However, doing so leads to a self-referential condition on $\theta_0(t)$ that is unwieldy. Hence one should solve these EOMs numerically in general, and when doing so it can be useful to work with the complex laser pulses $G_j e^{i \phi_j}$ directly instead of their polar decomposition.
Note that Ref.~\citenum{axline2018demand} get very similar EOMs for their complex control parameter $g(t)$ as compared to our $G_j e^{i \phi_j}$, see their supplemental Eq.~(10).
However, we can still make progress analytically by considering a subset of all possible $\beta_1$, namely those with either a certain `nice' phase
or
those that are slowly varying in senses that we will make precise.

\textbf{$\boldsymbol{\beta_1}$ with simple phase}.
One way to guarantee that we satisfy the consistency condition that $G_1$ is real is to set the numerator and denominator of Eq.~(\ref{eq:genG1fromBeta1}) to be real independently.
Doing so and using the decomposition $\beta_1(t) = B(t) e^{i b(t)}$ such that $\dot\beta_1(t) = \br{\dot{B}(t) + i \dot{b}(t) B(t)} e^{i b(t)}$ we have from the numerator that
\be\label{eq:GSqNumCond}
e^{2 i (\theta_0 + b)} \br{ \dot{B}  + \frac{1}{2} \g_1 B + i \p{\dot{b} + d_1} B }^2 
\in \mathbb{R}
\quad \implies \quad
2  (\theta_0 + b) + 2 \arctan\p{\frac{\dot{b} + d_1}{\dot{B}/B  + \g_1/2}} \equiv 0 \mod 2 \pi.
\ee

Note the denominator must be real for all $t \geq t_p$ so at $t = t_p$ we have that $\alpha_1^2(t_p) e^{2i \theta_0(t_p)} \in \mathbb{R}$, where $|\alpha_1(t_p)|$ should be 1 for an appropriate preparation time so we consider the phases to be such that  $\alpha_1^2(t_p) e^{2i \theta_0(t_p)} = 1$. Thus, for the denominator to be real at subsequent times the argument of the integral must be real, which along 
with the above condition from the numerator implies
\be
\arctan\p{\frac{\dot{b} + d_1}{\dot{B}/B  + \g_1/2}} \equiv 0 \mod 2 \pi
\ee
such that $\dot{b} = -d_1$ and hence $b(t) = b_0 - d_1 t$. Inserting this back into Eq.~(\ref{eq:GSqNumCond}) we find that $\theta_0 + b_0 - d_1 t \equiv n\pi$ for $n$ an integer, which then implies that $\theta_0(t) = d_1 t + c_0$ with $c_0 = n\pi - b_0$ some constant. 
Hence $\dot\phi_1(t) 
= d_1 -\delta\om_1(t)$ for which the $d_1$ shift acts as a base frequency for $\phi_1$, see Eq.~(\ref{eq:HjExtended}).
With these simplifications we have
\be\label{eq:G1InTermsOfB}
G_1(t) = \frac{
\dot{B}(t) + \tfrac{1}{2} \g_1 B(t)  
}{ \sqrt{
1 - 2 \int_{t_p}^t dt'
\br{
	\dot{B}(t') + \tfrac{1}{2} \g_1 B(t')  
} B(t')
}}, 
\ee
which is the same pulse we obtain in the case where the `nice' laser pulse are phase are used such that $\alpha_1$ and $\beta_1$ have a constant phase, so we can take them to be real.

Hence we have a closed form (integral) expression for $G_1$ in terms of the class of wave packets with $\beta_1(t) = B(t) e^{-i d_1 t}$ (dropping the constant phase $b_0$) for which the amplitude $B(t)$, which specifies the wave packet shape, satisfies the relation
\be\label{eq:beta1Ineq}
2 \int_{t_p}^t dt' \p{\dot{B} + \frac{\gamma_1}{2} B} B \leq 1 \quad \forall~t > t_p.
\ee
Note that this condition is more easily violated than the one for $\alpha_1$. For instance, there is no pulse $G_1$ that will produce the, seemingly reasonable, normalized Gaussian wave packet corresponding to  $B(t) = \p{\g_1 \sigma \sqrt{\pi} }^{-1/2} e^{-t^2/2 \sigma^2}$, which violates inequality (\ref{eq:beta1Ineq}). 
To see this we consider the inequality of Eq.~(\ref{eq:beta1Ineq}) for this $B$ taking $t_p \ra -\infty$: 
\be\label{eq:IGauss}
I_\textrm{Gauss}(t) :=
1- 2 \int_{-\infty}^t dt' \p{\dot{B} + \frac{\gamma_1}{2} B} B
=  \frac {\textrm{erfc}(t/\sigma)} {2} -  \frac {e^{-t^2/\sigma^2}} {\sqrt{\pi}\g_1 \sigma}
\overset{?}{\geq} 0,
\ee 
where 
\be
\textrm{erfc}(x) := \frac{2}{\sqrt{\pi}} \int_x^\infty dy~e^{-y^2}
\ee
is the complimentary error function. 
We can compute the extrema of $I_\textrm{Gauss}(t)$ using the dimensionless time $\tau = t/\sigma$ 
and decay rate $g = \g_1 \sigma/2$
\be
0 = \frac{d I_\textrm{Gauss}}{d \tau}  
= \frac{d}{d \tau} \p{
\frac {\textrm{erfc}(\tau)} {2} -  \frac {e^{-\tau^2}} {\sqrt{\pi} 2 g}
}
=  \frac{1}{\sqrt{\pi}} e^{-\tau^2} \p{\frac {\tau }{g} - 1}
\implies 
\tau = g > 0,
\ee
which is a minimum as
\be
\frac{d^2 I_\textrm{Gauss}}{d \tau^2}  \bigg|_{\tau = g}
= \frac{1}{\sqrt{\pi}} e^{-\tau^2}
\br{ \frac {1}{g} - 2 \tau  \p{\frac {\tau }{g} - 1} } \bigg|_{\tau = g}
= \frac{1}{\sqrt{\pi} g} e^{-g^2} > 0. 
\ee
Hence the minimum value is
\be
\min\br{I_\textrm{Gauss}(\tau)} 
= I_\textrm{Gauss}(\tau = g) = 
\frac{1}{2} \p{
\textrm{erfc}(g) -  \frac {e^{-g^2}} {\sqrt{\pi} g} 
}
< 0
\ee
as for $x > 0$
\be
\textrm{erfc}(x) 
< \frac{2}{\sqrt{\pi}} \int_x^\infty dy~\frac{y}{x} e^{-y^2}
= \frac{1}{\sqrt{\pi} x} \int_{x^2}^\infty d(y^2) e^{- y^2}
= \frac{1}{\sqrt{\pi} x } e^{-x^2}.
\ee
Hence the inequality of Eq.~(\ref{eq:IGauss}) is violated for all $g$ and hence all $\sigma$ as $I_\textrm{Gauss}$ goes negative for some times $t$, which entails that there is no pulse that will produce Gaussian wave packet with any standard deviation $\sigma$. 
Many other wave packets shapes do not have this drastic behavior that the Gaussian has, where they can never be produced ideally. For instance, the symmetric, normalized $B(t) = \p{2\g_1 \sigma}^{-1/2} \sech(t/\sigma)$ can be ideally produced for $\sigma \geq 2/\g_1$ setting a lower bound for such a photons temporal width. 
To show this one can consider the inequality of Eq.~(\ref{eq:beta1Ineq}) for this $B$, analogous to our treatment of the Gaussian case in Eq.~(\ref{eq:IGauss}):
\be\label{eq:ISech}
I_\textrm{sech}(t) :=
1- 2 \int_{-\infty}^t dt' \p{\dot{B} + \frac{\gamma_1}{2} B} B
=  \frac{1}{2} \p{ 1 - \tanh(t/\sigma) - \frac{\sech^2(t/\sigma)}{\g_1 \sigma} }.
\ee 
Then, one can easily show that $I_\textrm{sech}(t)$ goes negative for some $t$ if $\g_1 \sigma < 2$, whereas it is entirely positive for $\g_1 \sigma \geq 2$.

Note, however, we can rescale the desired $\beta_1$ by a weight $0 < w < 1$ such that we produce a subnormalized wave packet, i.e., a superposition of the vacuum and the wave packet.\footnote{
This can be done for any wave packet envelope provided the integral expression on the LHS of the inequality of Eq.~(\ref{eq:beta1Ineq}), which we will refer to as $\mathcal{I}(t)$, is finite $\forall~t$. Then one can simply choose $w^2 \leq 1/\max\{\mathcal{I}(t)\}$ to force the inequality to hold. 
}
Then, modifying the above Gaussian case, 
we have that if the quantity
\be
I_\textrm{Gauss}^{(w)}(\tau) 
=  1 - w^2 
+ \frac{w^2}{2} \p{
\textrm{erfc}(\tau)- \frac {e^{-\tau^2}} {\sqrt{\pi} g}
}
\ee 
is positive $\forall \tau$ then the corresponding $B$ can be produced. Using the above analysis this function has a minimum at $\tau = g$ of
\be
\min\br{I_\textrm{Gauss}^{(w)}(\tau)}
=  1 - w^2 
+ \frac{w^2}{2} \p{
\textrm{erfc}(g)- \frac {e^{-g^2}} {\sqrt{\pi} g}
}
\ee 
and as  
\be
\frac {-1} {\sqrt{\pi} g} <
-x_g \equiv
\textrm{erfc}(g) - \frac {e^{-g^2}} {\sqrt{\pi} g}
< 0,
\quad \textrm{for } g > 0 
\ee
it follows that
\be
\min\br{I_\textrm{Gauss}^{(w)}(\tau)} \geq 0 
\quad \textrm{if} \quad
w^2 \leq w_\textrm{max}^2 
\equiv \frac{1}{1 +  x_g/2}.
\ee
Hence, any Gaussian wave packet can be probabilistically produced with a probability of at most 
\be
P_\textrm{Gauss-prod} = w_\textrm{max}^2 =
\br{ 1 +    \frac {e^{-(\g_1 \sigma/2)^2}} {\sqrt{\pi} \g_1 \sigma} - \frac{1}{2} \textrm{erfc}(\g_1 \sigma/2) }^{-1}
\ee
depending on the Gaussian's standard deviation $\sigma$. Complimentarily, with probability $1 - P_\textrm{Gauss-prod}$, no photon will be produced and the excitation will remain in atom 1 [so a $\Upsilon_2$ error would occur in Eq.~(\mainref{eq:HeffMap})].
To actually produce the desired wave packet one could employ the requisite laser pulse as determined by Eq.~(\ref{eq:G1InTermsOfB}) with an appropriately subnormalized $B$ and herald the success of the production by a measurement on atom 1. Specifically, by measuring if atom 1 is in the state $\ket{e_1}$, if so the excitation has remained in atom 1 and no photon has been produced so the atom must be reset and the procedure repeated until atom 1 undergoes the transition $\ket{e_1} \ra \ket{g_1}$ accompanied by the emission of the desired wave packet. 
Thus, in principal one could produce a very narrow wave packet with $\sigma \ll \g_1^{-1}$ provided they are willing to run the protocol enough times to combat the low probability of production, e.g., a given production attempt can obtain $\sigma = 10^{-10}/\g_1$ with the dismal probability of $w_\textrm{max}^2 \approx 2 \cdot 10^{-10}$. Note, however, one can obtain a broader wave packet with more reasonable chance, e.g., $w_\textrm{max}^2 \approx 0.83$ for  $\sigma = \g_1^{-1}$.

We acknowledge that this ability to probabilistically produce a desired photon wave packet is not very useful in our case here where we want to implement deterministic QST between two nodes whose Hamiltonians are equivalent up to parameter differences that can be circumvented using our unitary.
However, we still include this discussion as this ability and related questions like `What photon wave packets can be produced?' are interesting and not fully explored topics that may be applicable in different contexts. 
For instance, such probabilistic photon production may be of interest in heralded remote entanglement generation experiments (see Sec.~\hyperref[subsec:heraldedProtocols]{A7}), where the ultimate remote entangled state fidelity is largely determined by the spectral overlap of the photons being sent by two separate nodes. In such a case, one can have additional control of the generated photon wave packet, potentially leading to higher quality entanglement generation, at the expense of a decrease in rate due to the probabilistic production.
The lack of utility in using such probabilistic production in our scheme comes from doing so being unnecessary: there are wave packets we can produce deterministically (or at least with higher probability, perhaps limited by the emitter itself) so why degrade the protocol by limiting the photon production. 

Moreover, such a probabilistically produced wave packet, like the Gaussian example, will not be probabilistically absorbed with same probability it was emitted. Namely, the absorption probability will be lower than $w^2$ if we initialize node 2 in the ground state $\ket{g_2}$ as usual, $\alpha_2(-\infty) = 0$, as then the absorption process will not simply be the effective time-reversed process to emission (which here would require that $|\alpha_2(-\infty)| = |\alpha_1(+\infty)|$).
In principle, such errors could naturally be corrected using the schemes discussed in Sec.~\mainref{subsec:errorCorrection} and in Sec.~\hyperref[sec:ECZprotocols]{D}, though this would be at the expense of more protocol repetitions that could be avoided by using a deterministically producible wave packet.
If we know we are only probabilistically producing a photon, and assuming we can calculate the value of $\alpha_1$ for long times, we could alternatively prepare the state $ \ket{\tilde{g}_2 } \equiv \ket{e_2}\alpha_1(\infty) + \sqrt{1 - \alpha_1^2(\infty)} \ket{g_2}$ in the attempt to improve absorption. However, this leads to other issues, principally, bit-flip errors can occur both ways: $\ket{\tilde{g}_2 } \leftrightarrow \ket{\tilde{e}_2 }$ (where $\ket{\tilde{e}_2}$ is an orthonormal state to $\ket{\tilde{g}_2}$ in the $\{\ket{g_2}, \ket{e_2}\}$ subspace), which renders this approach not useful as the ECZ error correction methods cannot be used.

\begin{figure*}[ht] 
\includegraphics[width=\linewidth]{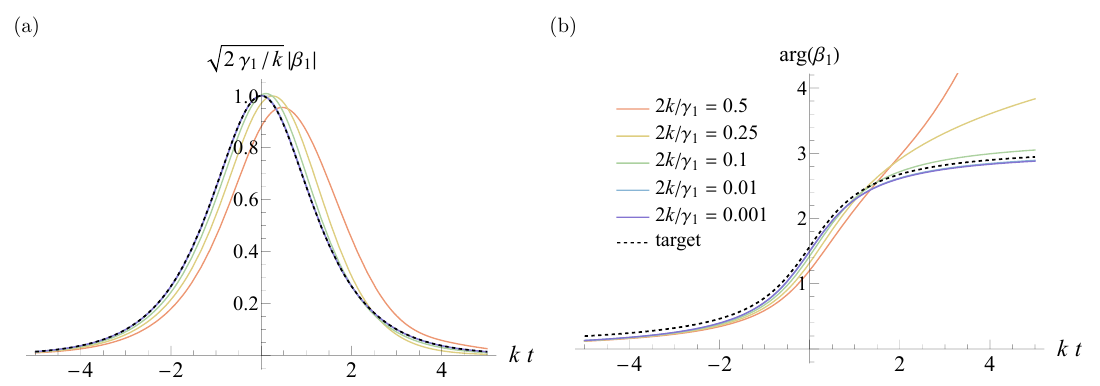} 
\caption{
Plots of (a) scaled magnitude and (b) phase of the amplitude $\beta_1$ in several instances.
For the colored lines corresponding to various $k$ values, $\beta_1$ is determined by solving the amplitude EOMs of Eq.~(\ref{eq:generalAmpEOMs}) directly for the laser pulse magnitude and phase specified by Eq.~(\ref{eq:G1SlowlyVaryingB}) and $\theta_0 = -\theta_\beta$, respectively, 
when trying to obtain the target amplitude
$\beta^{\textrm{target}}_1(t)$ (dashed black lines).  
This plot shows that as $2k/\g_1$ (which is equal to $r^{-1}$ from previous subsections) decreases the slowly varying solution quite accurately matches the target amplitude (and hence wave packet shape).
Note there appears to be a small residual phase offset for small $k$ though this will not affect the overlap $\mathcal{I}(k)$ in Fig.~\ref{fig:beta1Overlap} below.
}
\label{fig:beta1MagAndArg}
\end{figure*}

\begin{figure*}[ht] 
\includegraphics[width=\linewidth]{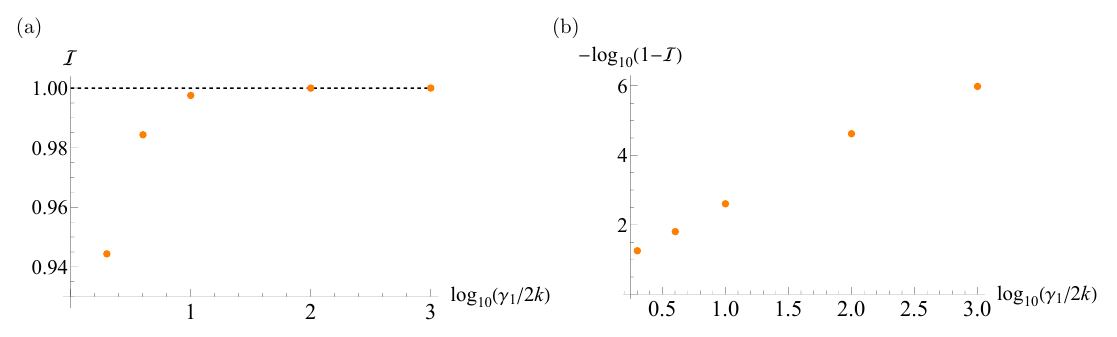} 
\caption{
Plot of the overlap $\mathcal{I}(k) \equiv \left| \int_{-t_m}^{t_m} dt~\g_1 \beta^{\textrm{target}}_1(t) \beta_1^*(t; k) \right| $ versus $\g_1/2k$ on (a) linear-log and (b) log-log scale.
In $\mathcal{I}(k)$ the function $\beta_1(t; k)$ corresponds to the colored line in Fig.~\ref{fig:beta1MagAndArg} with the corresponding value of $k$. See the Fig.~\ref{fig:beta1MagAndArg} caption for how these $\beta_1(t; k)$ are calculated.
An overlap of one entails that the emitted wave packet $\sqrt{\g_1} \beta_1$ has the target shape and phase. We see that this overlap monotonically increases approaching 1 as $k/\g_1$ decreases, this serves to demonstrate the validity of the slowly varying approximation.
Here we take the integration bounds to be $\pm t_m = \pm 15/k$ for our numerics.
}
\label{fig:beta1Overlap}
\end{figure*}

\textbf{Slowly varying $\boldsymbol{\beta_1}$ limit.}
In the different regime where $\beta_1$ is slowly varying, i.e., $|\dot\beta_1(t)| \ll |\p{i d_1 + \g_1/2} \beta_1(t)|$ we can simplify Eq.~(\ref{eq:genG1fromBeta1}) making the consistency condition of $G_1 \in \mathbb{R}$ tractable.
Using the polar decompositions $\beta_1(t) = B(t) e^{i \theta_\beta(t)}$\footnote{
We use a different symbol to denote the angle here, $\theta_\beta$, to avoid confusion with the case above with angle $b$.
}
and $i d_1 + \g_1/2 = \sqrt{d_1^2 + \g_1^2/4}~e^{i \arctan(2 d_1/\g_1) } \equiv D e^{i \theta_d}$ 
and assuming we are in the slowly varying limit so that we can neglect $\dot\beta_1(t)$ terms we have
\be
G_1^2(t) \approx \frac{ D^2 e^{2i \br{ \theta_0(t) + \theta_\beta(t)  + \theta_d}
}  B^2(t) }{ 
c - 2 D e^{-i \theta_d}\int_{t_p}^t dt' e^{2i \br{ \theta_0(t') + \theta_\beta(t') + \theta_d}}   B^2(t')
}
\ee
with $c = \alpha_1^2(t_p) e^{2i \theta_0(t_p)}$.
Then by picking $\theta_0$ such that $\theta_0(t) + \theta_\beta(t)  + \theta_d = \eta/2 = \textrm{constant}$ we have
\be
G_1^2(t) \approx \frac{ D^2  B^2(t) }{ 
e^{i ( \arg{c} - \eta)} |c| - 2 D e^{- i \theta_d}\int_{t_p}^t dt'   B^2(t') },
\ee
which will be real provided $\eta = \arg{c}$ and $\theta_d = 0 \implies d_1 = 0$\footnote{
Note that although $\theta_d = \pi$ would naively be less restrictive as then the denominator would always be positive, this is not possible in fact $|\theta_d| \leq \pi/2$ as the quantity $\g_1/2 + i d_1$, which $\theta_d$ is the argument of, has positive real component.
}
such that $\theta_0(t)  = \arg{c}/2 -  \theta_\beta(t) $. Thus for early $t_p$ for which $|\alpha_1(t_p)| \ra 1$ we have
\be
G_1^2(t) \approx \frac{ \g_1^2  B^2(t)/4 }{ 
1 - \g_1 \int_{t_p}^t dt'   B^2(t') }.
\ee
Moreover, taking $t_p \ra -\infty$, as the contribution from $B(t)$ should be negligible before $t_p$, we can use the normalization condition on $\Psi(t) = \sqrt{\g_1} \beta_1(t)$ of
$\int_{t_p}^{\infty} dt'  |\Psi(t')|^2 = 1$
to obtain 
\be\label{eq:G1SlowlyVaryingB}
G_1(t) \approx \frac{ \g_1  B(t)/2 }{ \sqrt{ \g_1 \int_{t}^{\infty} dt'  B^2(t')} }
= \frac{\sqrt{\g_1}}{2} \frac{ |\Psi(t)| }{ \sqrt{\int_{t}^{\infty} dt'  |\Psi(t')|^2} },
\ee
This suggests we can pick any phase for the wave packet we like $\theta_\beta$ provided we work in the slowly varying limit  $|\dot\beta_1(t)| \ll \frac{\g_1 }{2} |\beta_1(t)|$, tune $d_1 = 0$, and select the laser phase such that $\dot\phi_1 =  -\dot\theta_\beta(t) - \delta\om_1$. 
Note that sufficiently fast varying phases, which would correspond to large variations about the photon wave packets central carrier frequency, will ultimately be incompatible with the Markov approximation \cite{penas2023improving} needed to obtain the amplitude EOMs we started this section with. This issue will be limited by the appropriate selection of wave packets satisfying the slowly varying condition, which for the polar decomposition of $\beta_1$ is that 
$(\dot{B}/B)^2 + (\dot\theta_\beta)^2  \ll \g_1^2/4$. However, analysis of such deviations from Markovianity is beyond the scope of this work.
As an example let us consider the normalized target amplitude
$\beta^{\textrm{target}}_1(t) = \sqrt{\frac{k}{2 \g_1}} \sech(k t)e^{i \br{\arctan(k t) + \pi/2}} := B(t) e^{i \theta_\beta(t)}$, which has a nice symmetric envelope and its phase smoothly goes from $0$ to $\pi$. We can do quite well in generating a wave packet of this form by putting ourselves in the slowly varying condition, which in this case amounts to $k \ll \g_1/2$ as illustrated in Figs.~\ref{fig:beta1MagAndArg} and \ref{fig:beta1Overlap}. 

\textbf{Comparison to Ref.~\citenum{gorshkov2007photon}.}
We note that Refs.~\citenum{gorshkov2007photon} and subsequently \citenum{morin2019deterministic} perform similar analysis identifying the requisite pulses to produce certain photon wave packets and both report laser pulses of the same form as Eq.~(\ref{eq:G1SlowlyVaryingB}).
Refs.~\citenum{gorshkov2007photon} provide further analysis and conditions for when adiabatic elimination of the excited state, $\ket{r}$ in our notation, is possible in a three-level $\Lambda$-type atom (beyond just the large detuning case we consider in this work). They analyze the single photons that can be produced by an atomic ensemble of $N$ such atoms in a cavity. Ref.~\citenum{morin2019deterministic} extend aspects of the work of Ref.~\citenum{gorshkov2007photon} by accounting for the impact of additional atomic structure including higher excited states for the case of a single $\Lambda$-type atom in a cavity. 
We emphasize that their reported complex Rabi frequencies will only produce the intended wave packet if it is sufficiently broad wave in time in both magnitude and phase (i.e., is slowly varying), as is the case for Eq.~(\ref{eq:G1SlowlyVaryingB}). They both mention such limitations in their work, though we think it is appropriate to reiterate that these results are limited in that not any single photon shape can be produced.

We will now explicitly demonstrate the correspondence of the Rabi frequency determined by Ref.~\citenum{gorshkov2007photon} and that which we determine.
Specifically, we start with Eq.~(21) of Ref.~\citenum{gorshkov2007photon}, which gives the necessary  complex (slowly-varying) Rabi frequency to emit a wave packet $e(t)$ (in their notation) to be
\be\label{eq:GorshkovOm}
\Omega(t) = -\frac{\g (1+C) + i \Delta}{\sqrt{2\gamma(1+C)}}
\frac{e(t)}{\sqrt{\int_t^\infty dt' |e(t')|^2}}
e^{\frac{i \Delta }{2 \g(1+C) } \ln\p{1 - \int_0^t dt'  |e(t')|^2} },
\ee
where they have the cooperativity $C = g^2 N/\kappa \g$.
In their notation $\g$ 
is the decay rate of the optical coherence of the upper level (our $\ket{r}$), 
$2 \kappa$ is the cavity decay rate (our $\g_j$ for node $j=1,2$), and as in our case $g$ is coupling of an atom to the cavity mode, and $\Delta$ is the laser detuning though they use the opposite sign convention from us.
To make this case align with ours we work with special cases of both results. Note that their result, summarized in Eq.~(\ref{eq:GorshkovOm}), accounts for the impact of cavity and atomic parameters on photon production (as quantified by the cooperativity $C$ relative to the detuning $\Delta/\g$) more carefully than our work does (as they have different scopes). The special case of our work with a slowly varying $\beta_1$, as given in Eq.~(\ref{eq:G1SlowlyVaryingB}), should align with the special case of their result with $N=1$ atom (per cavity), a large cooperativity $C \gg 1$, and a far off-resonant Raman scheme $|\Delta| \gg \g C$.
In this case their result, Eq.~(\ref{eq:GorshkovOm}), becomes
\be
\Omega(t) = -i \Delta \sqrt{ \frac{  \kappa}{2 g^2} }
\frac{e(t)}{\sqrt{\int_t^\infty dt' |e(t')|^2}}
e^{\frac{i \Delta \kappa}{2 g^2 } \ln\p{\int_{t}^\infty dt'  |e(t')|^2} },
\ee
as they work with the starting time $t_p = 0$. 
Note that formally this large $C$ limit could be determined by taking $\g \ra 0$, noting that $\g C = g^2/\kappa$ is independent of $\g$, though one needs to be careful with the corresponding interpretation.\footnote{\label{suppressingSeFootnote} 
	One may be tempted to think that as `spontaneous emission from the upper level can be suppressed by using an off-resonant Raman scheme' (as is often stated, e.g., see Ref.~\citenum{cirac1997qst,vogell2017deterministic}), one should be able to achieve $\g = \Gamma_{sd}/2 \ra 0$ `simply' by using a large detuning $\Delta$.  
	However, this is not the case: the effects of spontaneous decay cannot be eliminated nor does $\Gamma_{sd}/2$ approach 0 for large detuning (it is a constant). Accordingly, statements such as the one above should be made with caution to avoid giving a misleading impression.
	This idea comes from the fact that the \textit{effective} spontaneous decay rate of the upper level $\Gamma_\textrm{eff}$ is suppressed in a Raman process that is far detuned relative to the drive strength $\Omega$ as $\Gamma_\textrm{eff}(t) = \Gamma_{sd} [\Omega(t)/2\Delta]^2$, as shown in Eq.~(\ref{eq:sdEOMaLargeDetuning}) \cite{vogell2017deterministic}. 
	However, the atom-cavity coupling is similarly modified (suppressed for large detuning) in this effective picture, i.e., $g \ra G(t) = g \Omega(t)/2\Delta$ and these two effects balance out so the cooperativity is left unchanged: $C_\textrm{eff} = \frac{2 G^2(t)}{\kappa \Gamma_\textrm{eff}(t)} = \frac{2 g^2}{\kappa \Gamma_{sd}} = C$.
	A large detuning is indeed vital in the adiabatic elimination of the upper level in the Raman process and for effectively eliminating the factor $\Gamma_r$ in Eq.~(\ref{eq:ampEOMsWithDecay}). Nevertheless, the impact of cooperativity in Eq.~(\ref{eq:sdEOMaLargeDetuning}) remains even for large $\Delta$ and ultimately, as seen in Sec.~\hyperref[subsec:atomicDecay3]{B3}, gives rise to an upper bound for the efficiency $P_\textrm{em-max} = C/(1+C)$, which is independent of detuning \cite{gorshkov2007photon,vogell2017deterministic}.
	%
	Note that if multiple excited state levels are included that couple to the intended Raman process this bound does acquire a frequency dependence \cite{morin2019deterministic}.
}

Translating to our notation via appropriate relabelings of $\kappa \ra \g_1/2$ (for node $1$), $\Delta \ra - \Delta_1$, and $e(t) = \Psi(t) = \sqrt{\g_1} \beta_1(t)$ (taken to be a normalized wave packet) we have 
\be\label{eq:complexOmGorskovMatching}
\Omega(t) =  \frac{i  \sqrt{\g_1} \Delta_1 }{2 g_1} 
\frac{ \beta_1(t) }{ \sqrt{ \int_t^\infty dt' |\beta_1(t')|^2}}
e^{\frac{-i \Delta_1 \g_1}{4 g_1^2 } \ln\p{ \g_1 \int_{t}^\infty dt'  |\beta_1(t')|^2} }.
\ee
However, their $\Omega$ is defined as half the traditional Rabi frequency, which is $\Omega_1 e^{-i \phi_1}$ 
in our notation, hence we have the following correspondence to our Rabi frequency magnitude
\be
\Omega_1(t) \leftrightarrow  
2 |\Omega(t)| \approx \frac{\sqrt{\g_1} |\Delta_1|}{g_1} 
\frac{|\beta_1(t)|}{\sqrt{\int_t^\infty dt' |\beta_1(t')|^2}}.
\ee
Meanwhile using definition of $G_1$ in terms $\Omega_1$,\footnote{
Note the $\Delta_1$ in the proportionality constant between $\Omega_1$ and $G_1$ should technically have a modulus on it in order for $\Omega_1$ to be a magnitude, with $G_1 > 0$. This can be adapted by including the sign of $\Delta_1$ into the phase $\phi_1$ via the argument of $\Delta_1$ (either $0$ or $\pi$), which contributes a constant that can be lumped into the constant of Eq.~(\ref{eq:phaseCorrespondance}).
}
and using the slowly varying pulse Eq.~(\ref{eq:G1SlowlyVaryingB}) we have 
\be\label{eq:Om1SlowlyVaryingB}
\Omega_1(t) = \frac{2 |\Delta_1|}{g_1} G_1(t)
\approx \frac{\sqrt{\g_1} |\Delta_1|}{g_1} 
\frac{  |\beta_1(t)| }{ \sqrt{ \int_{t}^{\infty} dt' |\beta_1(t')|^2 } },
\ee
which precisely matches the result of Ref.~\citenum{gorshkov2007photon} in the appropriate limit, i.e., the $2 |\Omega(t)|$ above.
For completeness we must also compare the phases, we should have the following phase correspondence:
\be
-\phi_1(t) \leftrightarrow \arg{\Omega(t)}.
\ee
Note the minus sign on $\phi_1$ in the equation above, which is due to our phases $\phi_j$ corresponding to a lowering operator, $\ketbra{e_j}{r_j}$, in the original node Hamiltonian, whereas their complex $\Omega$ corresponding to a raising operator.
We analyze this by first considering the $\delta\om_1(t) = \Omega^2_1(t)/(4 \Delta_1)$ integral term in $\phi_1$ with $\Omega_1$ given by Eq.~(\ref{eq:Om1SlowlyVaryingB}):
\be
\int dt~\delta\om_1(t) 
=  \frac{\g_1 \Delta_1}{4 g_1^2} \int dt~ 
\frac{ \g_1 |\beta_1(t)|^2 }{ \g_1 \int_{t}^{\infty} dt'~|\beta_1(t')|^2  }
= -\frac{\g_1 \Delta_1}{4 g_1^2} \ln\p{ \g_1 \int_{t}^{\infty} dt'~|\beta_1(t')|^2  }
\ee
(the $\g_1$ is included in the logarithm so that it has appropriate dimensions).
Meanwhile from Eq.~(\ref{eq:complexOmGorskovMatching}) we have
\be
\arg{\Omega(t)} = -\frac{\pi}{2} + \theta_\beta(t) - \frac{\g_1 \Delta_1}{4 g_1^2 } \ln\p{ \g_1 \int_{t}^\infty dt'~|\beta_1(t')|^2},
\ee
where we are again using the notation $\theta_\beta = \arg{\beta_1}$.
Thus, we see that by integrating the condition $\dot\phi_1 =  -\dot\theta_\beta(t) - \delta\om_1$ we have 
\be\label{eq:phaseCorrespondance}
-\phi_1(t) 
= \theta_\beta(t) - \frac{\g_1 \Delta_1}{4 g_1^2} \ln\p{ \g_1 \int_{t}^{\infty} dt'~|\beta_1(t')|^2  } + \textrm{constant},
\ee
which is specified up to a constant encoding the initial condition.
We indeed find that $-\phi_1(t) = \arg{\Omega(t)}$ (up to the unphysical constant) so both the phase and magnitude of our Rabi frequency match that of Ref.~\citenum{gorshkov2007photon} in the appropriate limit.	

\setcounter{equation}{0}
\renewcommand{\theequation}{C\thesection\arabic{equation}}
\renewcommand{\theHequation}{C\thesection\arabic{equation}} 
\phantomsection
\section*{C.~Index of Separability} \label{sec:sepIndex} 
Here we elaborate on our methods for computing the index of separability $\mathcal{S}$ for various two-variable joint probability distributions as defined in Eq.~(\mainref{eq:indexofSep}); this includes stability analysis accomplished by changing the input values used. 
For each of the error variable pairs, we compute the separability index on evenly spaced $n$ by $n$ grids over the scaled versions of the regions illustrated in Fig.~\mainref{fig:PsuccVsXiandT}, which we will collectively refer to as $R = \left\{ T, \xi, \om_0 : 
|\g_2 \Delta T| \leq 7.5, |\Delta\ell_\xi| \leq 6, |\Delta\om_0/\g_2| \leq 3
\right\} $, picked based on widths of the single error variable $\Psucc$ distributions. We refer to the region with each parameters domain scaled up by $s$ as $R_s$. In Table \ref{sepIndexTable} we report $\mathcal{S}$ values for different regions and different matrix sizes (corresponding to different spacings). We consider  $\mathcal{S}$ values computed using the singular values of the original matrices, $A$, as well as their zero-mean counterparts, $A^{(0)}$, defined for an $m \times n$ matrix $A$ as 
$A^{(0)}_{i j} \equiv A_{i j} - \frac{ 1 }{n m} \sum_{k=1}^m \sum_{l=1}^n A_{k l}$. We see that the original (nonzero mean) values are relatively stable (at least within a given region). The values reported in Eqs.~(\mainref{eq:STxi}-\mainref{eq:Som0T}) are given in the first row (in the nonzero-mean case).
The intuition behind being able to change the region size is that these functions decay to zero for large errors and padding a matrix with zeros does not changes its nonzero singular values.

\setlength\tabcolsep{1em}
\begin{table}
\begin{tabular}{cccccccc} \toprule
\multicolumn{2}{c}{\underline{~Case description~}} \vspace{1mm} & \multicolumn{6}{c}{\underline{~Indices of separability~}} \vspace{1mm}  \\ 
Region & $n$ & $S_{T, \xi}$ & $S_{\om_0, \xi}$ & $S_{\om_0, T}$ &
$S^{(0)}_{T, \xi}$ & $S^{(0)}_{\om_0, \xi}$ & $S^{(0)}_{\om_0, T}$ \\ \midrule
$R_2$ & 121 & 0.869 & 0.870 & 0.9976548 & 0.800 & 0.804 & 0.969 \\ 
$R_2$ & 61 & 0.869 & 0.869 & 0.9976548 & 0.801 & 0.805 & 0.970 \\ 
$R_2$ & 41 & 0.868 & 0.869 & 0.9976547 & 0.801 & 0.805 & 0.970 \\ \midrule
$R_{4/3}$ & 81 & 0.886 & 0.886 & 0.9976548 & 0.765 & 0.770 & 0.942 \\ 
$R_{4/3}$ & 41 & 0.886 & 0.885 & 0.9976548 & 0.767 & 0.771 & 0.943 \\ \midrule
$R$ & 61 & 0.902 & 0.901 & 0.9976551 & 0.726 & 0.732 & 0.908 \\ 
$R$ & 31 & 0.901 & 0.900 & 0.9976550 & 0.729 & 0.735 & 0.910 \\ 
\bottomrule
\end{tabular}
\caption{\label{sepIndexTable} 
Index of separability $\mathcal{S}$ for each possible pairs of errors in the parameters $\om_0$, $\xi$, and $T$ with the same parameter values used in the main text plots ($k=2$, $\om_{0 i} = 50$, $\xi_i = 1/2$, and  in units where $\g_2 =1$). Here we vary  
i) whether we work with the original matrix ($A$) or its zero-mean counterpart ($A^{(0)}$) indicated by the presence of a $(0)$ superscript, 
ii) what region the value is computed using,
and iii) what matrix size is used for sampling over the region (in each case we chose the grid spacing to obtain a square $n \times n$ matrix).
We include enough digits to show differences between the values.
Calculation methods are discussed at the start of this section, which includes the region sizes considered. In each case, we effectively take $l$ to be infinite, i.e., we consider the entire transformed wave packet. This does not significantly change the results, for instance changing to $l = 10 c/\g_2$ (assuming $t_f = t_s + t_l/\xi_i$ still) in the case shown in the top row gives no change in the computed values to the reported precision. 
Note that $S_{\om_0, T} $ is more stable than $S_{T,\xi}$ and $S_{\om_0,\xi}$ because it goes to zero more quickly over the considered regions.
}
\end{table}

We will now explain why additionally considering $\mathcal{S}$ for the zero-mean counterpart of $A$, $A^{(0)}$, is valuable. If the original matrix under consideration $A$ has nonzero-mean element $\mu$, it is a mean-zero perturbation of the rank 1 matrix whose entries are all $\mu$ and hence there should be a corresponding singular value that is close to $|\mu|$. For appreciable $|\mu|$, relative to variations in the matrix elements, this singular value will be quite large (it can be the maximum) and will skew the separability $\mathcal{S}$ towards large values (closer to 1). 
For instance, one can show that for a large $n \times n$ matrix $U$ with i.i.d.~elements from a uniform distribution on $[0, 1]$, the expectation value of $\mathcal{S}( U^{(0)} )$ tends to 0 as $n \ra \infty$, 
whereas without shifting to a zero-mean matrix $\mathcal{S}(U)$ would tend to 3/4 as $n \ra \infty$. 
(This can easily be checked numerically or more carefully using random matrix theory.) In the $61 \times 61$ case (to align with the values computed in Table \ref{sepIndexTable}), such a matrix has mean $0.758$ with standard deviation $0.004$. Hence the given nonzero-mean $\mathcal{S}$ values should be compared to $\sim 0.75$ for a random matrix instead of their global minimum of $1/n \sim 0$.\footnote{
Comparing against this $0.75$ value specifically, which comes from a $\mu = 1/2$ case, is of course a drastic simplification. It is somewhat reasonable for the sizes of matrices we consider which have appreciable means (on the order of $0.5$). However, this effect will become less and	less of an issue as we consider $\Psucc$ matrices over larger regions as they decay to zero away from their peaks and hence their means will approach zero.
Accordingly, $S$ tends to decrease as a function of $s$ (in the region size $R_s$).
\label{note:SNote}
}
Accordingly, we report both the zero and nonzero-mean cases as we want to show that the ordering of the separability indices  
is not determined by the means and  because the  zero-mean matrices are less stable under changes to the region size,\footnote{
For an original matrix $A$ the edge of the considered regions (for which the errors are large) will be nearly zero, yet after subtracting the mean this value will be take on a values around $- \mu$ (which will go to zero for large grids but much slower than the original case). Padding by the necessary nonzero values will change the singular values of $A^{(0)}$ under increases to the region size, quantified by $s$. Specifically, the maximum singular value (relative to the others) will tend to increase as a function of $s$ and hence $S^{(0)}$ will as well. For instance, for small $s$, $\Psucc$ should be approximately constant so each of the zero-mean matrix approximations will be close to the constant matrix with all zero entries for which $S^{(0)} =0$. As $s$ increases and more structure is revealed a dominant singular value will thus emerge.
\label{note:S0Note}
}
respectively. Moreover, (in light of the phenomena discussed in footnotes \ref{note:SNote} and \ref{note:S0Note}) the original and zero-mean matrix separability indices, $S$ and $S^{(0)}$, are upper and lower bounds for the separability index approached for large grid sizes, as can be seen in Fig.~\ref{fig:r0SepIndPlots} below. (We do not go to larger $s$ values here to avoid additional complications with numerical integration of sharply localized and highly oscillatory functions.)

The perceptive reader may have noticed that $S_{T,\xi} \approx S_{\om_0, \xi}$ in each of the above cases and wondered if this is a coincidence or due to some underlying structure. The answer is partially both, in that this behavior does not hold as closely for other emitted wave packets and corresponding amplitudes $\beta_1$ yet often these values will be close to one another.
To gain some intuition behind this, we note that Eq.~(\mainref{eq:PsuccDerived}) can be expressed in terms of the dimensionless error variables $x = \Delta\om_0/\g_2$, $y= \Delta \ell_\xi$, and $z = \g_2 \Delta T$ (see footnote \ref{note:repSymbol}), in the $l \ra \infty$ case where no matter what unitary errors occur all of the wave packet is transformed, as
\begin{align}\label{eq:PsuccXYZ}
\Psucc(x, y, z) &= \bigg|
\int_{-\infty}^\infty dt \underbrace{\sqrt{\g_2} e^{-i \om_{0 i} (T_i - t) } \beta_1^*\p{\xi_i (T_i - t)}}_{\Phi^*(t)}
\underbrace{\sqrt{\g_1 \xi} e^{ i \om_0 (T - t) } \beta_1\p{\xi(T - t)}}_{\Psi(t)}
\bigg|^2 \nonumber \\
&= 2^y \bigg|
\int_{-\infty}^\infty d\tau~e^{ i x \tau } \beta_{\g 1}^*\p{\tau} \beta_{\g 1}\p{ 2^y (\tau + z) }
\bigg|^2 
\end{align}
where in the last line we changed to the dimensionless time variable $\tau = \g_2 (T_i - t)$ and defined $\beta_{\g 1}(\tau) := \beta_1\p{\tau/\g_1}$.
Note this clearly shows that for a given shape of the emitted (not transformed) wave packet, which is proportional to the amplitude $\beta_{\g 1}$, the maximum value of $\Psucc$ (which will be reduced due to other errors including those caused by finite $l$)  is determined by the three errors parameters $x$, $y$, and $z$. 
For instance, regardless of what wave packet is emitted from node 1, a stretching error is always determined by the parameter $y = \Delta\ell_\xi = \log_2(\xi/\xi_i)$. Thus if we consider a different implementation of node 2, called node $2'$, with larger coupling $\g_2' = 100 \g_2$ such that $\xi_i' = 100 \xi_i$ (for fixed $\g_1$), then to achieve the same stretching error we would need $\xi'  = 100 \xi$ (the exact factor does not matter). Thus, the results for given parameter values, e.g., $\zeta = \om_{0 i}$ and $\g_2$, can clearly be generalized to different values via scaling. This entails that plots like Figs.~\mainref{fig:PhiAndPsiWavePackets}-\mainref{fig:PsuccVsXiandT}, which do assume a certain emitted wave packet shape, can be applied to a node 2 with any parameter values (assuming it can be modeled by the same effective Hamiltonian considered in the main text and discussed in Sec.~\hyperref[subsec:modelGenerality]{A2}).

Now, noting that Fig.~\mainref{fig:PsuccVsXiandT} (a) looks quite similar to Fig.~\mainref{fig:PsuccVsXiandT} (b) after the mapping $y \ra -y$, we compute $P_{z, y} \equiv \Psucc(0, y, z)$ and $P_{x, -y} \equiv \Psucc(x, -y, 0)$ so we can compare them. Defining $\varpi$ to be the Fourier conjugate variable to $\tau$ with
$\beta_{\g 1}(\tau) = \int_{-\infty}^\infty \frac{d\varpi}{2 \pi} e^{ -i \varpi \tau } \tilde\beta_{\g 1}(\varpi)$, we have that (with all integrals taken from $-\infty$ to $\infty$)
\be
P_{z, y} = 2^y \bigg|
\int d\tau~\beta_{\g 1}^*\p{\tau} \beta_{\g 1}\p{ 2^y (\tau + z) }
\bigg|^2
= 2^y \bigg|
\int d\tau 	\int \frac{d\varpi}{2 \pi} 
\beta_{\g 1}^*(\tau) \tilde\beta_{\g 1}(\varpi)
e^{ -i 2^y (z + \tau) \varpi } 
\bigg|^2
\ee
and using $\tau' = 2^{-y} \tau$  in an intermediate step
\begin{align}
P_{x, -y}  &=  2^{-y} \bigg|
\int d\tau~e^{ i x \tau } \beta_{\g 1}^*(\tau) \beta_{\g 1}\p{ 2^{-y} \tau }
\bigg|^2
=  2^y \bigg|
\int  d\tau'~e^{ -i x 2^y \tau' } \beta^*_{\g 1}\p{ \tau' } \beta_{\g 1}(2^y \tau') 
\bigg|^2 \nonumber \\
&= 2^y \bigg|
\int  d\tau  \int \frac{d\varpi}{2 \pi} \beta^*_{\g 1}\p{ \tau } \tilde\beta_{\g 1}(\varpi) 
e^{ -i 2^y (x + \varpi ) \tau } 
\bigg|^2,
\end{align}
which evidently takes on a similar form except for differences in the phase. Namely, the phases are equal under the exchanges $x \leftrightarrow z$ and $\varpi \leftrightarrow \tau$. Thus, if $ \beta^*_{\g 1}(v) \approx \tilde\beta_{\g 1}(v)$ as functions, then $P_{z, y}  \approx P_{x, -y}$ and it would follow that $S_{T,\xi} \approx S_{\om_0, \xi}$.
We can now confirm whether this intuition is valid in our case here, where $\beta_1(t)$ is given by Eq.~(\ref{eq:beta1rOfOneHalf}). More precisely, from Fig.~\mainref{fig:PsuccVsXiandT} we seem to have the scaling $2.5 x \leftrightarrow z$ so we actually anticipate $P_{z=2.5x, y} \approx P_{x, -y} $ which would imply that
\be
\int dv  dv'
\beta_{\g 1}^*(2.5 v ) \tilde\beta_{\g 1}(v'/ 2.5)
e^{ -i 2^y ( x +  v)  v' } 
\approx
\int  dv' dv \beta^*_{\g 1}(v') \tilde\beta_{\g 1}(v) 
e^{ -i 2^y (x + v ) v' }, 
\ee
which clearly has $\beta_{\g 1}^*(2.5 v )  =  \tilde\beta_{\g 1}(v) $ as an exact  solution (in which case the approximation would become an equality). This argument could clearly be applied to scalings other than the 2.5 value considered.
We can quantify the degree to which the approximation $\beta_{\g 1}^*(2.5 v ) \approx  \tilde\beta_{\g 1}(v)$ holds by computing the magnitude of the inner product of the two functions. Doing so numerically in our case with $\beta_1(t)$ given by Eq.~(\ref{eq:beta1rOfOneHalf}), we find 
\be
\left| \int dv~\hat\beta_{\g 1}(2.5 v )  \hat{\tilde\beta}_{\g 1}(v) \right| = 0.96,
\ee
where a function with a hat indicates that it is unit normalized as
\be
\hat{f}(v) = \frac{f(v)}{\sqrt{ \int dv' |f(v')|^2  }}.
\ee
This inner product magnitude will be 1 if the functions are the same (up to a proportionality constant) so this large value of 0.96 indicates the approximation $\beta_{\g 1}^*(2.5 v ) \approx  \tilde\beta_{\g 1}(v)$ is quite good, from which it follows that $S_{T,\xi} \approx S_{\om_0, \xi}$ as we found.

Note that this is not always the case. A simple example illustrating this is the $r \ra 0$ limit given in Eq.~(\ref{eq:b1Smallr}), for which 
$\beta_{\g 1}(\tau) = \Theta(\tau) e^{-\tau/2}$ and from Eq.~(\ref{eq:PsuccXYZ}) it directly follows that
\be\label{eq:Psuccr0}
\Psucc^{r \ra 0}(x,y,z) = \frac{ 2^{y+2} }{ (1+2^y)^2 + (2 x)^2 } \times
\begin{cases}
e^z, & z \leq 0 \\
e^{- 2^y z}, & z > 0
\end{cases}.
\ee
In this case, because we have an easy to work with analytical formula for $\Psucc$, we can plot how the separability indices change for significantly larger regions, see Fig.~\ref{fig:r0SepIndPlots}. We again consider scaling the bounds of the rectangular base region $R$ by $s$ in each direction ($\om_0$, $\xi$, and $T$) yielding the region $R_s$. We see that for each parameter pair the $\mathcal{S}$ values of both the original and zero-mean matrices converge to the same value, namely, $S_{\om_0,T} = 1.00 > S_{\om_0,\xi} =  0.85 > S_{T,\xi} = 0.78$ (with uncertainties of at most $0.01$ in each case).
Note that for $y=0$ the function $\Psucc^{r \ra 0}(x,0,z) = e^{- |z| }/(1 + x^2)$ is evidently separable (it is a product of a function of $x$ alone and a function of $z$ alone) so an index of separability of $S_{\om_0,T} = 1$ is fitting, in fact, requisite. We note that the ordering of the separability indices remains consistent for large grids and $S_{T,\xi} \not\approx S_{\om_0, \xi}$ here.\footnote{
This is intuitive here as the exponentially decaying $\beta_{\g 1}(\tau)$ is quite different than its Fourier transform, whose modulus squared is a Lorentzian.
} 
This aligns with our main point here, that $\om_0$ and $T$ errors are nearly separable while the other errors pairs are less so. 

\begin{figure*}[ht] 
\includegraphics[width=0.7\linewidth]{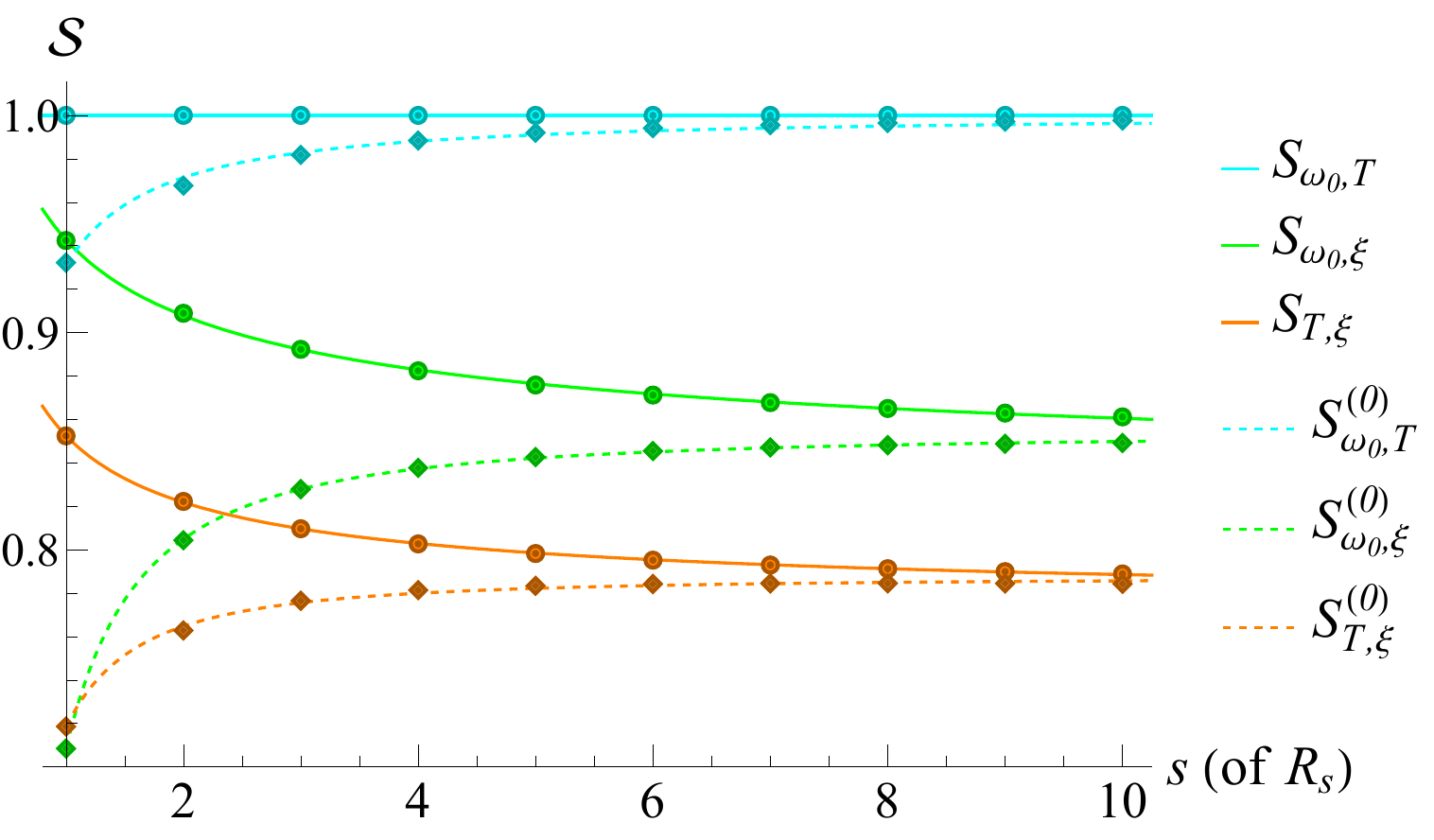} 
\caption{
Plots of the indices of separability $\mathcal{S}$ for each error parameter pair as a function of region size as quantified by the scaling $s$ corresponding to region $R_s$.
This is done for both the original nonzero-mean matrices (solid lines) and their zero-mean counterparts (dashed lines), which serves as upper and lower bounds for the corresponding $\mathcal{S}$ value, respectively. These bounds approach one another as $s$ increases and so $\mathcal{S}$ closes in to a fixed value (as reported above).  Thus, although either matrix (original or zero mean) can itself be used for a separability estimate, they are best used in concert.
The shown (circular and kite shaped) points are calculated for integer values of $s$ from $1$ to $10$ with by evaluating Eq.~(\ref{eq:Psuccr0}) on a matrix (original and zero-mean counterpart) evaluated over the corresponding region $R_s$, where the spacing are kept constant for each error variable at the original region dimensions of $R$ (from end to end)  divided by 100.
The underlaid curves are a power-law fit to the corresponding data points.
}
\label{fig:r0SepIndPlots}
\end{figure*}

As mentioned in the main text, we do not account for implementation specific  dependencies of the parameters of $U$ (and hence of the corresponding errors).
We will now give a brief example of the role such additional dependencies play in analyzing multiple errors.
In the sum frequency generation based implementation proposed by Ref.~\citenum{timereversal}, the unitary parameter $\om_0$ would generically not be experimentally correlated with the parameters $\xi$ and $T$. However, $\xi$ and $T$ may be correlated depending on the linear dispersion of the material used to implement the transformation. In this implementation, a frequency error could occur due to inaccurate characterization of or the inability to produce the frequencies of the pump pulse driving the transformation as well as the Raman lasers driving both systems. Furthermore, we have assumed these lasers to be monochromatic, where in reality they will have some natural linewidth leading to $\om_0$ errors. A stretching error would occur due to an incorrect group slowness (inverse group velocity) $\beta'_j$ in one of the bands $j \in \{1, 2, p\}$ as $\xi = \frac{\beta'_p - \beta'_2}{\beta'_1 - \beta'_p} > 0$. 
Note that $\beta_j(\omega)$ describes the linear dispersion 
of the medium so such an error could occur due to incorrect characterization of or defects in the material. Furthermore, higher-order dispersion, which Ref.~\citenum{timereversal} assumed to be negligible, could lead to an effective broadening of $\xi$ that would induce errors. Here the timing $T$ and corresponding errors are determined by the pump lasers timing relative to the wave packet emitted by system 1. Similar to $\xi$ errors, this would also be affected by incorrect group slownesses $\beta'_j$.
Accordingly, $\xi$ and $T$ errors would also correspond to changes in the realized value of $l$ for a given nonlinear medium length $L$ (see App.~B of Ref.~\citenum{randles2023quantum}) due to changes in the linear dispersion of the medium.

\setcounter{equation}{0}
\renewcommand{\theequation}{D\thesection\arabic{equation}}
\renewcommand{\theHequation}{D\thesection\arabic{equation}} 
\phantomsection
\section*{D.~Error analysis and correction} \label{sec:ECZprotocols} 
As discussed in Sec.~\mainref{sec:Discussion}, the practical utility and potential versatility of a QST scheme like ours comes in consonance with the use of error correction protocols.
Our scheme is designed to minimize the presence of errors, so that the relatively costly error correction protocols are subject to less overhead.
Here we will elaborate on the simplifying strategies and their corresponding standard errors considered in Sec.~\mainref{subsec:strategies}.
Then we will further analyze the ECZ protocols highlighted in the main text, which are two  error correction protocols that are suited to our QST scheme using a mode occupation number encoding \cite{van1997ideal,van1998photonic}.
One notable drawback of this encoding is that a logical zero (e.g., no photon $\ket{0}$) is not distinguishable from a photon loss error $\ket{1} \ra \ket{0}$.
While this is true in our scheme as presented, it can be rectified by using auxiliary atoms at each node that serve as a form of redundancy in the case of such a photon loss error. 
Then such photon loss errors and phase errors can be detected and if an error occurs, the state of an auxiliary atom at the sending node will be projected onto the desired state to be transferred (at least after an appropriate local quantum computation) and an attempt of transferring the state can be repeated.
One necessary condition for the ECZ protocols, as mentioned in the main text, is that we are in a regime where no relevant photon production can occur (and that input field is in the vacuum state), i.e., the mapping $\ket{0} \ra \ket{1}$ is not possible.
This condition allows the ECZ protocols to be relatively simple and typically have less overhead as compared to other standard error correction schemes, which have to be able to correct for both $\ket{1} \ra \ket{0}$ and $\ket{0} \ra \ket{1}$ logical errors (which are both possible in a polarization encoding).

\begin{figure*}[h]
\includegraphics[width=0.6\linewidth]{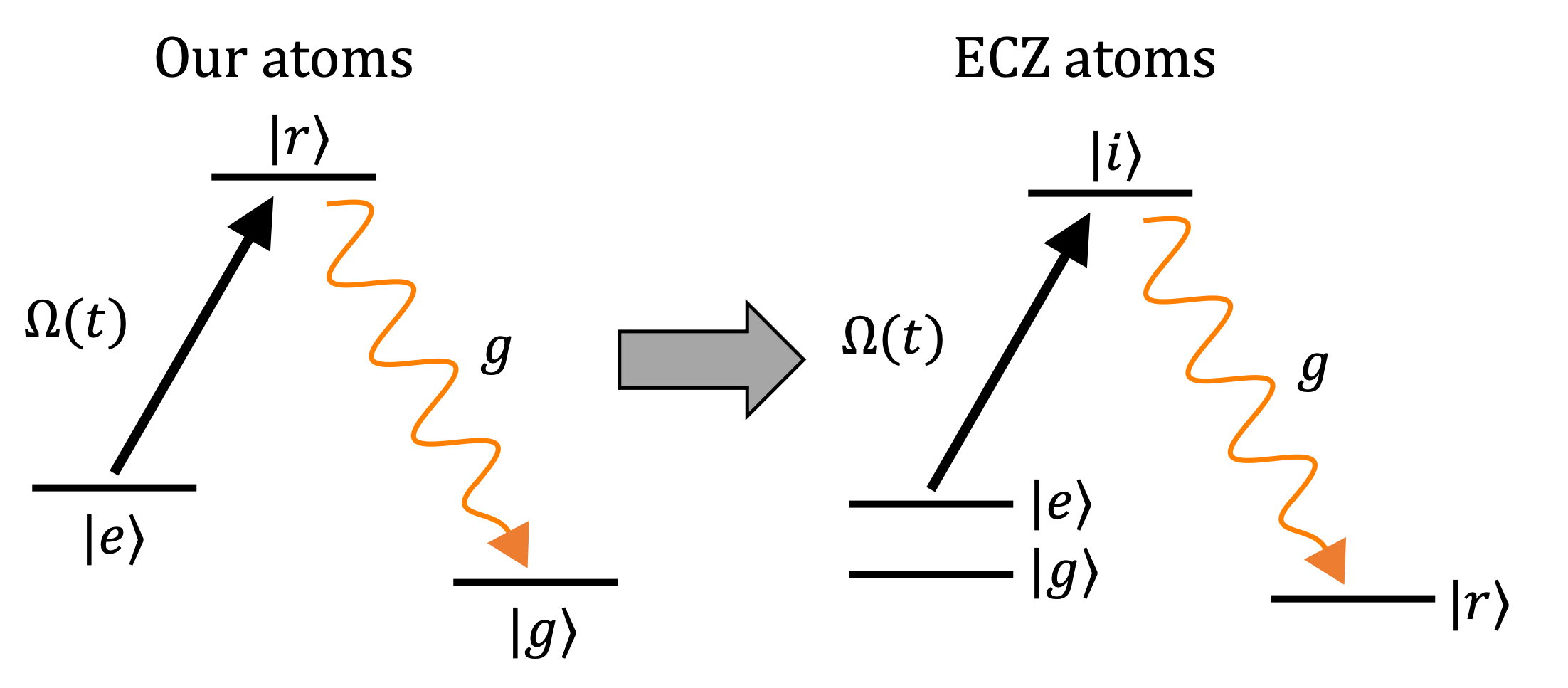}
\caption{Comparison of atomic configurations in our scheme (left) versus that used in the earlier ECZ error correction protocol (right) \cite{van1997ideal}.}
\label{fig:atomConfigs}
\end{figure*}

We will primarily focus on the earlier ECZ protocol \cite{van1997ideal} for concreteness and consistency with the exemplar atom in cavity systems we consider (though these protocols apply for other types of nodes provided the various steps can be implemented), mentioning differences in the later protocol \cite{van1998photonic} when applicable.
Their basic setup is similar to our work except there are two atoms in each cavity to act as a form of redundancy and each atom has slightly different level structure due to one additional ground state (see Fig.~\ref{fig:atomConfigs}).
One technical complication of using multiple atoms in a cavity is in accounting for the impact of the spatial dependence of the atom cavity coupling $g_j(\vec{x})$. Such position dependence is also present in the single atom case, though the interaction of the multiple atoms (or emitters more generally), especially as local operations are performed on the atoms,
will increase the variation of $g_j(\vec{x})$. This will lead to some errors due to the chosen Raman pulses $G_j$ being slightly incompatible with the actual values of the couplings $g_j(\vec{x})$.
Both our scheme and their protocol use states named $\ket{g}$ and $\ket{e}$ to encode the qubit.  However, in our scheme these are the only two (long-lived) levels and they are joined by a Raman transition. Whereas in this ECZ protocol the `excited' qubit state $\ket{e}$ is linked  via a Raman transition to an auxiliary level $\ket{r}$, not to be confused with our upper level of the same name.
In this case, ECZ's nominal ground state $\ket{g}$ is isolated in terms of transitions to the other levels during the Raman process, though it could possibly amplitude damp or phase shift 
due to interacting with the environment [e.g., due to photon absorption or spontaneous emission (to another level outside the levels considered)]. 
Despite these differences, our protocol can easily be mapped to their level structure with the EOMs we derive for our $\ket{e} \leftrightarrow \ket{g}$ Raman transition simply corresponding to  $\ket{e} \leftrightarrow \ket{r}$ for ECZ.

\phantomsection
\subsection*{1.~Accounting for standard errors}\label{subsec:standardErrors}
Here we elaborate on how the losses we set aside in our simplified treatment of Sec.~\mainref{subsec:strategies} can be incorporated into the analysis.
For clarity of exposition we will focus on our `exemplar case' in which each node consists of a $\Lambda$-type atom (or ion) in an optical cavity and they are coupled via optical fiber. In Table \hyperlink{extendedParameterTable}{ \ref*{tab:extendedParameterTable}} we list the attained values, across multiple such experiments, of several parameters that are used to quantify the atom- (or ion-) cavity and cavity-channel losses of strategies 1) and 2), respectively.
(Similar ideas should apply to other node and channel implementations though the formulas used will not necessarily be of the same form.)
We refer the reader to other work (via citations) for a more thorough treatment of these issues, chiefly see Refs.~\citenum{vogell2017deterministic}, \citenum{gorshkov2007photon}, and \citenum{morin2019deterministic}.

\textbf{Strategy 1).}	
We distinguish strategy 1) from 2) as even if we consider an ideal cavity [strategy 2)], the maximum efficiency of emission into (or absorption out of) a specific cavity mode is limited by the atomic structure.
For instance, the spontaneous decay rate $\Gamma_{sd}$ of the $\Lambda$ system's upper level (relative to parameters of the driving laser and cavity) limits the probability of being able to produce the desired photon in the Raman transition. 
Refs.~\citenum{gorshkov2007photon}, \citenum{morin2019deterministic}, and \citenum{vogell2017deterministic} 
show this -- that the optimal emission and absorption efficiencies are bounded above --  in contexts similar to our exemplar case, 
finding an upper bound for $\mathcal{P}_\textrm{em}$ based on the emitter cooperativity parameter $C_\textrm{em}$ as
\be\label{eq:PemOrabs}
\mathcal{P}_\textrm{em} = \frac{C_\textrm{em}}{1 + C_\textrm{em}}
\textrm{\quad with \quad}
C_\textrm{em} \equiv \frac{4 g^2}{\g \Gamma_{sd}}
\ee
as seen in Sec.~\hyperref[subsec:atomicDecay3]{B3}.
This cooperativity is written in our notation\footnote{\label{symbolNotationFootnote2}
	See footnote \ref{symbolNotationFootnote1}. Additionally, our use of $\Gamma_{sd}$ to denote the spontaneous decay rate of an emitter was chosen to avoid ambiguity with the  node-transmission-line couplings $\g_j$ and with the Green's function $\Gamma$ introduced in Sec.~\mainref{subsec:UParameterErrs}. In particular, note that the full cavity and atomic linewidths of $\g$ and $\Gamma_{sd}$ in our notation are often reported as $2 \kappa$ and $2 \gamma^\textrm{(atom)}$ [the superscript is used here to distinguish this parameter from our $\g$] in other work, respectively. In such other works, the corresponding half-linewidths $\kappa$ and $\gamma^\textrm{(atom)}$ are typically referred to as the cavity field decay rate (or photon exit rate) and the atomic polarization decay rate, respectively (e.g., see Refs.~\citenum{gorshkov2007photon} and \citenum{vogell2017deterministic}). For instance, our emitter cooperativity $C_\textrm{em}$ is equal to $4 \mathcal{C} = \frac{2 g^2}{\Gamma \kappa}$ in the notation of Ref.~\citenum{vogell2017deterministic} and corresponds to $\frac{g^2}{\g \kappa}$ in the notation of Ref.~\citenum{gorshkov2007photon} (in which the parameter $\g = \g_e/2 + \g_\textrm{deph}$ accounts for the possibility of an additional dephasing mechanism on top of spontaneous emission of the upper state). 
	See Sec.~\hyperref[subsec:modelGenerality]{A2} and the comparison to Ref.~\citenum{gorshkov2007photon} in Sec.~\hyperref[subsec:wavePacketShape5]{B5} for some additional discussion.
	} 
with the nodal subscripts omitted.
We do not expect such a bound to significantly limit us here as we are chiefly concerned with nodes in the strong-coupling of cavity QED limit with large atom-cavity couplings compared to the cavity linewidth and spontaneous emission rate, i.e., $2 g \gtrsim \g, \Gamma_{sd}$.
Note that this cooperativity based bound \textit{cannot} be surpassed by `simply' using a far off-resonant Raman process (see footnote \ref{suppressingSeFootnote}). 
However, significant coupling to other excited states could change this, and indeed understanding the impact of the multilevel structure of realistic atoms in cavity quantum electrodynamics (QED) is an active area of research, wherein optical pumping offers a potential means of controlling decay channels to other states thereby potentially still achieving controlled interactions of single-photons with material qubits \cite{morin2019deterministic,samutpraphoot2020strong,suarez2023collective}.
This entails that, within our model, certain wave packets can be emitted (or absorbed) with efficiency $\mathcal{P}_\textrm{em}$, which is near unity for large $C_\textrm{em}$. Notably, not all wave packet shapes can be produced (even up to a scaling factor). The class of wave packets that can be produced at or near this efficiency is determined by the control drives $G_j(t)$ and the cavity linewidths $\g_j$ (see Sec.~\hyperref[sec:wavePacketShape]{B}).

\textbf{Strategy 2).}
In terms of cavity loss, we note that although the probability of loss per single reflection from the cavity can be very low (parts per million), the number of round-trips the photon makes is comparably large for a high-$Q$ cavity so cavity loss becomes significant. In our exemplar case an asymmetric cavity is employed, which uses mirrors with different transmissivities to preferentially couple the cavity output to the transmission line. 
We denote the cavity mirror coupled to the transmission line as the inner one and the ideally perfectly reflecting mirror as the outer one (the light and dark gray mirrors in Fig.~\ref{fig:networkZoomedIn}, respectively).
Thus, we can calculate the probability of cavity light being emitted in the correct mode given the mirrors' respective transmissivities 
$\mathcal{T}_i$ and $\mathcal{T}_o$ as 
\be\label{eq:cavloss}
\mathcal{P}_\textrm{cav} = \frac{\mathcal{T}_i}{\mathcal{T}_i + \mathcal{T}_o + \mathcal{L}}
= \frac{C_\textrm{cav}}{1 + C_\textrm{cav}}
\textrm{\quad with \quad}
C_\textrm{cav} \equiv \frac{\mathcal{T}_i}{\mathcal{T}_o + \mathcal{L}},
\ee 
where $\mathcal{L}$ accounts for any additional coupling of the cavity to wrong output modes, e.g., due to absorption and scattering loss 
and in the second equality we put the equation in terms of a `cavity cooperativity'\footnote{
	The probability of an excitation transfer (be it atom to cavity, cavity to transmission line, vice versa, etc.) can be expressed schematically as $\mathcal{P} = (\textrm{desired rate})/(\textrm{total rate})$, where total rate $=$ desired rate $+$ undesired rate. Thereby, one can naturally define a corresponding cooperativity parameter $C = $ (desired rate)$/$(undesired rate), which quantifies how well something (e.g., an emitter or cavity) couples to a desired output mode, such that $\mathcal{P} = C/(1+C)$.
} $C_\textrm{cav}$.
Symmetrically, such a degradation of the overall success probability also occurs due to loss during absorption at cavity 2.

\begin{table}[htp]
\hypertarget{extendedParameterTable}{}
\begin{tabularx}{0.93\linewidth}{ccccccc}
\multicolumn{7}{c}{Panel A} \\
\toprule[1pt]
\multirow{2}{*}{Emitter type} &
\multirow{2}{*}{Reference} &
\multicolumn{5}{c}{Emitter (atomic) decay \vspace{1mm}} \\
&
&
$g/2\pi$ (MHz) &
$\g/2\pi$ (MHz) &
$\Gamma_{sd}/2\pi$ (MHz) &
$C_\textrm{em}$ &
$\mathcal{P}_\textrm{em}$ (\%) \\ \midrule[0.5pt]
\multirow{7}{*}{Neutral   atom} &
\cite{ritter2012elementary} & 5 & 6 & 6 & 2.78 & 73.5$^\dagger$ \\
& \cite{reiserer2013nondestructive, reiserer2014quantum} & 6.7 & 5 & 6 & 5.99 & 85.7 \\
& \cite{chibani2016photon} & 20$^\ddagger$ & 4     & 6     & 66.67 & 98.5 \\
& \cite{morin2019deterministic}             & 4.9   & 5.4   & 6.06  & 2.93  & 74.6 \\
& \cite{daiss2021quantum} A                 & 7.6   & 5     & 6     & 7.70  & 88.5 \\
& \cite{daiss2021quantum} B                 & 7.6   & 5.6   & 6     & 6.88  & 87.3 \\
& \cite{deist2022mid}                       & 2.7   & 1.06  & 6     & 4.58  & 82.1 \\ \midrule[0.5pt]
\multirow{6}{*}{Ion} &
\cite{keller2004continuous} &
0.92 &
2.4 &
1.69 &
0.83 &
45.5 \\
& \cite{steiner2014photon}                  & 1.6   & 50    & 4.22  & 0.05  & 4.6  \\
& \cite{begley2016optimized}                & 0.9   & 0.47  & 22.3  & 0.31  & 23.6 \\
& \cite{krutyanskiy2023entanglement} A      & 0.77  & 0.137 & 21.48 & 0.81  & 44.7 \\
& \cite{krutyanskiy2023entanglement}   B    & 1.2   & 0.14  & 21.48 & 1.92  & 65.7 \\ 
& \cite{takahashi2020strong} 				& 12.3 & 8.2 & 23.0 & 3.21 &	76.2
\\ \bottomrule[1pt]
\end{tabularx}

\bigskip

\begin{tabularx}{0.93\linewidth}{p{0.1\textwidth}cccccc}
\multicolumn{7}{c}{Panel B} \\
\toprule[1pt]
\multirow{2}{*}{$\quad\cdots$} &
\multicolumn{5}{c}{Cavity   loss} &
Combined loss \\
&
$\mathcal{T}_i$ (ppm$^*$) &
$\mathcal{T}_o$ (ppm$^*$) &
$\mathcal{L}$ (ppm$^*$) &
$C_\textrm{cav}$ &
$\mathcal{P}_\textrm{cav}$ (\%) &
$P_\textrm{tot}$ (\%) \\ \midrule[0.5pt]
\multirow{7}{*}{$\quad\cdots$} & 100  & 6          & \textcolor{gray}{5.1} & \textcolor{gray}{9.00} & \textcolor{gray}{90.0} & 43.8 \\
& 95   & \multicolumn{2}{c}{8}   & 11.88 & 92.2 & 62.5 \\
& 17.8 & 2.5        & 11         & 1.32  & 56.9 & 31.4 \\
& $\propto$ 2.4  & \multicolumn{2}{c}{$\propto$ 0.3} & 8.00  & 88.9 & 44.0 \\
& $\propto$ 2.3  & \multicolumn{2}{c}{$\propto$ 0.2} & 11.50  & 92.0 & 66.3 \\
& $\propto$ 2.4  & \multicolumn{2}{c}{$\propto$ 0.4} & 6.00   & 85.7 & 56.0 \\
& -   & -         & -         & -    & -   & -   \\ \midrule[0.5pt]
\multirow{6}{*}{$\quad\cdots$} & 600  & 6          & NR         & \textcolor{rust}{100.00}   & \textcolor{rust}{99.0} & \textcolor{rust}{20.3} \\
& 100  & 10         & 200        & 0.48  & 32.3 & 0.02 \\
& 100  & 5          & \textcolor{gray}{9.9}        & \textcolor{gray}{6.69}  & \textcolor{gray}{87.0}$^\mathsection$ & 4.2  \\
& 13   & 1.3        & \textcolor{gray}{50.7} & \textcolor{gray}{0.25}  & \textcolor{gray}{20.0} & 0.8  \\
& 90   & 2.9  & \textcolor{gray}{22.5} & \textcolor{gray}{3.55}  & \textcolor{gray}{78.0} & 26.3 \\
& 25 & 25 & 75 & 0.25 & 20.0$^\mathparagraph$ & 2.3
\\ \bottomrule[1pt]
\end{tabularx} 
\caption{\label{tab:extendedParameterTable}
	Extended version of Table \mainref{tab:cooperativityTableAbridged} of the main text (split into two panels with consistent row ordering), which includes the additional parameters $g, \g, \Gamma_{sd}$ (all reported over $2\pi$ as is standard) for atomic decay (Panel A) and $\mathcal{T}_i, \mathcal{T}_o, \mathcal{L}$ for cavity loss (Panel B).
	Some references report a combined undesired coupling rate $\mathcal{T}_o + \mathcal{L}$, shown as a centered value under the two corresponding columns.
	These parameter values were found in the accompanying reference(s), translated to our notation (see footnote \ref{symbolNotationFootnote2}), reported to the accuracies given in their respective original work(s), and then used to calculate the cooperativities and probabilities via  Eqs.~(\ref{eq:PemOrabs}) and (\ref{eq:cavloss}). The combined probability of the full emission process at node 1 and symmetrically of absorption at node 2, assuming identical nodes, is $P_\textrm{tot} = (\mathcal{P}_\textrm{em}  \mathcal{P}_\textrm{cav})^2$.
	[We do not consider nor propagate the errors for these values, the cooperativities are shown to two decimal places and the probabilities ($\mathcal{P}_\textrm{em}, \mathcal{P}_\textrm{cav}, P_\textrm{tot}$) are shown to the nearest $0.1\%$ (or 
	the leading nonzero digit)].
	Each of the neutral atom experiments reported uses a single $^{87}$Rb atom in a cavity (accordingly they share the same $\Gamma_{sd}$ value).
	The reported trapped ion experiments use various emitter(s) (in a cavity in each case): Ref.~\citenum{steiner2014photon} uses a single $^{174}$Yb$^+$ atom, Ref.~\citenum{begley2016optimized} use up to 5 $^{40}$Ca$^+$ ions, and the others use a single $^{40}$Ca$^+$ ion. \vspace{2pt} \\
	\footnotesize \textit{Legend and additional notes.} 
	In some references, no value of $\mathcal{L}$ (nor the combined $\mathcal{T}_o + \mathcal{L}$) is given yet a quantity analogous to $\mathcal{P}_\textrm{cav}$ is reported. In these cases we infer the values of $C_\textrm{cav}$ and $\mathcal{L}$ (indicated via \textcolor{gray}{gray} coloring) by backtracking. 
	NR stands for not reported and is used when neither $\mathcal{L}$ nor $\mathcal{P}_\textrm{cav}$ is given, in which case we take $\mathcal{L} = 0$ in the ensuing (over)estimations of $C_\textrm{cav}, P_\textrm{tot}$, and $\mathcal{P}_\textrm{cav}$ (indicated via \textcolor{rust}{brown} coloring). 
	Dashes, -, are used for Ref.~\citenum{deist2022mid}'s cavity parameters as their setup is used for cavity enhanced measurement of one of two optical-tweezer-trapped atoms rather than controlled photon production out an asymmetric cavity.
	The references and table values with NR and - qualifications are omitted in the abridged main text Table \mainref{tab:cooperativityTableAbridged}.
	$*$: The values of $\mathcal{T}_i, \mathcal{T}_o, \mathcal{L}$ are reported in ppm unless otherwise noted using the $\propto$ symbol. In these other cases the respective values reported are the `correct' and `lossy' cavity decay rates $\kappa_c$ and $\kappa_l$ divided by $2 \pi$ in MHz, which satisfy $\g = 2(\kappa_c + \kappa_l)$ [$=2\kappa$ in many references]. Importantly, $\kappa_c \propto \mathcal{T}_i$ and $\kappa_l \propto \mathcal{T}_o + \mathcal{L}$ with the same proportionality constant (which depends on the speed of light and length of the cavity) \cite{vogell2017deterministic,morin2019deterministic} 
	and thus $C_\textrm{cav} = \kappa_c/\kappa_l$. 
	$\dagger$: Ref.~\citenum{ritter2012elementary} reports an analog of $\mathcal{P}_\textrm{em}$, their `photon production efficiency,' to be $60\%$, which is lower than our calculated value, potentially due to their reported $g$ value being a maximum, parameter uncertainties, or other degradations. 
	Using this value for $\mathcal{P}_\textrm{em}$ results in $P_\textrm{tot} =29.2\%$.
	$\ddagger$: We note that Ref.~\citenum{chibani2016photon} is able to obtain a large coupling $g$ 
	yet in a different context than our work of cavity electromagnetically induced transparency 
	(not single photon emission and absorption). 
	Similar emitter values are reported in subsequent work from this group, e.g., Ref.~\citenum{hamsen2017two} have the same emitter decay parameters (though $g/2\pi$ drops slightly 
	due to cavity driving). 
	$\mathsection$: This value of $\mathcal{P}_\textrm{cav}$ is reported in Table I of Ref.~\citenum{vogell2017deterministic}. 
	$\mathparagraph$: The fiber-cavity setup of Ref.~\citenum{takahashi2020strong} is 
	used to achieve a state-of-the-art strong coupling $g$ and large $C\textrm{cav}$ (for ion-cavity-QED experiments). Their setup is not designed for single-photon production preferentially out of one of the mirrors so the corresponding $C_\textrm{cav}$ and $\mathcal{P}_\textrm{cav}$ reported here are strong lower bounds for the achievable values in an asymmetric cavity setup (e.g., for QST or remote entanglement generation). 
	}
\end{table}

\textbf{Strategy 3).}
Generally, photon loss in a transmission line, e.g., fiber, occurs exponentially with length. 
That is, the probability that an initial photon will have not been absorbed after propagating a distance $x$  in the transmission line is  
\be\label{eq:transloss}
P_3(x) = e^{-x/x_\textrm{tl}},
\ee
where $x_\textrm{tl}$ is a (frequency dependent) attenuation distance in a particular transmission line (e.g., see the quantum repeater references mentioned in Sec.~\mainref{subsec:BackgroundAndScope}). 
We simply state these standard results here and refer the reader to Ref.~\citenum{vogell2017deterministic} for further discussion and analysis of such fiber and cavity losses in deterministic QST schemes. 

\textbf{Strategy 4).}
The strength and phase of the drives [laser pulses $G_j(t)$ in our case] used to control the systems can only themselves be controlled up to some precision set by their implementation. 
However, they can typically be controlled on short enough timescales relative to the cavity decay rate, e.g., a pulse duration on the order of $\g^{-1} \sim 1$ $\mu$s is typical in optical systems. Thus, good drive control is typically a modest assumption, yet we reiterate that the relative timing of the drives on sub $\g^{-1}$ time scales is also crucial.
See Ref.~\citenum{morin2019deterministic} for an example of the precision with which such drives can be controlled on sub-$\mu$s time scales.
There they report a photon production fidelity $\mathcal{F}$, 
given by the norm squared temporal overlap of the actual and target temporal modes, of $90 \%$, which they suspect could be increased to $98\%$ by suitably tuning the frequency of the local oscillator used.
This indicates that errors in the drives of order $1 - \mathcal{F} \lesssim 1\%$, in terms of producing a photon wave packet with a particular desired shape and realizable duration (based on cavity and system parameters), should be achievable in a well-engineered setup, at least for atom in a cavity type nodes. 
Ref.~\citenum{keller2004continuous} demonstrated a similar ability to generate single photons of $\sim 1-5$ $\mu$s widths with controllable envelope shape (on a sub $\mu$s timescale) using a trapped ion in a cavity yet they did not report target-actual state fidelities. 

\textbf{Strategy 5).}
Dispersion in the channel leads to a change in the shape and phase of the photon wave packet $\Psi(t)$, its impact can thus be accounted for by Eq.~(\ref{eq:PsuccSynopsis}). 
Additionally, the wave packet will experience a group time delay due to its propagation in the transmission line which we assume is accounted for in the relative timing of $U$ and of the drives $G_j(t)$.
Material dispersion in the transmission line occurs due to the different frequency components of a wave packet traveling at different speeds due to a frequency dependent index of refraction $n_\textrm{tl}(\om)$.
Hence, its impact can be lessened by using photon wave packets with small frequency bandwidths 
compared to the frequency scale over which $n_\textrm{tl}(\om)$ is approximately linear (relative to a given central frequency) (see Sec.~\hyperref[subsec:generalNodeH]{B1}). 
Accordingly, strategy 5), like 4), is less appropriate in cases where fast varying drives are desired (as we have alluded to this is an interested regime for achieving faster rates yet it comes with a multitude of complications).

If the dispersion is known and systematic (say by characterizing the channel) it can be (at least partially) compensated for by adjusting $U$. For instance, the stretching parameter $\xi$ can be used to mitigate wave packet broadening issues due to dispersion (see Sec.~\hyperref[subsec:timeReversal]{A4}).
For a given implementation, accounting for the impacts of dispersion is crucial for matching the actual (dispersed) wave packet $\Psi(t)$ to the target shape $\Phi(t)$, as is necessary to obtain $\Psucc$ near $1$. 
However, such analysis is beyond the scope of this work and we therefore recommend its further consideration by the community, both generally (e.g., see Ref.~\citenum{penas2023improving}) as well as in particular implementations.

\textbf{Summary.} 
Thus, $\Psucc$ of the main text, which accounts for the impact of the unitary transformation, can be modified to account for the impacts of strategies 1-3) through the post hoc multiplication of their respective survival probabilities as given by Eq.~(\mainref{eq:tildePsuccGen}). 
In cases where strategies 4-5) are not appropriate, the actual wave packet $\Psi$ will be intimately effected and the corresponding impacts must be accounted for in  $\Psucc$ itself, similar to how we analyzed errors in $U$.
A full analysis accounting for the impact of additional structure (energy levels) of the emitter in strategy 1), laser errors in strategy 4), and channel induced distortion in strategy 5) would require additional model details about the nodes being used and the drives used to control them and so is not considered here. Moreover, there may be additional degradation factors depending on the precise setup, e.g., loss due to a nonideal cavity-transmission-line mode-matching. 

\phantomsection
\subsection*{2.~First ECZ protocol summary}\label{subsec:ECZ97Summary}
\begin{figure*}[h]
\includegraphics[width=0.5\linewidth]{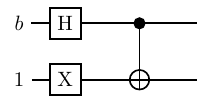}
\caption{The gate $U_{b1}$ as a circuit diagram.}
\label{fig:CNOTcirc}
\end{figure*}

We will now summarize the earlier ECZ protocol \cite{van1997ideal}. We include this in the SM for reference and so that we can analyze the expected number of repetitions of a primitive of the protocol, $E[n]$, necessary for successful QST. This is accomplished by explicitly  tracking the normalization factors.
To our knowledge, this calculation of $E[n]$ and our commentary is novel, though we want to emphasize that the protocol itself is fully credited to Ref.~\citenum{van1997ideal}.
We use their labeling for the atoms, cavity 1 contains two atoms 1 and $b$ (for \textit{backup}) while cavity 2 contains two atoms 2 and $a$ (for \textit{auxiliary}). The cavities are coupled to a transmission line in a unidirectional manner analogous to our scheme that allows a photon emitted by system 1 to propagate to and potentially be absorbed by system 2. Their protocol consist of five steps:

(i) ``Local redundant encoding.'' The state we want to transfer is initially encoded in atom 1 as $\ket{\psi_1} = c_0 \ket{g_1} + c_1 \ket{e_1}$ with atom $b$ in $\ket{g_b}$ and the  system 2 atoms both in the auxiliary level, $\ket{r_2} \ket{r_a}$, such that the total state is $\ket{\psi_\textrm{tot}} = \ket{g_b} \ket{\psi_1} \ket{r_2} \ket{r_a}$. 
Then we perform an entangling operation $U_{b1} = \textrm{CNOT}_{b1} (H_b \otimes X_1)$ (with $b$ acting as the control qubit for the CNOT, as depicted in Fig.~\ref{fig:CNOTcirc}) on the system 1 qubits, where $H$ and $X$ are the Hadamard and not gates. This yields the state
\begin{equation}\label{eq:PsiTotI}
\ket{\psi_\textrm{tot}}  \overset{U_{b1}~}{\longrightarrow} 
\frac{1}{\sqrt{2} }\p{
\ket{{\overbar\psi}_b} \ket{g_1}
+ \ket{\psi_b} \ket{e_1}
} \ket{r_2} \ket{r_a}
= \frac{1}{\sqrt{2} }\p{
\ket{{\overbar\psi}_b}  \ket{\Phi_i} 
+ \ket{\psi_b} \ket{\Psi_i} 
} 
\end{equation}
with $\ket{\overbar\psi_j} = X \ket{\psi_j} = c_1\ket{g_j} + c_0\ket{e_j}$ for $j$ one of the atom labels, $\ket{\Phi_i} = \ket{g_1} \ket{r_2} \ket{r_a}$, and $\ket{\Psi_i} = \ket{e_1} \ket{r_2} \ket{r_a}$ with $i$ denoting the step of the protocol. We will track how these states $\ket{\Psi}$ and $\ket{\Phi}$ (not to be confused with our actual and ideal wave packets),  evolve over the steps of the protocol, notably they will stay orthogonal $\braket{\Psi}{\Phi} = 0$ for each step (denoted in their subscript). Here atom $b$ at system 1 acts as a form of redundancy in case an error occurs, it does not explicitly participate in the transmissions.

(ii) ``Transmission from atom 1 to 2.'' 
We now implement our QST scheme between atom 1 and atom 2. The possibility of photon loss and spontaneous emission to other atomic levels during the transfer leads to dissipative dynamics which can be described in the quantum trajectory formalism \cite{dalibard1992wave,gardiner1992wave,molmer93}. In this description, for long times the system will either undergo a dissipative quantum jump due to loss, causing the state of the atoms to leave the desired subspace, with probability $P_\textrm{jump}$
or it will undergo smooth evolution (no quantum jump occurs) with probability $1 - P_\textrm{jump}$.
If a quantum jump does not occur, the states evolve according to an effective Hamiltonian $H_\textrm{eff}$ for large times according to Eq.~(\mainref{eq:HeffMap}) in the main text, which can be restated in the language of ECZ yielding their Eq.~(3) \cite{van1997ideal}
\be\label{eq:HeffMapECZ}
\begin{matrix}
\ket{g_1} \ket{r_2} \\
\ket{e_1} \ket{r_2}
\end{matrix}
\overset{H_\textrm{eff}~}{\longrightarrow} 
\begin{matrix}
\alpha \ket{g_1} \ket{r_2}\\ 
\beta \ket{r_1} \ket{e_2} + \Upsilon_1 \ket{r_1} \ket{r_2} + \Upsilon_2 \ket{e_1} \ket{r_2}
\end{matrix}.
\ee
(Here we try to use Ref.~\citenum{van1997ideal}'s notation including state labels, except we use $\Upsilon_{1,2}$ instead of $\g_{1,2}$ to avoid confusion with our node-channel couplings $\g_j$).
The amplitudes $\alpha$ and $\beta$ are ideally 1, their departure from unity corresponds to amplitude damping and phase errors.
An error in $\alpha$ would typically be a phase error in the ground states $\ket{g}$ and $\ket{r}$. This could be a (likely small) phase error due to expected imperfections in accounting for the phases imparted to these states by the ac Stark shifts during the Raman process or due to magnetic field noise (e.g., in a Zeeman or hyperfine qubit). 
Additionally, (typically with much lower probability) $\alpha$ errors could occur as a result of more dramatic effects such as transition to a state out of the qubit manifold or auxiliary state $\{\ket{g}, \ket{e}, \ket{r} \}$, e.g., by spontaneous emission, followed by repumping to $\ket{g_1} \ket{r_2}$ \cite{van1998photonic}. 
Thus, $\alpha$ will typically be a pure phase though its value depends strongly on which states are used to encode the material qubit. 
%
The amplitude $\beta$ accounts for an excitation being successfully transferred. An error in $\beta$ could occur due to loss of the intermediate photon or an error in the local oscillators used to track the phase.
Meanwhile, $\Upsilon_2$ corresponds to an error where the excitation to be transferred undesirably remains at system 1, say due to an incorrect first Raman pulse that does not induce the transition $\ket{e_1} \ra \ket{g_1}$.
Finally, $\Upsilon_1$ corresponds to photon loss during the transmission (photon absorption, leaking out of cavity 2, spontaneous emission during the first Raman process, an incorrect second Raman pulse that does not induce the transition $\ket{r_2} \ra \ket{e_2}$, etc.). 

For this transmission attempt we will have 
$|\beta|^2 + |\Upsilon_1|^2 + |\Upsilon_2|^2 = 1$, 
where we assume that we either stay in the single-excitation subspace ($\beta$ and $\Upsilon_2$ terms) or atom 1 undergoes the Raman transition from $\ket{e_1}$ to $\ket{r_1}$ yet the photon is lost so atom 2 stays in $\ket{r_2}$ ($\Upsilon_1$ term), which corresponds to a quantum jump mechanism.
Note that $|\alpha| \leq 1$ due to spontaneous emission to other levels ($\alpha$ is treated differently than $\beta$ and $\Upsilon_{1,2}$ as said other levels are outside the subspace we have considered so we do not write their amplitudes explicitly).
In this case where no quantum jump occurs, we need to renormalize the entire state by dividing by $\mathcal{N}_{ii} = \sqrt{ (1 + |\alpha|^2)/2 }$. 
Thus, under the above map for a transmission attempt, the states become
\be
\ket{\Phi_{ii}} = \alpha' \ket{g_1} \ket{r_2} \ket{r_a}
\ee
and
\be\label{eq:Psi_ii}
\ket{\Psi_{ii}} = \p{\beta' \ket{r_1} \ket{e_2} + \Upsilon_1' \ket{r_1} \ket{r_2} + \Upsilon_2' \ket{e_1} \ket{r_2}} \ket{r_a}
\ee
where we define the primed amplitudes as their original counterparts renormalized, e.g., $\alpha' = \alpha/\mathcal{N}_{ii}$ and likewise for $\beta'$, $\Upsilon_1'$, and $\Upsilon_2'$, such that $|\alpha'|^2 + |\beta'|^2 + |\Upsilon_1'|^2 + |\Upsilon_2'|^2 = 2$  [with the normalization due to the convention of Eq.~(\ref{eq:PsiTotI})]. 

Alternatively, a quantum jump can occur due to either the loss of an excitation with probability\footnote{
Mapping this to our earlier notation $|\beta| \ra |\alpha_2(t_e)|$ and $|\Upsilon_2| \ra |\alpha_1(t_e)|$, we see that this is the same jump probability one would obtain from equation (52) of Ref.~\citenum{randles2023quantum}, accounting for the possibility that atom 1 stays excited, which contributes to an error (via $\Upsilon_2$) but not an jump.
}
$P_\textrm{jump}^{(1)} = 1 - \p{|\beta|^2 + |\Upsilon_2|^2} = |\Upsilon_1|^2$
or due to amplitude damping of the ground states 
with probability $P_\textrm{jump}^{(0)}  = 1 - |\alpha|^2$. 
Due to original amplitudes of the initial state in Eq.~(\ref{eq:PsiTotI}), where effectively $|c_e| = |c_g| = 1/\sqrt{2}$ in our notation, we have that 
\be\label{eq:PJumpCoef}
P_\textrm{jump} 
= \frac{1}{2} \p{ P_\textrm{jump}^{(0)} + P_\textrm{jump}^{(1)} }
=  \frac{1}{2} \p{ 1 - |\alpha|^2 + |\Upsilon_1|^2 }.
\ee
As noted by Ref.~\citenum{van1997ideal}, such a quantum jump is a special case of Eq.~(\ref{eq:HeffMapECZ}) with $\alpha = \beta = \Upsilon_2 = 0$, the corresponding error can be detected in step (iv) (detailed below) \cite{van1997ideal}. Importantly, the protocol still works if a quantum jump occurs during one transmission but not the other; this is notable because the amplitudes will be different in this case yet the error can still be detected.

Next after the transmission attempt, we measure whether atom 1 is in state $\ket{e_1}$. If yes, an error has occurred where no photon has been emitted, and the state collapses to
\be
\ket{\psi_\textrm{tot}} \ra \ket{\psi_b} \ket{e_1} \ket{r_2} \ket{r_a}.
\ee
In this case, the backup atom $b$ has recovered the original state so, after resetting the other atoms, steps (i-ii) can then be repeated (with the roles of atoms $1$ and $b$ switched) until no $\Upsilon_2$ error occurs. The probability of having to stop and repeat the protocol at step (ii) is
\be\label{eq:Pii}
P_{ii} = \frac{|\Upsilon_2'|^2}{2}
= \frac{|\Upsilon_2|^2}{1 + |\alpha|^2},	  
\ee
which should be very small (under appropriate driving, $G_1$).
Otherwise, if atom 1 is not found to be in $\ket{e_1}$, then the corresponding term in Eq.~(\ref{eq:Psi_ii}) is projected out and the state is renormalized, e.g., $\alpha' \ra \alpha'' = \alpha'/\sqrt{1 - \tfrac{1}{2} |\Upsilon_2'|^2}$ with $\beta''$ and $\Upsilon_1''$ defined similarly.

(iii)  ``Symmetrization.'' Perform a local operation on atom 1 that maps $\ket{r_1} \ra \ket{g_1}$ and $\ket{g_1} \ra \ket{e_1}$, such that
\be
\ket{\Phi_{iii}} = \alpha'' \ket{e_1} \ket{r_2} \ket{r_a}
\ee
and
\be\label{eq:Psi_iii}
\ket{\Psi_{iii}} = \ket{g_1}  \p{\beta'' \ket{e_2} + \Upsilon_1'' \ket{r_2}} \ket{r_a}.
\ee

(iv) ``Transmission from atom $1$ to $a$.'' 
The states evolve according to the map above, Eq.~(\ref{eq:HeffMapECZ}), with $2 \ra a$, with potentially different amplitudes (denoted by tildes). If no quantum jump occurs, with probability $1 - \tilde{P}_\textrm{jump}$, this yields the renormalized states
\be
\ket{\Phi_{iv}} = \alpha''
\p{\tilde\beta' \ket{r_1} \ket{e_a} + \tilde\Upsilon_1' \ket{r_1} \ket{r_a} + \tilde\Upsilon_2' \ket{e_1} \ket{r_a}}
\ket{r_2} 
\ee
and
\be\label{eq:Psi_iv}
\ket{\Psi_{iv}} =  \tilde\alpha' \ket{\Psi_{iii}} ,
\ee
where, as in step (ii), we define the normalized amplitudes by dividing the original tilded ones by
\be
\mathcal{N}_{iv}  = \sqrt{ \frac{  |\alpha|^2 + |\tilde\alpha|^2 \p{ 1  - |\Upsilon_2|^2 } }{1 + |\alpha|^2 - |\Upsilon_2|^2} },
\ee
e.g., $\tilde\beta' = \tilde\beta/\mathcal{N}_{iv}$.
Alternatively, a jump will occur with a probability analogous to Eq.~(\ref{eq:PJumpCoef}) with $\tilde{P}_\textrm{jump}^{(1)} = |\tilde\Upsilon_1|^2 |\alpha''|^2$
and
$\tilde{P}_\textrm{jump}^{(0)} = \p{1 - |\tilde\alpha|^2} \p{|\beta''|^2 + |\Upsilon_1''|^2}$
(which take on their previous form multiplied by the relative probability of actually being in that state) so that 
\be\label{eq:tildePJump}
\tilde{P}_\textrm{jump} = \frac{1}{2} \br{\p{1 - |\tilde\alpha|^2} \p{|\beta''|^2 + |\Upsilon_1''|^2}
+  |\tilde\Upsilon_1|^2 |\alpha''|^2}.
\ee
Then we measure if the atoms at node 2 are in the state $\ket{r_2} \ket{r_a}$.\footnote{
Note Ref.~\citenum{van1997ideal} says to measure if atom 1 is in $\ket{e_1}$ before this, though we do not need to as the same error can be detected by this measurement. 
} 
If yes, the states are (up to a common normalization factor)
\be
\ket{\Phi_{iv}} \ra \alpha
\p{\tilde\Upsilon_1 \ket{r_1} + \tilde\Upsilon_2 \ket{e_1}}
\ket{r_2} \ket{r_a}
\ee
and
\be
\ket{\Psi_{iv}} \ra  \tilde\alpha \Upsilon_1 \ket{g_1} \ket{r_2} \ket{r_a}
\ee
and an error has occurred and we measure the state of atom 1. If $\ket{g_1}$ is measured, atom $b$ is in the state $\ket{\psi_b}$, otherwise if we measure $\ket{r_1}$ or $\ket{e_1}$, then atom $b$ is in the state $\ket{\overbar{\psi}_b}$ and a NOT gate, $X$, should be performed. Either way, atom $b$ ends in the state to be transferred $\ket{\psi_b}$, and the process should be repeated (after resetting the other atoms). 
Similar to step (ii), the probability having to stop and repeat the protocol here is
\be\label{eq:Piv}
P_{iv} = \frac{1}{2} \br{  |\alpha''|^2 (|\tilde\Upsilon_1'|^2 + |\tilde\Upsilon_2'|^2) + |\tilde\alpha'|^2 |\Upsilon_1''|^2 }.
\ee
If we do not measure $\ket{r_2} \ket{r_a}$, the states are
\be\label{eq:Phi4}
\ket{\Phi_{iv}} \ra \alpha \tilde\beta \ket{r_1} \ket{r_2} \ket{e_a}
\ee
and
\be\label{eq:Psi4}
\ket{\Psi_{iv}} \ra  \tilde\alpha \beta \ket{g_1}  \ket{e_2} \ket{r_a}
\ee
(up to a common normalization factor). 
Hence under the assumption that the $\alpha$ and $\beta$ errors are systematic and hence the same for both transmissions, i.e., $\alpha = \tilde\alpha$ and $\beta = \tilde\beta$, we have (accounting for the normalization) 
\be\label{eq:systematicPsiTot}
\ket{\psi_\textrm{tot}} 
=  \frac{e^{i \arg(\alpha \beta)}}{\sqrt{2}}
\p{
\ket{{\overbar\psi}_b}   \ket{r_1} \ket{r_2} \ket{e_a}
+ \ket{\psi_b} \ket{g_1}  \ket{e_2} \ket{r_a}
},
\ee
where we can neglect the phase as it is global. 

(v) ``Teleportation.'' We first perform a local operation that takes $\ket{r_2} \ra \ket{g_2}$. Then we make several measurements of the state of: (a) atom $b$ in the standard basis, (b) atom $1$ in the $\ket{\pm_1} = (\ket{g_1} \pm \ket{r_1})/\sqrt{2}$ basis, and (c) atom $a$ in the $\ket{\pm_a} = (\ket{e_a} \pm \ket{r_a})/\sqrt{2}$ basis. 
We summarize the results in a piecewise function, where in each line the right-most part, after the comma, summarizes the measurement outcomes
\be
\ket{\psi_\textrm{tot}} \ra
\begin{cases}
c_1 \ket{g_2}  + c_0 \ket{e_2}, &  \ket{g_b} \ket{+_1} \ket{+_a} \\
c_1 \ket{g_2} - c_0 \ket{e_2}, &  \ket{g_b} \ket{+_1} \ket{-_a} \\
- c_1  \ket{g_2}  + c_0  \ket{e_2}, & \ket{g_b} \ket{-_1} \ket{+_a} \\
- c_1  \ket{g_2} -c_0  \ket{e_2},   &  \ket{g_b} \ket{-_1} \ket{-_a} \\
c_0  \ket{g_2} + c_1 \ket{e_2}, & \ket{e_b}  \ket{+_1} \ket{+_a} \\
c_0  \ket{g_2} - c_1 \ket{e_2}, & \ket{e_b}  \ket{+_1} \ket{-_a} \\
-c_0  \ket{g_2} + c_1 \ket{e_2},  & \ket{e_b} \ket{-_1} \ket{+_a} \\
-c_0  \ket{g_2} - c_1 \ket{e_2},  & \ket{e_b} \ket{-_1} \ket{-_a}
\end{cases}.
\ee
These resulting states of atom 2 can clearly all be changed to the desired state $\ket{\psi_2} = c_0  \ket{g_2} + c_1 \ket{e_2}$ using standard local operations. Hence, we have effectively teleported the state from atom 1 to 2, even in the presence of errors and  noise.

\phantomsection
\subsection*{3.~Scope of ECZ protocols}\label{subsec:errorCorrectionScope}
This protocol \cite{van1997ideal} can lead to ideal QST after an appropriate number of process repetitions (trials), assuming the errors in the transmission are systematic within a given trial and the system channel interaction is described by a Markovian process. 
Specifically, it is assumed that the parameters $\alpha$ and $\beta$, which account for phase shift and amplitude damping of the atomic qubit encoding, are the same in both transmissions in a given trial, e.g., steps (ii) and (iv).
If these parameters are not systemic and hence vary in a given trial, the fidelity of the resulting tilded state $\ket{\tilde\psi_\textrm{tot}}$ [with $\alpha \neq \tilde\alpha$ and $\beta \neq \tilde\beta$ in Eqs.~(\ref{eq:Phi4}) and (\ref{eq:Psi4}) yet accounting for normalization] with the target state $\ket{\psi_\textrm{tot}}$ of Eq.~(\ref{eq:systematicPsiTot}) is
\be
\mathcal{F}_\textrm{tot} = \left| \langle \psi_\textrm{tot} \ket{\tilde{\psi}_\textrm{tot}}  \right|^2
= \frac{\left| 
\alpha \tilde\beta + \tilde\alpha \beta \right|^2}{2 (|\tilde\alpha|^2 |\beta|^2 + |\alpha|^2 |\tilde\beta|^2)}.
\ee
This total state fidelity limits the ultimate fidelity of the state transferred in step (v).
To quantify the discrepancy we define $\tilde\alpha = r_a e^{i\delta_a} \alpha$, $\tilde\beta = r_b e^{i \delta_b} \beta$, and $\delta_{ab} \equiv \delta_a - \delta_b$ such that
\be
\mathcal{F}_\textrm{tot} 
= \frac{ \left| 
r_b  + r_a e^{i\delta_{ab}} \right|^2}{2 (r_a^2  + r_b^2 )}
= \frac{1}{2} + \frac{r_a r_b }{r_a^2  + r_b^2} \cos(\delta_{ab}),
\ee
which oscillates about $1/2$ as a function of $\delta_{ab}$ with amplitude $0 \leq r_a r_b/(r_a^2 +r_b^2) \leq 1/2$ with the maximum amplitude occurring at $r_a = r_b$ (note $r_a \approx 1$ typically).
This clearly emphasizes the need to suppress phase errors as, no matter what $r_a$ and $r_b$ are, this fidelity is maximum for $\delta_{ab} = 0$ and	
averages to the futile value of $1/2$ over all phase errors $0 \leq \delta_{ab} < 2 \pi$. Importantly, one only needs $\delta_{ab}$ to be small over subsequent transmissions in a given trial, not necessarily between trials.
Thus, in cases where systematic errors are dominant 
$\mathcal{F}_\textrm{tot} \approx 1$ and this protocol is appropriate.

Note that the later ECZ protocol \cite{van1998photonic} extends the former by being able to correct for more general random errors, namely $\alpha \neq \tilde{\alpha}$ or $\beta \neq \tilde{\beta}$, in a noisy channel with potentially non-Markovian decoherence.
To make this possible, one has to repeat  certain `purification' process that causes the joint state of one qubit at node 1 and one at node 2 to iteratively approach a target entangled state (e.g., a Bell state) with a fidelity that should exponentially approach 1. Then quantum state teleportation can be used to implement QST between the nodes \cite{bennett1993teleporting,bennett1996purification}.
Note that both of the ECZ protocols assume that local operations can be performed perfectly on the ground states of a single atom (emitter more generally), including the auxiliary ground state for Ref.~\citenum{van1997ideal}, and for entanglement operations on two atoms in a given node.
This condition of error-free local gates could of course be approached using quantum error correction schemes at the nodes themselves. However, this would be at the cost of a large overhead in physical qubits with long coherence times that are acted on by high fidelity gates (i.e., in the  fault-tolerant limit), which even if possible would likely make interfacing with the communication channel more difficult. Hence, as discussed in the Sec.~\mainref{sec:Discussion}, a compromise needs to be made between the potential benefit of an error correction protocol and the additional overhead and errors that will accompany it.

Now that we have kept track of the probabilities of having to restart the earlier ECZ protocol \cite{van1997ideal} due to a quantum jump occurring or after detecting an error in the measurements of steps (ii) and (iv), we can compute the expected number of trials of the protocol $E[n]$.
The probability that an entire trial is successful is
\be
P_s = (1 - P_\textrm{jump})   (1 - P_{ii}) (1 - \tilde{P}_\textrm{jump}) (1 - P_{iv})
\ee
and the corresponding probability of a given trial failing (having to be stopped and repeated) is $P_f = 1 - P_s$. Thus the expected number of trials (repetitions of the protocol) until success is the standard
\be
E[n] =  P_s \sum_{n=1}^{\infty}  n P_f^{n-1}
= \frac{P_s}{(1-P_f)^2} 
= \frac{1}{P_s}.
\ee
[In the language of statistics, the probability distribution underlying such a process is called the geometric distribution (assuming independent trials with the same $P_s$) and its properties are well known, e.g., its standard deviation is $\sqrt{1-P_s}/P_s = E[n] \sqrt{1 - 1/E[n]}$.]
Now our task is to carefully compute $P_s$; to do so we must keep track of the scaling of the renormalized variables. Doing so, using Eq.~(\ref{eq:PJumpCoef}), (\ref{eq:Pii}), (\ref{eq:tildePJump}), and (\ref{eq:Piv}), and proceeding under the assumption that all the errors are systematic and hence the same for both transmissions (including the $\Upsilon_{1,2}$ ones for simplicity, though this is not strictly necessary), we have 
\begin{align}
P_s &= \frac{1}{2} \p{ 1 + |\alpha|^2 - |\Upsilon_1|^2 } 
\p{1 - \frac{|\Upsilon_2|^2}{1 + |\alpha|^2}  } 
\p{ 1 -  \frac{ 1 -  |\beta|^2 |\alpha|^2  - |\Upsilon_2|^2  }{1 + |\alpha|^2 - |\Upsilon_2|^2} }
\p{1 - \frac{ 2 |\Upsilon_1|^2 + |\Upsilon_2|^2  }{ 2 - |\Upsilon_2|^2  } } \nonumber \\
&=  \frac {|\alpha|^2 |\beta|^2 (1 + |\beta|^2) (1 + |\alpha|^2 - |\Upsilon_1|^2) }{(1 + |\alpha|^2 ) (2 -  |\Upsilon_2|^2 )},
\end{align}
where we used the relation $|\beta|^2 + |\Upsilon_1|^2 + |\Upsilon_2|^2 = 1$. Hence, in this systematic case
\be
E[n] = \frac{1}{P_s} 
=  \frac{(1 + |\alpha|^2 ) (2 -  |\Upsilon_2|^2 )}{|\alpha|^2 |\beta|^2 (1 + |\beta|^2) (1 + |\alpha|^2 - |\Upsilon_1|^2) }.
\ee
If we take $|\alpha| = 1$, as was the case in the main text, we find
\be
P_s = \frac{ |\beta|^2 \p{1 +  |\beta|^2 } \p{ 2 - |\Upsilon_1|^2  }  }{ 2 \p{2 - |\Upsilon_2|^2}},
\ee
which is clearly zero only if $\beta = 0$ as one may expect. 
Moreover, if we assume that a photon is already created at node 1, i.e., there are no errors in the first Raman pulse, then $\Upsilon_2 = 0$ so the above expression simplifies to
\be
P_s = \frac{ |\beta|^2 \p{1 +  |\beta|^2 }^2  }{ 4 }. 
\ee

Although we expect to be dominated $\Upsilon_1$ errors with $|\alpha| \approx 1$ and $\Upsilon_2 \approx 0$, suppose that each variable suffers an order $\varepsilon$ error. Specifically, we take
$|\Upsilon_1|^2 = x \varepsilon$
and
$|\Upsilon_2|^2 = (1 - x) \varepsilon$
with $ 0 \leq x \leq 1$ such that $|\beta|^2 = 1 - \varepsilon$
and for simplicity we also take $|\alpha|^2 = 1 - \varepsilon$ as an effectively worst case scenario (as generically $\beta$ errors will be much more significant than those in $\alpha$). In this model, the lowest values of $P_s$ all occur for $x=1$ and in that worst case we obtain
\be
E[n] =  \frac{1}{ (1 - \varepsilon)^3 },
\ee
which is small (near $1$) for $\varepsilon$ near $0$ yet diverges toward $+ \infty$ as $\varepsilon \ra 1$. This divergence is similar to the case considered in the main text with $|\alpha| =1$, see Eq.~(\mainref{eq:WorstEnForUnitAl}), yet occurs more rapidly (as one would expect, as we are considering a case with both $\alpha$ and $\beta$ amplitude damping errors). 

\newpage
\twocolumngrid

\end{document}